\documentclass[iop,appendixfloats]{emulateapj}

\newcommand{\figurepath}{.}

\newcommand{\bit}[1]{ \textbf{\emph{#1}} }

\newcommand{\dS}{d\bit{S}}

\newcommand{\ri}{r_{\rm 0}}
\newcommand{\Ri}{R_{\rm 0}}

\newcommand{\Mr}{M_{R}}
\newcommand{\Mth}{M_{\theta}}

\newcommand{\Macc}{\dot{M}_{\rm acc}}
\newcommand{\Mej}{\dot{M}_{\rm ej}}

\newcommand{\Jacc}{\dot{J}_{\rm acc}}
\newcommand{\Jjet}{\dot{J}_{\rm jet}}
\newcommand{\JJkin}{\dot{J}_{\rm jet,kin}}
\newcommand{\JJmag}{\dot{J}_{\rm jet,mag}}
\newcommand{\JAkin}{\dot{J}_{\rm acc,kin}}
\newcommand{\JAmag}{\dot{J}_{\rm acc,mag}}

\newcommand{\Eacc}{\dot{E}_{\rm acc}}
\newcommand{\Ejet}{\dot{E}_{\rm jet}}
\newcommand{\EJkin}{\dot{E}_{\rm jet,kin}}
\newcommand{\EJmag}{\dot{E}_{\rm jet,mag}}

\newcommand{\EJthm}{\dot{E}_{\rm jet,thm}}
\newcommand{\EJgrv}{\dot{E}_{\rm jet,grv}}
\newcommand{\EAkin}{\dot{E}_{\rm acc,kin}}
\newcommand{\EAmec}{\dot{E}_{\rm acc,mec}}
\newcommand{\EAmag}{\dot{E}_{\rm acc,mag}}
\newcommand{\EAthm}{\dot{E}_{\rm acc,thm}}
\newcommand{\EAgrv}{\dot{E}_{\rm acc,grv}}

\newcommand{\Msun}{{M}_{\odot}}

\newcommand{\Rschw}{{\rm R}_{\rm S}}

\newcommand{\etatens}{\overline{\overline{\eta}}}

\newcommand{\Grav}{\Phi_{\rm g}}

\newcommand{\Cs}{C_{\rm s}}
\newcommand{\CsT}{C_{\rm s}^{\rm T}}

\newcommand{\Vtot}{\bit{V}}

\newcommand{\Vp}{V_{\rm P}}
\newcommand{\Va}{V_{\rm A}}
\newcommand{\Vr}{V_{R}}
\newcommand{\Vth}{V_{\rm \theta}}
\newcommand{\Vphi}{V_{\rm \phi}}
\newcommand{\Vk}{V_{\rm K}}

\newcommand{\Btot}{\bit{B}}
\newcommand{\Bp}{B_{\rm P}}
\newcommand{\Br}{B_{\rm R}}
\newcommand{\Bth}{B_{ \theta}}
\newcommand{\Bphi}{B_{ \phi}}

\newcommand{\Jtot}{\bit{J}}

\newcommand{\Jr}{J_{ R}}
\newcommand{\Jphi}{J_{ \phi}}
\newcommand{\Jth}{J_{ \theta}}

\newcommand{\Fphi}{F_{\rm \phi}}

\newcommand{\ass}{{\alpha}_{\rm ssm}}
\newcommand{\am}{{\alpha}_{\rm m}}

\newcommand{\Ht}{ H_{\rm T} }

\newcommand{\Fe}{ F_{\rm \eta} }

\newcommand{\MUzero}{ \mu_{\rm 0} }
\newcommand{\MUact}{ \mu_{\rm act} }

\newcommand{\Ep}{ \eta_{\rm P}}
\newcommand{\Et}{ \eta_{\rm T}}

\newcommand{\AU}{ {\rm AU}}
\newcommand{\AUyso}{ {\rm 0.1}}
\newcommand{\AUbd} { {\rm 0.01}}

\newcommand{\YSO}{ {\rm (YSO)}}
\newcommand{\BD}{ {\rm (BD)}}
\newcommand{\AGN}{ {\rm (AGN)}}

\newcommand{\Myso}{ \left( \frac{M}{\Msun} \right) }

\newcommand{\Magn}{ \left( \frac{M}{10^8 \Msun} \right) }

\newcommand{\Ryso}{ \left( \frac{\Ri}{\AUyso \AU}  \right) }

\newcommand{\Ragn}{ \left( \frac{\Ri}{10 \Rschw} \right) }

\newcommand{\Vx}{{\rm km\,s^{-1}}}
\newcommand{\DNx}{\rm g\,cm^{-3}}
\newcommand{\Mx}{ {\rm \Msun\,yr^{-1}} }
\newcommand{\Jx}{ {\rm dyne\,cm} }
\newcommand{\Ex}{ {\rm erg\,s^{-1}} }

\newcommand{\MUx}{ \left(\frac{\mu}{0.1}\right) }
\newcommand{\EPSx}{ \left(\frac{\epsilon}{0.1}\right) }

\newcommand{\DNyso}{ \left( \frac{\rho_0}{10^{-10} \rm \frac{g}{cm^{3}}}\right) }

\newcommand{\DNagn}{ \left(\frac{\rho_0}{10^{-12} \rm \frac{g}{cm^{3}}}\right) }

\newbox\grsign \setbox\grsign=\hbox{$>$}
\newdimen\grdimen \grdimen=\ht\grsign
\newbox\laxbox \newbox\gaxbox
\setbox\gaxbox=\hbox{\raise.5ex\hbox{$>$}\llap
     {\lower.5ex\hbox{$\sim$}}}\ht1=\grdimen\dp1=0pt
\setbox\laxbox=\hbox{\raise.5ex\hbox{$<$}\llap
     {\lower.5ex\hbox{$\sim$}}}\ht2=\grdimen\dp2=0pt
\newcommand{\gax}{$\mathrel{\copy\gaxbox}$}

\shorttitle{Disks and Jets}
\shortauthors{Stepanovs and Fendt}

\usepackage{ctable} 
\usepackage{epstopdf}
\begin{document}
\title{Modelling MHD accretion-ejection - from the launching area to propagation scales}

%

\author{Deniss Stepanovs, Christian Fendt}
\affil{Max Planck Institute for Astronomy, K\"onigstuhl 17, D-69117 Heidelberg, Germany}
\email{stepanovs@mpia.de, fendt@mpia.de}

\begin{abstract}
We present results of axisymmetric magnetohydrodynamic (MHD) simulations investigating the 
launching of jets and outflows from a magnetically diffusive accretion disk.
The time evolution of the disk structure is self-consistently taken into account.
In contrast to previous works we have applied {\em spherical} coordinates for the numerical
grid, implying substantial benefits concerning the numerical resolution and the stability
of the simulation.
Thanks to the new setup we were able to run simulations for more than 150,000 dynamical times 
on a domain extending 1500 inner disk radii with a resolution of up to 24 cells per 
disk height in the inner disk.
Depending on the disk magnetization, jet launching occurs in two different but complementary
regimes - jets driven predominantly by centrifugal or magnetic forces. 
These regimes differ in the ejection efficiency concerning mass, energy and angular momentum.
We show that it is the {\em actual} disk magnetization and not so much the initial 
magnetization which describes the disk-jet evolution best.
Considering the actual disk magnetization we also find that simulations starting with different
initial magnetization evolve in a similar - typical - way as due to advection and diffusion
the magnetic flux in the disk evolves in time.
Exploring a new, modified diffusivity model we confirm the self-similar structure of the
global jet-launching disk, obtaining power laws for the radial profiles of the disk physical 
variables such as density, magnetic field strength, or accretion velocity.
\end{abstract}

\section{Introduction}
%
%
Astrophysical jets as highly collimated beams of high velocity material and outflows of
less degree of collimation and lower speed are an ubiquitous phenomenon in a variety of
astrophysical sources.
The role of magnetic fields in the realm of jets and accretion disks cannot
be underestimated.
It is crucial for the launching, acceleration, and collimation of jets
(see, e.g. \citealt{1982MNRAS.199..883B, 1983ApJ...274..677P, 1992ApJ...394..117P, 1990RvMA....3..234C,
1994A&A...287..893S, 2007prpl.conf..277P}).
However, due to the complexity of the physical problem, the exact time evolution and geometry of these
processes is still under debate.

Jets and outflows from young stellar objects (YSO) and active galactic nuclei (AGN) clearly affect their
environment,
and, thus, at the same time the formation process of the objects that are launching them
%
(see, e.g., \citealt{2007ApJ...668.1028B, 2012MNRAS.425..438G}).
However, in order to quantify the feedback phenomenon - namely to specify how much mass, angular momentum, and energy
is being ejected into the surrounding via the outflow channel - it is essential to model the physics in
the innermost launching area of the disk-jet system with a high resolution.
It is commonly accepted that ejection and accretion are tightly connected to each other
\citep{1995ApJ...444..848L, 1995A&A...295..807F}.
The study of these phenomena is motivated also by the observed correlation between accretion and ejection
signatures \citep{1990ApJ...354..687C}.

Our paper deals exactly with these topics - we will provide a relation between actual magnetization within
the disk, and the ejection to accretion ratio for mass and energy.

The first numerical simulations of this kind were presented by \citet{2002ApJ...581..988C, 2004ApJ...601...90C}, 
who demonstrated how an outflow can be self-consistently launched out of the accretion disk,
acclerated to high velocity and collimated in a narrow beam.
Later \citet{2006A&A...460....1M} studied in particular the impact of a central
stellar wind on the accretion disk magnetic field inclination.
The work by \citet{2007A&A...469..811Z} revealed the great importance of the underlying disk diffusivity,
namely the strength of diffusivity and its directional anisotropy.
Studying two limits of rather high and low diffusivity, and keeping the same (about equipartition) magnetic
field strength and field structure, the authors found that a steady state of the simulation could not be reached
for an arbitrary combination of these parameters.
\citet{2009MNRAS.400..820T} in particular found that the efficiency of the launching mechanism is strongly dependent on
the disk magnetization.

A common assumption was that in order to launch jets, the magnetic field should be rather strong,
somewhat about the equipartition value.
This question was investigated in detail by \citet{2010A&A...512A..82M}, demonstrating that even with a weak 
magnetization of $\mu \approx 0.002$ jets could be driven.

So far, the general mechanism of jet launching from magnetized disks have been studied by a number of authors.
However, due to the complexity of the problem, the combined action of the various processes engaged could not
be easily disentangled.
Another problem arises if only a short-term evolution of the system is considered, as this will be 
strongly dependent on the initial conditions. 
What is somewhat complicating the interpretation of simulations in the literature is that usually the model
setup is categorized by the {\em initial} parameters, 
and not by the {\em actual} quantities such as the actual disk magnetization, accretion velocity etc. at
a certain evolutionary time. 
The latter was first discussed by \citet{2012ApJ...757...65S}, however, the parameter space applied in 
those simulations was rather limited.
In the present study we will show that it is the actual disk properties, in particular the disk magnetization, 
that govern the accretion and ejection.

Accretion disks are considered to be highly turbulent for any degree of the disk magnetization. 
The source of the turbulence is still debated, however, a great variety of unstable modes in magnetized 
accretion disks exists \citep{2002ApJ...569L.121K}. 
In case of moderately magnetized disks, the main candidate is the magneto-rotational instability (MRI)
\citep{1991ApJ...376..214B, 2013EAS....62...95F}. 
Highly magnetized disks are subject to the Parker instability \citep{2010MNRAS.405...41G, 2008A&A...490..501J} 
and the trans-slow Alfv\'en continuum modes \citep{2004PhPl...11.4332G}.
The puzzling question is what is the effective diffusivity and viscosity the disk turbulence provides, 
and how can these effects be properly implemented under the mean field approach.

As shown by \citet{1995ApJ...440..742H} and later adapted by \citet{2004ApJ...601...90C, 2006A&A...460....1M},
the turbulent energy and angular momentum flux is dominated by the magnetic stress rather than the Reynolds stress. 
Thus, in the presence of a moderately strong magnetic field the Reynolds stress becomes less important. 
In order to disentangle the complex behaviour and keep the simulations more simple, we explore only non-viscous disks.

Considering the accretion-ejection scenario, we are convinced that before any general relation between 
physical quantities can be claimed, it is essential that the system itself has dynamically evolved over
a sufficiently extended period of time.
We have therefore evolved our simulations for at least 10.000 dynamical times.
We will show that such a long simulation requires that the advective and diffusive processes 
have to be well in balance.

In our paper we apply the following approach.
First, we consider the standard diffusivity model (see e.g. \citealt{2007A&A...469..811Z}). 
After having obtained a near equilibrium solution with advection and diffusion in balance, we closely examine the 
state of the system. 
Essentially, we will argue that it is the balance between diffusion and advection that governs the strength
of the {\em actual} disk magnetization.
The latter appears to be the key ingredient for the evolution of the whole system.
Exploring a wide range of the actual disk magnetization 
allows us to derive a general correlation between the actual disk magnetization and major quantities of
the disk-jet system, such as the mass and energy ejection efficiencies.

We also present a model setup which is well suited for a long-term evolution study of the jet launching 
problem. 
The use of spherical geometry provides a high resolution in the inner region of the disk - the site of
jet launching, and a low resolution for the outer regions, where the physical processes typically evolve on
much longer time scales.

Our paper is organized as follows. 
Section 2 describes the numerical setup, the initial and boundary conditions of our simulations.
In Section 3 we discuss our reference simulation, that is characterized by the balance between advection and diffusion, and uses
the standard diffusivity model. 
In Section 4 we present a detailed analysis of jet launching disks, 
revealing the major role of the disk magnetization in the disk-jet evolution.
In Section 5 we discuss simulations applying a new diffusivity 
model that essentially overcomes the accretion instability observed
in the previous simulations. 
This allows us to follow substantially longer the evolution of the disk-jet system. 
Finally we summarize our results in Section 6

\section{Model approach}
We apply the MHD code PLUTO \citep{2007ApJS..170..228M}, version 4.0,
solving the time-dependent, resistive MHD equations on a spherical grid $(R,\Theta)$. 
We refer to $(r,z)$ as cylindrical coordinates.
The code numerically solves the equations for the mass conservation,
\begin{equation}
\frac{\partial\rho}{\partial t} + \nabla \cdot(\rho \Vtot)=0,
\end{equation}
with the plasma density $\rho$ and flow velocity $\Vtot$,
the momentum conservation,
\begin{equation}
\frac{\partial \rho \Vtot}{\partial t} + 
\nabla \cdot \left[ \rho  \Vtot \Vtot +
\left(P + \frac{\Btot \cdot \Btot}{2}\right)I - 
\Btot\Btot \right] +\rho \nabla \Grav = 0
\end{equation}
with the thermal pressure $P$ and the magnetic field $\Btot$. 
The central object of point mass $M$ has a gravitational potential 
$\Grav=- G\,M/R$.
Note that equations are written in non-dimensional form, and as usual the
factor $4\pi$ is neglected.
We apply a polytropic equation of state,
$P \propto \rho^{\gamma}$,
with the polytropic index $\gamma=5/3$.

The code further solves for the conservation of energy,
\begin{equation}
\frac{\partial e}{\partial t} +  \nabla \cdot \left[ \left( e + P + \frac{\Btot \cdot \Btot}{2} \right) \Vtot
                              -  (\Vtot \cdot \Btot)\Btot + \etatens \Jtot \times \Btot \right] = -\Lambda_{\rm{cool}},
\end{equation}
with the total energy density,
\begin{equation}
e = \frac{P}{\gamma -1} + 
    \frac{\rho \Vtot \cdot \Vtot}{2} + 
    \frac{\Btot \cdot \Btot}{2} + \rho \Grav,
\end{equation}
given by the sum of thermal, kinetic, magnetic, and gravitational energy,
respectively. 
The electric current density is denoted by $\Jtot=\nabla \times \Btot$.
As shown by \citet{2013MNRAS.428.3151T}, cooling may indeed play a role for jet launching, 
influencing both jet density and velocity.
For the sake of simplicity we set the cooling term equal to Ohmic heating,
$\Lambda_{\rm{cool}} = -\etatens \Jtot \cdot \Jtot$.
Thus all generated heat is instantly radiated away.

The magnetic field evolution is governed by the induction equation
\begin{equation}
\frac{\partial \Btot}{\partial t} = 
\nabla\times (\Vtot \times \Btot  -  \etatens \Jtot),
\end{equation}

Our simulations are performed in 2D axisymmetry applying spherical coordinates. 
We use the Harten-Lax-van Leer (HLL) Riemann solver together with a third-order order 
Runge-Kutta scheme for time evolution and the PPM (piecewise parabolic method) reconstruction 
of \citep{1984JCoPh..54..174C} for spatial integration.
The magnetic field evolution follows the method of Constrained Transport \citep{2004JCoPh.195...17L}.

\subsection{Numerical grid and normalization}\label{sec:numgrid}
No physical scales are introduced in the equations above. 
The results of our simulations will be presented in 
non-dimensional units. We normalize all variables, 
namely $P, \rho, \Vtot, \Btot$, to their values at the inner disk radius $\Ri$.
Lengths are given in units of $R_0$, corresponding to inner disk radius. 
Velocities are given in units of $V_{\rm K,0}$, corresponding to the Keplerian speed at $R_0$. 
Thus $2\pi T$ corresponds to one revolution at the inner disk radius. 
Densities are given in units of $\rho_0$, corresponding to $R_0$. 
Pressure is measured in $P_{0} =  \epsilon^2 \rho_{0} V_{0}^2$.

We thus may apply our scale-free simulations to a variety of jet sources.
In the following we show the physical scaling concerning three different object classes - brown 
dwarfs (BD), young stellar objects (YSO), and active galactic nuclei (AGN). 
In order to properly scale the simulations, we vary the following masses for the central object,
$M = 0.05 \Msun$ (BD), $M = 1 \Msun$ (YSO), $M = 10^8 \Msun$ (AGN), 
and, thus, define a scale for the inner disk radius of
\begin{eqnarray}
 \Ri & = & \AUyso\,\AU  \quad \YSO \nonumber \\ 
     & = & \AUbd\,\AU  \quad \BD \nonumber \\
     & = & 20\,{\rm AU} \Ragn \Magn \quad \AGN
,
\end{eqnarray}
where $\Rschw = 2GM/c^2$ is the Schwarzschild radius of the central black hole. 
For  consistency with our non-relativistic approach, we require $\Ri > 10 \Rschw$,
implying that we cannot treat highly relativistic outflows.
Table \ref{tbl:normalization} summarizes the typical scales for leading
physical variables. 
For more detailed scaling laws we refer to the Appendix \ref{app:units}.

\begin{table}
\caption{Typical parameter scales for different sources. Simulation results will be given 
in code units and can be scaled for astrophysical application.
}
\begin{center}
    \begin{tabular}{lllll}
    ~         & YSO       & BD         & AGN        & [unit] \\     
    \noalign{\smallskip}    \hline \hline    \noalign{\smallskip}
    $R_0$     & $\AUyso$  & $\AUbd$    & 20 & AU     \\
    $M_0$     & 1         & 0.05       & $10^8$     & $\Msun$  \\
    $\rho_0$  & $10^{-10}$& $10^{-13}$ & $10^{-12}$ & $\DNx$~  \\    
    \noalign{\smallskip}\hline \noalign{\smallskip}
    $V_0$     & 94        & 66         & $6.7 \times 10^{4}$ & $\Vx$  \\
    $B_0$     & 15        & 0.5        & 1000       & G       \\
    $T_0$     & 1.7       & 0.25       & 0.5        & days      \\
  $\dot{M}_0$ & $3\times10^{-5}$   & $2\times10^{-10}$ & $10$ & $\Mx$       \\
  $\dot{J}_0$ & $3.0\times10^{36}$ & $1.5\times10^{30}$ & $3\times10^{51}$ & $\Jx$       \\
  $\dot{E}_0$ & $1.9\times10^{35}$ & $6.7\times10^{29}$ & $2.6\times10^{46}$ & $\Ex$       \\
    \end{tabular}
\end{center}
\label{tbl:normalization}
\end{table}

We apply a numerical grid with equidistant spacing in $\theta$-direction, 
but stretched cell sizes in radial direction, considering $\Delta R = R \Delta\theta$.
Our computational domain of a size $R=[1, 1500 R_0],\theta=[0,\pi/2]$ is 
discretized with $(N_R \times N_\theta)$
grid cells. 
We use a general resolution of $N_\theta = 128$. 
In order to cover a factor 1500 in radius, we apply $N_R = 600$. 
This gives a resolution of 16 cells per disk height ($2 \epsilon$) in the general case. 
However, we have also performed a resolution study applying a resolution twice high (or low, respectively), 
thus using $256 \times 1200$ (or $64 \times 300$) cells for the whole domain, or 35 (9) cells per disk height.
We satisfy the Courant-Friedrichs-Lewy condition by using a CFL number of 0.4.
\subsection{Initial conditions}
For the initial conditions we follow a meanwhile standard setup, applied in a number of previous publications 
\citep{2007A&A...469..811Z, 2012ApJ...757...65S, 2013ApJ...774...12F}.
The initial structure of the accretion disk is calculated as the solution of the
steady state force equilibrium,
\begin{equation}
\nabla P  +\rho \nabla \Grav  
- \Jtot \times \Btot - \frac{1}{R} \rho \Vphi^2 
 ({\bit e}_R \sin\theta + {\bit e}_\theta \cos\theta) = 0.
\end{equation}
We solve this equations assuming radial self-similarity, i.e. assuming that
all physical quantities $X$ scale as a product of a power law in $R$ with 
some function $F(\theta)$,
\begin{equation}
X \equiv X_0 R^{\beta_X} F(\theta).
\end{equation}
Self-similarity requires in particular that the sound speed and the Alfv\'en speed scale as the Keplerian velocity, 
$\Vk \propto r^{-1/2}$, along the disk midplane. 
As a consequence, the power law coefficients $\beta_X$ are determined as follows,
$\beta_{\Vphi} = -1/2, \beta_{\rm P}= -5/2, \beta_\rho = -3/2$, and
$\beta_{\Br} = \beta_{\Bth} = \beta_{\Bphi} = -5/4$.

\begin{figure}
\centering

\includegraphics[width=5cm]{\figurepath/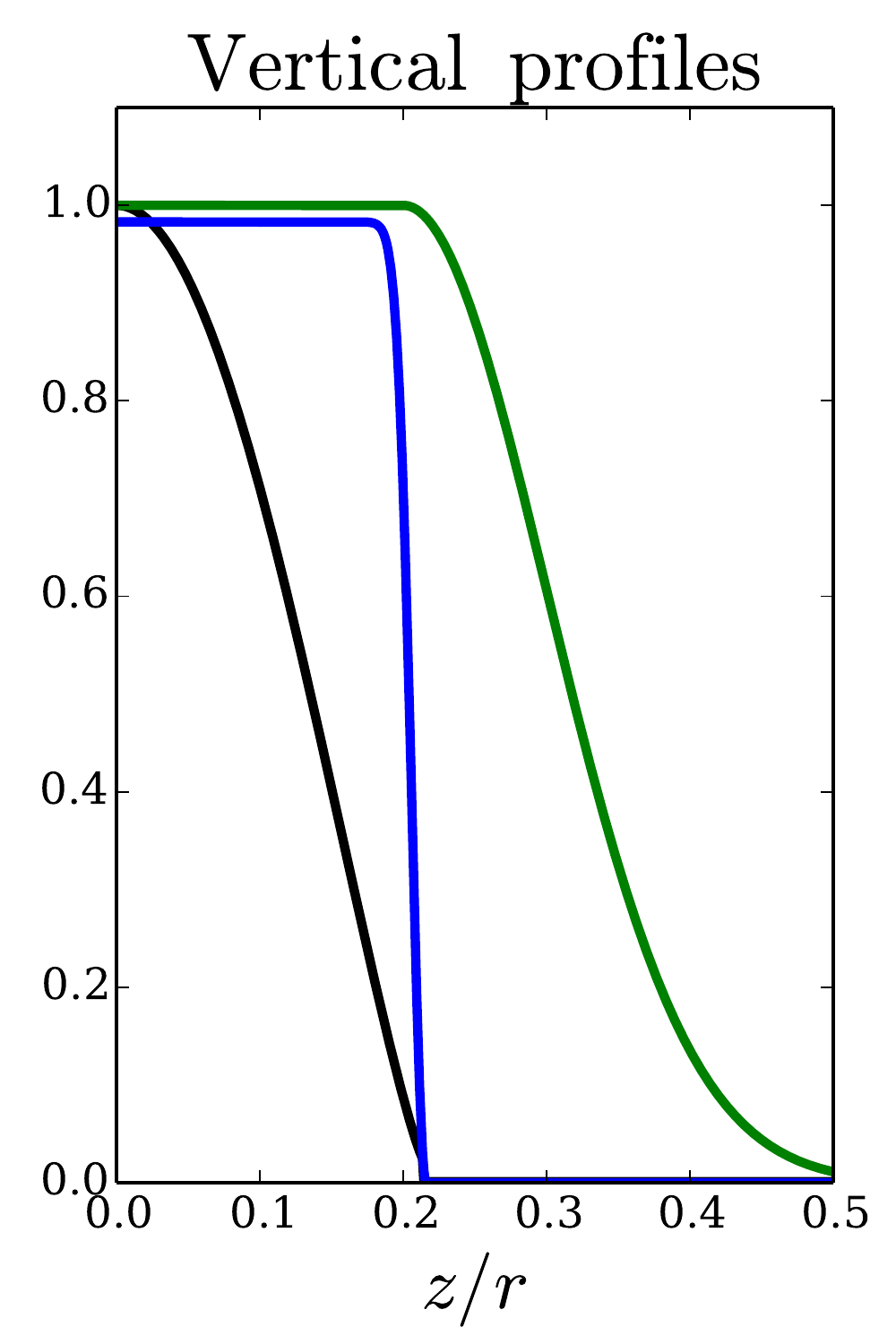}

\caption{Initial vertical profiles of the disk quantities: density (black, blue shaded), rotational velocity 
(blue), and magnetic diffusivity $F_\eta$ (green). The magnetic diffusivity profile is set constant in time.
}
\label{fig:main_profiles}
\end{figure}

An essential non-dimensional parameter governing the initial disk structure is 
the ratio $\epsilon$ between the isothermal sound speed $\CsT = \sqrt{P/\rho}$ and
the Keplerian velocity $\Vk = \sqrt{GM/r}$, evaluated at the disk midplane,
$ \epsilon = \left[ \CsT / \Vk \right]_{\,\theta=\pi/2}. $
This quantity determines the disk thermal scale height $\Ht = \epsilon r$. 
In our simulations we generally assume a thin disk with $\epsilon = 0.1$ initially. 
Note that for the rest of the paper, when discussing the dynamical properties of disk 
and outflow, we consider the {\em adiabatic sound speed} $\Cs=\sqrt{\gamma P/\rho}$.
The geometrical disk height, namely the region where the density and rotation 
significantly decrease, is about $2\epsilon$ (see Figure~\ref{fig:main_profiles}). 
We therefore define the geometrical disk height as $H \equiv 2 \epsilon r$.
 
\begin{figure}
\centering
\includegraphics[width=9cm]{\figurepath/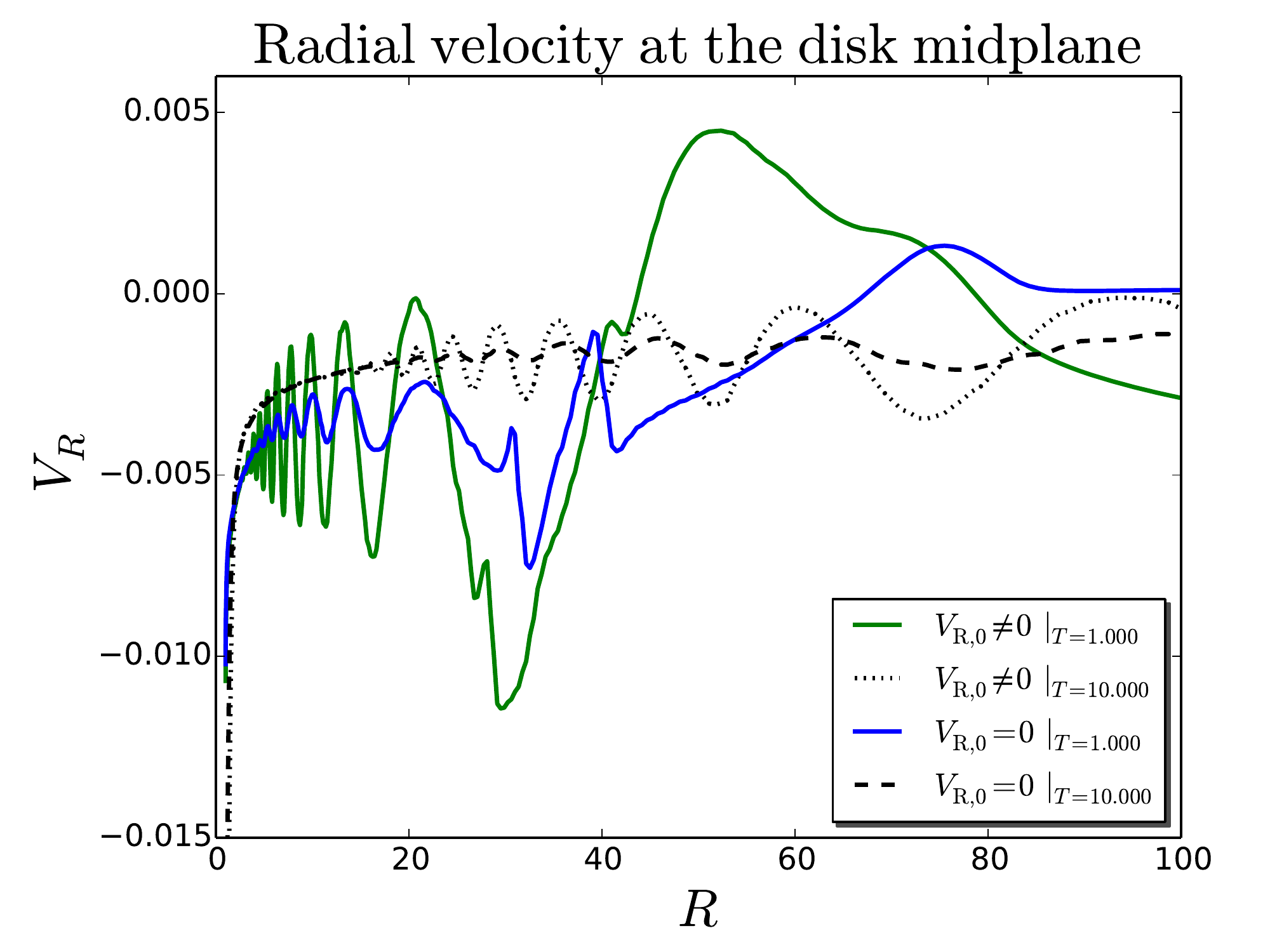}
\caption{Accretion velocity profile along the disk midplane at $T = 1000$ and $T = 10000$
for the cases of zero and non-zero initial radial velocity.}
\label{fig:def_vr0}
\end{figure}

\begin{figure}
\centering
\includegraphics[width=7cm]{\figurepath/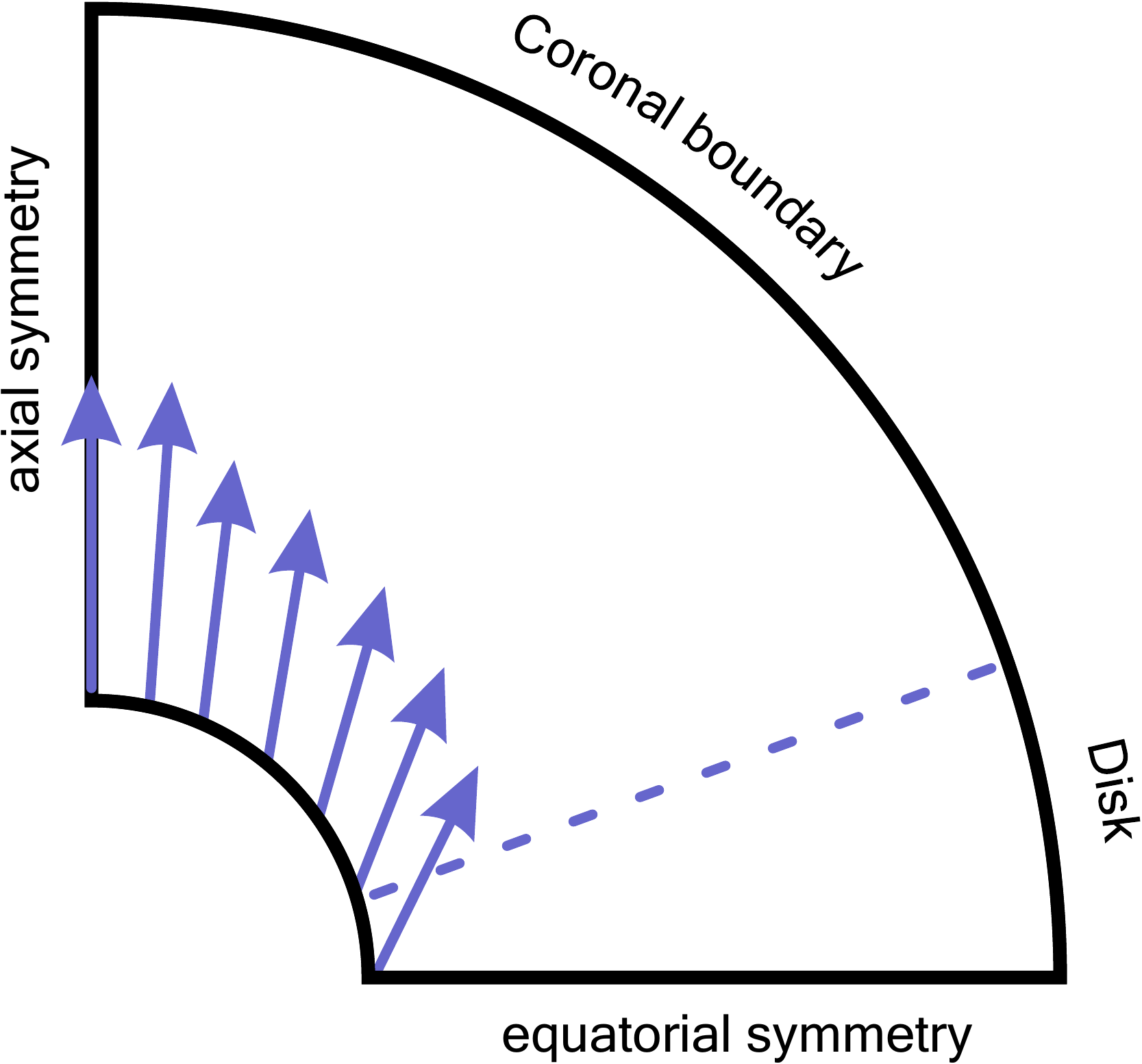}

\caption{Illustration of the boundary conditions. 
Along the inner and outer radial boundaries we distinguish two zones - the corona and the disk region. 
Arrows represent the magnetic field distribution along the inner boundary, which is preserved by our 
boundary condition.}
\label{fig:main_bc}
\end{figure}

Following \citet{2007A&A...469..811Z}, our reference simulation is initialized only with 
a poloidal magnetic field, 
defined via the vector potential $\Btot = \nabla \times A \bit{e}_{\phi}$, with
\begin{equation}\label{eq:Bpot}
A = \frac{4}{3} B_{\rm p,0} r^{-1/4} \frac{m^{5/4}}{ (m^2 + ctg^2\theta)^{5/8} }.
\end{equation}
The parameter $B_{\rm p,0} = \epsilon \sqrt{2\MUzero} $ determines the strength
of the initial magnetic field, 
while the parameter $m$ determines the degree of bending of the magnetic field lines.
For $m \to \infty$, the magnetic field is purely vertical. 
As we will see below, the long-term evolution of the disk-jet structure
is insensitive to this parameter, since due to advection and diffusion processes
and the jet outflow, the magnetic field structure is changed substantially over time.
We therefore set $m=0.5$ in general.

The strength of the magnetic field is governed by the magnetization parameter
\begin{equation}
\mu = \frac{\Bp^2}{2P}\, \vline_{\,\theta=\frac{\pi}{2}},
\end{equation}
the ratio between the poloidal magnetic field pressure and the thermal pressure, evaluated at the 
midplane, and is set to be constant with radius.
As we will see below, the magnetic field distribution substantially changes over time while
the disk-jet dynamics is governed by the {\em actual} disk magnetization.
Typically, the initial magnetization is $\MUzero \approx 0.01$ in this simulations.

Outside the disk the gas and pressure distribution is defined as hydrostatic ''corona'',
\begin{equation}
\rho_{\rm cor} = \rho_{\rm cor,0} R^{-1/(\gamma - 1) } ,
\quad\quad P_{cor} = 
\frac{\gamma -1}{\gamma} \rho_{\rm cor,0}  R^{-\gamma /(\gamma - 1)},
\end{equation}
where $\rho_{\rm cor,0} = \delta \rho_{\rm disk} (R=1,\theta=\pi/2)$ with $\delta = 10^{-3}$, 
%

Although it is common to define   the initial accretion velocity, balancing the imposed diffusivity 
$V_R = \eta J_\phi /B_\theta$, we find that in our set of parameters
this is not necessary. 
It can even disturb the initial evolution of the disk accretion.
Accretion requires corresponding torques to be sustained and since there is no initial poloidal 
current defined, $B_\phi$ = 0 (that takes time to build up from a weak poloidal field), 
a non-zero initial velocity will only lead to extra oscillations. 
Figure \ref{fig:def_vr0} illustrates this issue, showing two identical simulations with zero and 
non-zero accretion velocity. 
We found that the origin of the oscillations for both cases is the inner boundary. 
The first wave of accretion is somehow bounced outwards by the inner boundary and results in an
oscillatory pattern.
We believe that the bouncing at the inner disk boundary is a generic problem of most
accretion disk simulations resulting from subtle inconsistencies between the boundary conditions
and the intrinsic disk physics.
Although they both result in the same final profile (a steady state has been 
reached only for small disk radii), the simulations with zero velocity profile 
show fewer oscillations.

\begin{figure*}
\centering
\includegraphics[width=18cm]{\figurepath/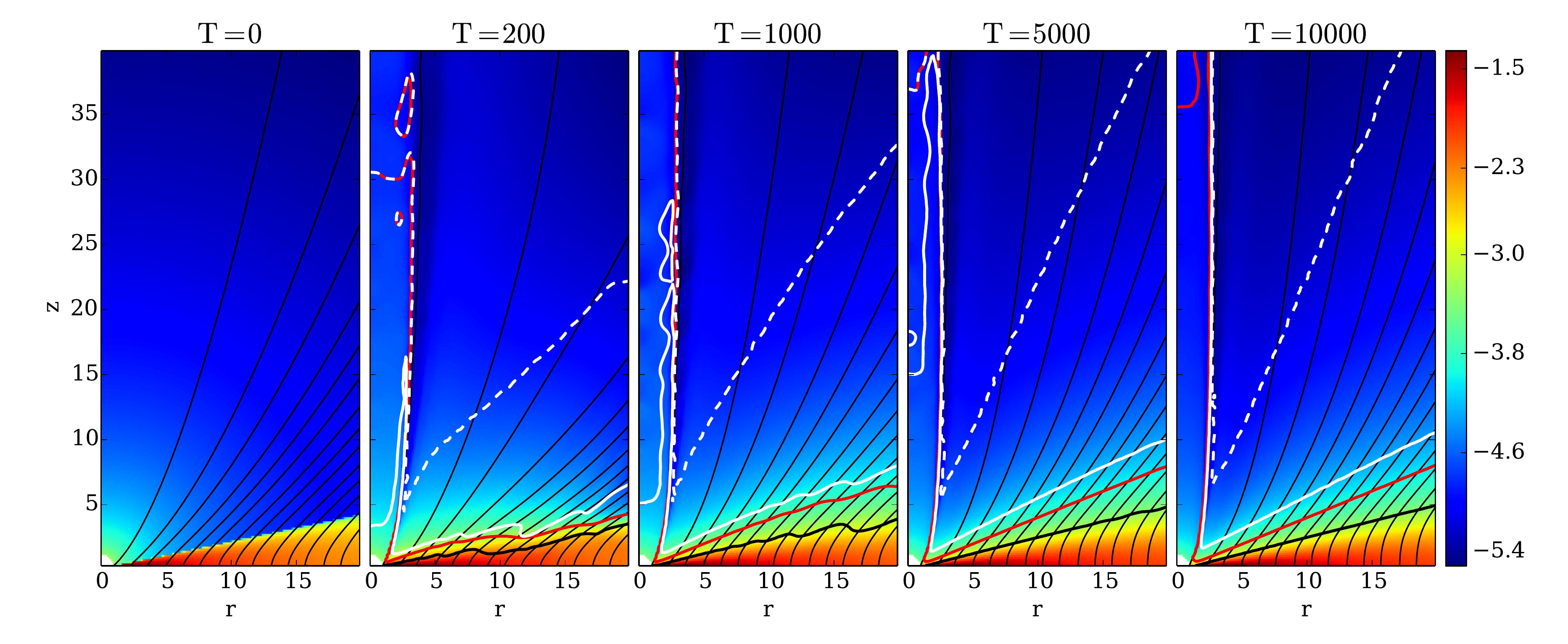}
\caption{Time evolution of the disk-jet structure for the reference simulation.
Shown is the evolution of the density (by colors, in logarithmic scale), 
the poloidal magnetic field lines (thin black lines), the disk surface (thick black line)
the sonic (red line), the Alfv\'en (white line), and the fast Alfv\'en (white dashed line) surfaces.}
\label{fig:def_tevol}
\end{figure*}

\subsection{Boundary conditions}
We apply standard symmetry conditions along the rotational axis and the equatorial plane.
Along the radial boundaries of the domain, we distinguish two different areas.
That is (i) a {\em disk boundary} for $\theta > \frac{\pi}{2}-2\epsilon$\footnote{Note, $2\epsilon \approx arctg(2\epsilon)$}, 
and (ii) a {\em coronal boundary} for $\theta<\frac{\pi}{2}-2\epsilon$,
and consider different conditions along them (see Figure~\ref{fig:main_bc}).

\begin{table*}
\caption{Inner and outer boundary conditions. Outflow is zero gradient condition, constant slope conditions are marked by slope in the table. }
\begin{center}
    \begin{tabular}{lllllllll}
    ~         & $\rho$       & P    & $\Vr$    & $\Vth$ & $\Vphi$ & $\Br$ &$\Bth$ & $\Bphi$ \\     
    \noalign{\smallskip}    \hline \hline    \noalign{\smallskip}
    inner disk  &$\sim r^{-3/2}$&$\sim r^{-5/2}$&$\sim r^{-1/2}, \leq 0$&0&$\sim r^{-1/2}$ &slope&slope&$\sim r^{-1}$  \\
    inner corona  &$\sim r^{-3/2}$&$\sim r^{-5/2}$&$0.2 cos(\varphi)$&$0.2 sin(\varphi)$&$\sim r^{-1/2}$ &slope&slope&$0$  \\

    outer disk  &$\sim r^{-3/2}$&$\sim r^{-5/2}$ &outflow, $\le 0$& outflow& outflow&div B =0&outflow&$\sim r^{-1}$  \\    
            
    outer corona  &$\sim r^{-3/2}$&$\sim r^{-5/2}$ &outflow, $\geq 0$& outflow& outflow&div B =0&outflow&$\sim r^{-1}$  \\    
    
    \noalign{\smallskip}\hline \noalign{\smallskip}
    
    \end{tabular}
\end{center}
\label{tbl:bc}
\end{table*}

Along the inner radial boundary for all simulations we impose 
a constant slope for the poloidal component of the magnetic field
\begin{equation}
\varphi = 70^\circ \left(1+ \exp( - \frac{\theta - 45^\circ}{15^\circ} \right)^{-1}
\end{equation}
where $\varphi$ is the angle with respect to unit vector ${\bit e}_R$.
The magnetic field direction is axial near the axis, $\theta=0$, while at the inner disk radius the
inclination is $70^\circ$ with respect to the disk surface. 
A smooth variation of the magnetic field direction is prescribed along the inner radial boundary. 
This is in concordance with \citet{1992ApJ...394..117P} who showed that for a warm plasma the 
maximum angle with respect to the disk surface necessary to launch outflows is about $70^\circ$, 
and slightly larger than for a cold plasma \citep{1982MNRAS.199..883B}.

The method of constraint transport requires the definition of only tangential component, thus to
prescribe $\Bth$ along the innermost boundary, while
the normal component $\Br$ follows from solving $\nabla \cdot \Btot =0$.
In order to implement the prescription of a constant magnetic field angle, we solve $\nabla \cdot \Btot =0$, 
taking into account the ratio of the cell-centered magnetic field components $\Bth / \Br = - \tan(\varphi)$. 
We start the integration from the axis ($\theta = 0$), where $\Bth = 0$. 
Thus, by fixing the slope of the magnetic lines, we allow the magnetic field strength to vary.

Along the inner coronal boundary, we prescribe a weak inflow into the domain with $\Vp = 0.2$. 
This is just applied to stabilize the inner 
coronal region between the rotational axis and the disk jet, since the interaction between the current 
carrying, magnetized jet and zero-$\Bphi$ coronal region may lead 
to some extra acceleration of the coronal gas.  
As it was shown by \citep{2006A&A...460....1M} the pressure of such an inflow (e.g. a stellar wind) may influence 
the collimation of the jet, changing the shape of the innermost magnetic field lines. 

In order to keep the influence of the dynamical pressure of the inflow similar during the whole evaluation 
(and also for different simulations),
we set the density of this inflow with respect to the disk density at the inner disk radius.
The density of the inflow corresponds to a hydrostatic corona  
$\rho_{\rm infl} = \rho_{\rm cor} = \rho_{\rm disk}|_{\rm midplane}(t) \cdot \delta$, 
where $\delta = 10^{-3}$.
The inflow direction is aligned with the magnetic field direction. 
By choosing a denser inflow we also increase the time step of our simulations by approximately 
three times, as the Alfv\'en speed in the coronal region lowers.

By varying the slope of the magnetic field $\varphi$ along this inner corona in the range of 
60 - 80 degrees, we found that it only slightly affects the slope of the innermost magnetic field 
lines. 
The global structure of the magnetic field is instead mainly governed  by the diffusivity model.
Since the inner boundary by design models the magnetic barrier of the star, we choose a rather steep
slope in order to avoid the disk magnetic flux entering the coronal region.

Across the inner disk boundary (that is the accretion boundary) density and pressure are both 
extrapolated by power laws, applying $\rho R^{-3/2} = const$, and $P R^{-5/2} = const$, respectively.
Both the toroidal magnetic field as well as the toroidal velocity components are set to vanish at the 
inner coronal boundary, $\Bphi = 0$, $\Vphi = 0$.
For the inner disk boundary, we further apply the condition $\Bphi \sim 1/r$ ($J_\theta = 0$), and extrapolate 
the radial and the toroidal velocity by power laws, $\Vr R^{-1/2} = const$, and $\Vphi R^{-1/2} = const$, 
respectively, while $\Vth = 0$.

For the inner disk boundary, only negative radial velocities are allowed, making the boundary to 
behave as a ''sink'', thus absorbing all material which is delivered by the accretion disk at the 
inner disk radius.

As the application of spherical coordinates provides an opportunity to use a much larger simulation 
domain compared to cylindrical coordinates, the outer boundary conditions have only little influence 
on the evolution of the jet launched from the very inner disk.
We therefore extrapolate $\rho$ and $P$ with the initial power laws and apply the standard PLUTO 
outflow conditions for $\Vr, \Vth, \Vphi$ at the outer boundary, thus zero gradient conditions.
We further require $\Bphi \sim 1/r$ ($\Jth = 0$) for the toroidal magnetic field component,
while a simple outflow condition is set for $\Bth$.
Again $\Br$ is obtained from the $\nabla \cdot \Btot =0$.

For the radial velocity component we distinguish between the coronal region, where we require positive
velocities $\Vr \geq 0$, and the disk region, where we enforce negative velocities $\Vr \leq 0$.

As our application of a spherical geometry is new in this context, we summarize the 
boundary conditions in the Table \ref{tbl:bc}.

\subsection{The magnetic diffusivity model}
Accretion disks are considered to be highly turbulent, subject to the magneto-rotational instability (MRI) in 
moderately magnetized disks \citep{1991ApJ...376..214B, 2013EAS....62...95F}, and the Parker instability
\citep{2010MNRAS.405...41G, 2008A&A...490..501J} for stronger magnetized disks.
It is believed that when the magnetic field becomes sufficiently strong the MRI modes become suppressed 
\citep{2013EAS....62...95F}. 
On the other hand, a strong magnetic field may become buoyant, leading to the Parker instability.
While the MRI is confined within the disk, the Parker instability operates closer to the surface 
of the disk where the toroidal magnetic field is stronger.

In order to extrapolate the results from a self-consistent, local treatment of turbulence to the mean field approach is not 
straightforward. 
In the local treatment the extraction of angular momentum is due to both turbulence - operating on small 
scales - and torques by the mean magnetic field on large scales. 
Thus, removal of angular momentum goes hand in hand with destroying of the turbulent magnetic field or 
the effective magnetic diffusivity.
In case of the mean field approach, there is no small scale turbulence and, thus, no angular momentum removal 
by local turbulent motions. 
Here, the diffusivity plays only a role for leveling out the magnetic field gradient, thus setting the 
overall structure of the magnetic field.
Unfortunately, we lack the complete knowledge of the disk turbulence, thus the connection between the mean magnetic field and the fluctuating part.
Or, in other words, the relation between the mean magnetic field and the effective torques, and the diffusivity and viscosity that turbulence provides.
We believe that when moving from a local turbulence approach to the mean field approach,
the relevance of the model should be approved by the relevance of the
magnetic field distribution itself, and not by the diffusivity model.
However, one should keep in mind that the magnetic field strength and structure of real accretion 
disks are also not known.
Therefore, when considering any simulation results, always the diffusivity model applied should be
taken into account.

A self-consistent study of the origin of the turbulence is beyond the scope of our paper.
We therefore prescribe a certain model of the magnetic diffusivity.
We apply an $\alpha$-prescription (\citet{1973shakuraetal}) for the magnetic diffusivity, implicitly assuming that the
diffusivity has a turbulent origin.
The diffusivity profile may extend up to one disk height above the disk surface 
(Figure~\ref{fig:main_profiles}).

Although we investigate different diffusivity models, all of them can be represented 
in a following form,
\begin{equation}
\Ep = \ass (\mu)  \Cs \cdot H \cdot \Fe(z),
\end{equation}
where the vertical profile of the diffusivity is described by a function
\[ \Fe(z) =   \left\{
\begin{array}{ll}
      1 & z\leq H \\
      \exp(- 2(\frac{z-H}{H})^2 ) & z > H, \\
\end{array} 
\right. \]
confining the diffusivity to the disk region.

Although this parametrization of diffusivity is commonly used (except the profile function $\Fe(z)$), 
there are no clear constraints upon the value $\ass$ may take. 
As an example, \citet{2007MNRAS.376.1740K} discuss a magnitude of the turbulent $\alpha$-parameter 
derived from observations and simulations, indicating observational values $\ass \simeq 0.1 .. 0.4$.
Numerical models with zero net magnetic field usually provide low numerical values $\ass \simeq 0.01$, reaching at most 
$\ass \simeq 0.03$ \citep{1996ApJ...463..656S, 2011MNRAS.416..361B, 2012MNRAS.422.2685S, 2013ApJ...763...99P}. 
On the other hand, numerical modelling of the MRI applying a non-zero net magnetic field 
\citep{2013ApJ...767...30B} indicate substantially higher values, $\ass \simeq 0.08 - 1.0$, 
with a corresponding magnetization $\mu= 10^{-4} , 10^{-2}$. 

Obviously, different functions of $\ass (\mu)$ will lead to different evolution.
We start from the well-known model for magnetic diffusivity applied by many authors before 
\citep{2004ApJ...601...90C, 2007A&A...469..811Z, 2012ApJ...757...65S},
\begin{equation}\label{eq:stdiff}
\Ep = \am \Va \cdot H \cdot \Fe(z)
\end{equation}
by applying $\ass = \am \sqrt{2\mu}$, where $\Va = \Bp / \sqrt{\rho}$ is the Alfv\'en speed,
and $\mu$, $\Cs$, and $H$ are the magnetization, the adiabatic sound speed and the local disk height,
respectively, measured at the disk midplane. 
We evolve $\ass$ and $\Cs$ in time, but for the sake of simplicity we keep $H$ and
$\Fe(z)$ constant in time, thus equal to the initial distribution. 
The main reason is to avoid additional feedback, which favors the accretion instability 
(see below, or e.g. \citet{2009MNRAS.392..271C}). 
Our test simulations evolving the disk height in time, in fact indicate the rise of such instability 
earlier than in case of a fixed-in-time disk diffusivity aspect ratio.

\subsection{Anisotropic diffusivity}\label{sec:andiff}
In general, the diffusivity tensor has diagonal non-zero components,
\begin{equation}
\eta_{\rm \phi\phi} \equiv  \Ep  \quad \eta_{\rm RR}=\eta_{\rm \theta\theta} \equiv  \Et,
\end{equation}
where we denote $\Ep $ as the {\em poloidal} magnetic diffusivity, and $\Et $ as the {\em toroidal} magnetic 
diffusivity, respectively.
The anisotropy parameter $\chi \equiv \Et/\Ep$ quantifies the different strength of diffusivity in poloidal and 
toroidal directions.

In the literature it is common to assume $\chi$ of order the of unity. 
Considering viscous disks, \citet{2000A&A...353.1115C} showed that there is a theoretical limit for $\Et$, namely $\Et > \Ep$.
Highly resolved disk simulations indeed suggest $\chi \simeq 2 ... 4$ \citep{2009A&A...504..309L},
implying that the toroidal field component is typically diffusing faster than the poloidal component.

The majority of simulations in the literature consider a magnetic field strength in equipartition with the
gas pressure.
However, studying also weakly magnetized disks, we find that there also exists an upper limit for the anisotropy 
parameter, above which the simulations show an irregular behaviour.
On the other hand, it was pointed out by several authors (see e.g. \citealt{2007A&A...469..811Z}) that in case of 
a very low anisotropy parameter (thus, a weak toroidal diffusivity) the simulations might suffer from instabilities 
caused by strong pinching forces.
Nonetheless, the existence of an upper limit for the anisotropy parameter was so far obscured by other processes.

Assuming a steady state and combining the poloidal component of the diffusion equation, $\Mr = \ass H J_\phi/\Bp$, 
with the relation for the Mach number $\Mr = 2/\sqrt{\gamma} H\Jr/\Bp \MUact$ (see below, \citealt{2011ppcd.book..283K}), 
an interrelation between the toroidal and poloidal electric currents can be derived,
\begin{equation}\label{eq:jrjphi}
\frac{\Jphi}{\Jr} = \frac{\sqrt{2\mu}}{\sqrt{\gamma}\am}.
\end{equation}
This relation has been proven to approximately hold for all of our simulations, thus indicating that a steady state has indeed 
been reached.
Since the only free parameter in this relation is $\am$, the choice of $\am$ governs the ratio
of the electric current components.

As shown previously \citep{1995A&A...295..807F, 1997A&A...319..340F, 2013MNRAS.428..307F} the toroidal component 
of the induction equation can be written as
\begin{equation}\label{eq:etjr}
\Et \Jr |_{\rm mid} = - R^2 \int_0^{2\epsilon} \mathbf{\Bp} \cdot \nabla \mathbf{\Omega} d\varphi - \Vth \Bphi
\end{equation}
(here expressed for spherical coordinates), where $\Jr$ is computed at the disk midplane. 
This equation essentially states that the induction of the toroidal magnetic field component (from twisting the poloidal component) 
is being balanced by the diffusion through the disk midplane and by escape of the flux through the disk surface.
We emphasize that we do {\em not} neglect the $\Vth \Bphi$ term, considering the assumption of a thin disk \citep{2013MNRAS.428..307F}, 
as we find it of key importance in our simulations, in particular in the regime of a moderately strong magnetic field, $\mu \geq 0.1$.

Assuming that the induction of the magnetic field is primarily due to radial gradients, the radial component of the magnetic
field can be approximated by a power law, $\Br \propto R^{-5/4}$,
and equation \ref{eq:etjr} may be transformed into $ \Et \Jr |_{\rm mid} \simeq \Br \Cs - \Vth \Bphi$, or
\begin{equation}
(\ass\chi - \Mth) \Jr \simeq \frac{\Br}{H},
\end{equation}
where 
\begin{equation}\label{eq:ejmach}
\Mth = - \Vth^{+}/\Cs
\end{equation}
is denoted as the ejection Mach number, where $\Vth^{+}$ is measured at the disk surface.

Using relation \ref{eq:jrjphi} between the poloidal and toroidal electric currents and defining the curvature 
part of the toroidal current $J_\phi^{\rm curv} \equiv \Br/H$, one may derive a constraint for the anisotropy parameter,
\begin{equation}\label{eq:curv}
\frac{J_\phi^{\rm curv}}{\Jphi} \simeq  \left(\am\chi - \frac{\Mth}{\sqrt{2\mu}}\right)\am \leq 1.
\end{equation}

Any magnetic field geometry which is outwardly bent and has a decreasing field strength in outward direction
has to satisfy this relation, as $\Jphi$ consists of two positive terms, the gradient and the curvature. 
In cases where the vertical velocity term can be neglected (e.g. for a very weak magnetic field with $\mu \leq 0.02$),
the anisotropy parameter is $\chi < 1/\am^2$, which for our choice of $\am$ is about 0.4. 
By probing the $\chi$ parameter space we found that in order to obtain a stable accretion-outflow configuration
for weakly magnetized disks, the $\chi$ should be in the range $0.3-0.7$. 
We therefore decided to apply $\chi = 0.5$ for all of our simulations.
We note that in simulations applying an anisotropy parameter $\chi \ge 0.7$, we faced the problem that the
poloidal magnetic field lines were ''moving'' rapidly, such that the bending of the field lines along the disk
midplane was actually inverted. 
This is a result of the combined effects of a strong outward diffusion and low torques at the midplane.
On the other hand, in case of a rather low anisotropy parameter, the accretion is rapid and the jet does establish
a steady behaviour.

The reason why the commonly chosen anisotropy $\chi > 1$ leads to a steady behaviour is rather simple. 
As the disk magnetization grows during advection, the ejection Mach number grows as well
(we find that $\Mth \propto 6 \mu$ saturating at a level of 0.8, see below).
Thus in case of a high disk magnetization - usually assumed in the literature - the above mentioned upper limit 
for the anisotropy parameter is satisfied.
In case of  weak magnetic field simulations, performed e.g. by \citet{2010A&A...512A..82M}, this limit is most likely
satisfied by additional viscous torques.

\subsection{Comparison to previous simulations}
In the introduction section we have already discussed the literature of accretion-ejection simulations.
Here we want to explicitly emphasize specific details in which our simulations differ from previous works.
\begin{itemize}
\item
A spherical grid has been applied, offering the opportunity of a much larger domain size as well as much higher 
resolution {\em in the inner part} of the disk.
A new set of boundary conditions is used that is adapted to the spherical grid.

\item
We were able to explore a continuous range of simulation parameters. In particular we were aiming to disentangle 
interrelations between the {\em actual} flow parameters, rather than an interrelation to the initial values.

\item
Altogether, our model setup allows very long term simulations on a large grid - so far we have run simulations for 
approximately $30,000$ time units for a standard diffusivity model, and more than $150,000$ time units for 
our modified strong diffusivity model.

\item
We allow the inflow (into the coronal region) density to vary in time, thus keeping the ratio between the 
inflow and the disk densities $\delta$ the same as initially. 

\end{itemize}

We have explored a vast range of the parameter space that covers the majority of simulations performed in the 
literature (see Table~\ref{tbl:compar}). 
Similar to all these papers we assume a thin disk $\epsilon = 0.1$. 
It is common to assume a magnetic diffusivity parameter $\am$ of about unity. 
In our case we apply values $\am = 1.1 ... 1.9 $.
For the magnetic field bending parameter we have chosen $m=0.5$, which is slightly higher than the values usually 
adapted, $m = 0.35 ... 0.4$.
Although we finally show that $m$ plays only a minor role, our choice is motivated by the fact that this value 
is more consistent with the inner boundary condition. 
The magnetic diffusivity anisotropy parameter $\chi$ is chosen smaller than unity, since it helps keeping sufficiently 
strong torques at the midplane and the bending of the magnetic field that supports launching.

Although we start our simulation from a moderately weak initial magnetic field, the actual field strength 
in the inner disk at a certain radius may vary substantially over time.
This allows us to study the interrelation between disk accretion and ejection physics and the
{\em actual} magnetic field strength (thus the actual disk magnetization).

\begin{figure}
\centering
\includegraphics[width=9cm]{\figurepath/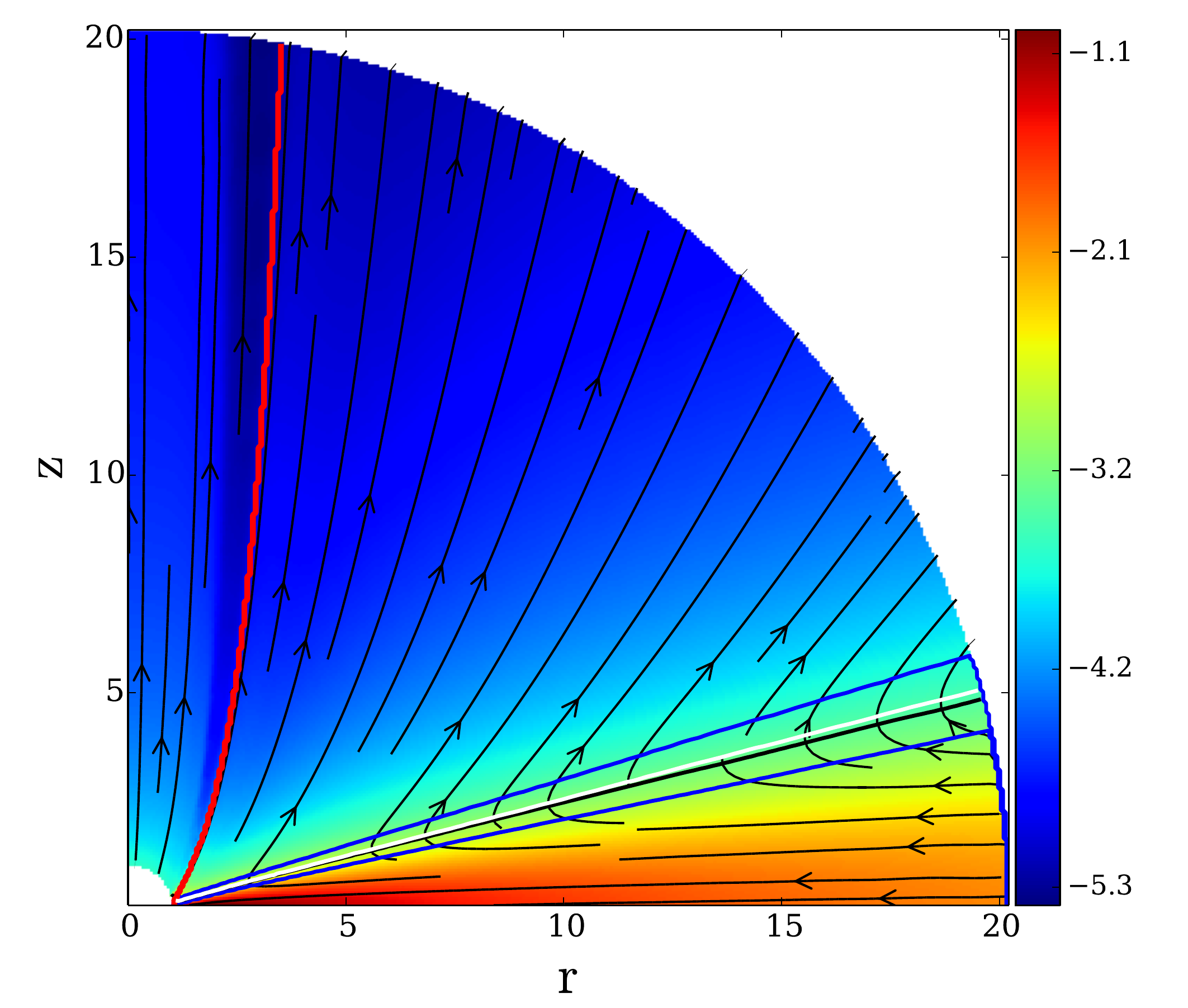}
\caption{Physically different regions of the disk-jet structure at $t=10,000$. 
Shown is the mass density (in logarithmic scale) and
streamlines of the poloidal velocity (black lines with arrows).
Red line marks the magnetic field line rooted at in innermost area of the midplane.
Upper blue line separates from the disk an area where $V_p || B_p$ 
The accretion and ejection areas are separated with white ($V_r = 0$) and black ($F_\phi = 0$) lines,
respectively.
The lower blue line separates the accretion area where $V_r >> V_\theta$ from the rest of the structure.}
\label{fig:def_struct}
\end{figure}

\section{A reference simulation}
In this section we present our reference simulation. 
Our aims were two-fold.
First, with our new setup we were able to increase both the period of time evolution and the spatial extension of 
jet launching conditions considerably compared to previous works.
Second, with our long-term evolution simulations we were able to investigate the interrelation between the {\em actual} 
disk properties such magnetization, the ejection to accretion ratios of mass and energy, jet velocity,
and others.
This has not been done in the past, as most papers have compared the {\em initial} parameters of the simulations. 
As shown by \citet{2012ApJ...757...65S} both the magnetization and diffusivity may substantially change 
during the disk evolution, and the parameters for the initial setup $\mu_0$ or $\ass$ are not sufficient to 
characterize the disk-jet system.

In order to uniquely specify the {\em initial conditions} for the simulation, we prescribe a number of non-dimensional 
characteristic parameters. 
The {\em initial} disk height is set by $\epsilon$. For all simulations we apply $\epsilon = 0.1$.
The {\em initial} strength and structure of the magnetic field is set by $\mu_0$ and $m$, respectively.
In all simulations we have chosen $m = 0.5$.

The initial disk magnetization does not play a major role in the {\em jet launching} process, 
but it is responsible for the overall disk torques.
The main reason is that the magnetization in the inner disk, from which the main jet is being launched, 
changes very quickly.
However, the magnetic field in the overall disk is primarily set by the initial magnetization.

In our simulations we examine $\mu_0 = (0.003, 0.01, 0.03)$. 

The model for diffusivity is chosen by selecting the distribution $\ass(\mu)$ and the anisotropy parameter $\chi$. 
We also set $\chi = 0.5 $ for all simulations.
We apply a standard diffusivity model (Equation~\ref{eq:stdiff}), thus the diffusivity is set by the $\am$ parameter.
As will be shown later, the simulations are very sensitive to this parameter. 
If not stated otherwise, $\am = 1.65$. 

We will refer to our reference simulation as to the setup with $\epsilon = 0.1, m = 0.5, \mu_0 = 0.01, \am = 1.65, \chi = 0.5$. 
Usually we run the simulations until $t = 10,000$, corresponding to about 1600 orbits at the inner disk radius.

\begin{table*}
\caption{Comparison of our simulations with simulations performed by other authors.
The resolution is estimated for the {\em inner} disk ($R=1$)
}
\begin{center}
    \begin{tabular}{llllllll}
    Reference & cell/2$\epsilon$ & $\epsilon$  & $m$ &   $\am$   & $\chi$ & $\mu_0$ & $\mu_{\rm act}$
\\  \noalign{\smallskip}    \hline \hline    \noalign{\smallskip}
    {\protect\citet{2004ApJ...601...90C}}& 0.5 & 0.1 &  ~   & $<1$        & 1      & $\simeq$ 1 & -   \\
    {\protect\citet{2007A&A...469..811Z}}& 2.5 & 0.1 & 0.35 & 0.1 ... 1.0 & 1, 3   & 0.3  & -   \\
    {\protect\citet{2009MNRAS.400..820T}}& 2.5 & 0.1 & 0.4  & 0.1 ... 1.0 & 3, 100 &  0.1-3.0   & -   \\
    {\protect\citet{2012ApJ...757...65S}}& 8.0 & 0.1 & 0.4  & 1.0 &   1/3, 3 &  0.002-0.1  & -   \\
    This work, reference simulation                 & 16 & 0.1 & 0.2-0.9  & 1.1-1.9   &  0.5  &  0.003-0.03 &  0.001-0.5   \\
    This work, resolution study & 24.5 & 0.1 & 0.2-0.9  & 1.1-1.9   &  0.5  &  0.003-0.03 &  0.001-0.5   \\

\noalign{\smallskip}\hline \noalign{\smallskip}
    \end{tabular}
\end{center}
\label{tbl:compar}
\end{table*}

Figure~\ref{fig:def_tevol} shows the time evolution of the disk-jet structure of the reference 
simulation. 
Note that here we present only a small cylindrical part of a much larger, {\em spherical} domain.
Although we explore a broad parameter space, the evolution for this simulation can be seen as a typical.

The first snapshot shows the initial state - a hydrodynamic disk in force-balance, the non-rotating
hydrostatic corona in pressure equilibrium with the disk, and the initial non-force-free magnetic 
field. 
After some 1000 revolutions, the inner parts of the disk-outflow reaches a quasi steady state.
However, it takes much longer time for the outer parts to reach such a state. 
The outflow, initiated already at early times and constantly accelerated, finally reaches super-fast 
magnetosonic speed. 
The outflow launching area along the disk surface grows with time.
However, the parts of the outflow being launched from larger disk radii are less powerful.

This reference simulation applying typical parameters from the literature can be re-established 
very well by our approach. 
Using a spherical setup, the resolution in the inner part of the disk is higher than in the literature, 
and the simulations run substantially longer than any other simulation published before.
Our simulations behave very robust.
We believe that there are two main reasons for that.
First, the spherical geometry does well resolve the inner part of the disk, from which the dominant
part of the jet is launched, but smooths out the small scale perturbations in the outer disk. 
By that, perturbations arising throughout the disk are diminished.
Second, our choice for the diffusivity parameter $\am$ allows to evolve the simulations into a quasi steady state in which 
advection is balanced by diffusion.
However, even with such an optimized numerical setup (the reference setup) our simulations show some 
irregular behaviour typically at about 30,000 time units.
The reason is that since the current diffusivity model is prone to the accretion instability
(see \citealt{1994MNRAS.268.1010L}, and section below), 
the simulations are always in a state of marginal stability.
As a consequence, the simulations evolve into a state of either high or low magnetization. 
In case of high magnetization the structure of the inner disk is being drastically changed 
and current model of diffusivity cannot be applied. 
In the opposite case of weak magnetization stable jets cannot be sustained.

In the following, we discuss different components of the disk-jet system and the jet launching and acceleration mechanism. 
We define how we measure certain disk properties, such as the location of the disk surface or the mass fluxes
in the physically different areas. 
We explore the role of the diffusivity, the strength and geometry of the magnetic field in respect to the outflow 
and accretion rates.

\subsection{Disk structure and disk surface}
We define the disk surface as a surface where the {\em radial velocity changes sign}. 
In a steady state the area where the radial velocity and the magnetic torque changes sign is almost identical.

Figure \ref{fig:def_struct} shows the typical structure of the disk-jet system. 
Several, physically different regions can be distinguished - the inflow area, the jet acceleration area, 
the launching area, and the accretion domain.
These are separated by colored lines.
White and black lines mark the disk surface that separates disk and corona regions. 
Two other lines separate the accretion, launching and acceleration areas. 
We define the {\em accretion region} as the area where velocity is mainly radial, $\Vth < 0.1 \Vr$, 
and the {\em acceleration region} as the area where the flow velocity 
is parallel to the magnetic field, $ \rm{\sin} ( {\rm angle}(\Btot,\Vtot)) < 0.1 $. 
As the {\em launching area} we characterize the region in between.

In our simulations, the position of the disk surface as defined above remains about constant in time.
This confirms our choice of the control volume (see below) and our
                  choice to fix the diffusive scale height during the simulation.

According to our boundary conditions we prescribe a weak inflow into the region between the inner disk 
radius and the rotational axis.
This inflow provides the matter content as well as the pressure balance along the rotational axis.
The astrophysical motivation can be the presence of a central stellar magnetic field or a 
stellar wind.
In Figure~\ref{fig:def_struct} the inflow area is the area between the rotational axis and the red line that 
marks the magnetic field line rooted in the inner disk radius at the midplane. 
In all figures below the magnetic field line closest to the axis always corresponds to the 
magnetic field line anchored at the inner disk radius\footnote{
Note that there are magnetic field lines which still penetrate the disk, but are 
not rooted at the disk midplane. 
These lines originate from inside the inner disk radius and are considered as intermediate between the axial 
coronal region and the main disk outflow. 
The pure inflow, which is prescribed from the coronal region along the inner boundary, is moving
with the injection speed, thus is not accelerated.
}.

\subsection{Launching mechanism}
Here we briefly comment on the jet launching mechanism in our reference simulation
that is - as in previous simulations - the Blandford-Payne magneto-centrifugal driving.

\begin{figure}
\centering
\includegraphics[width=8cm]{\figurepath/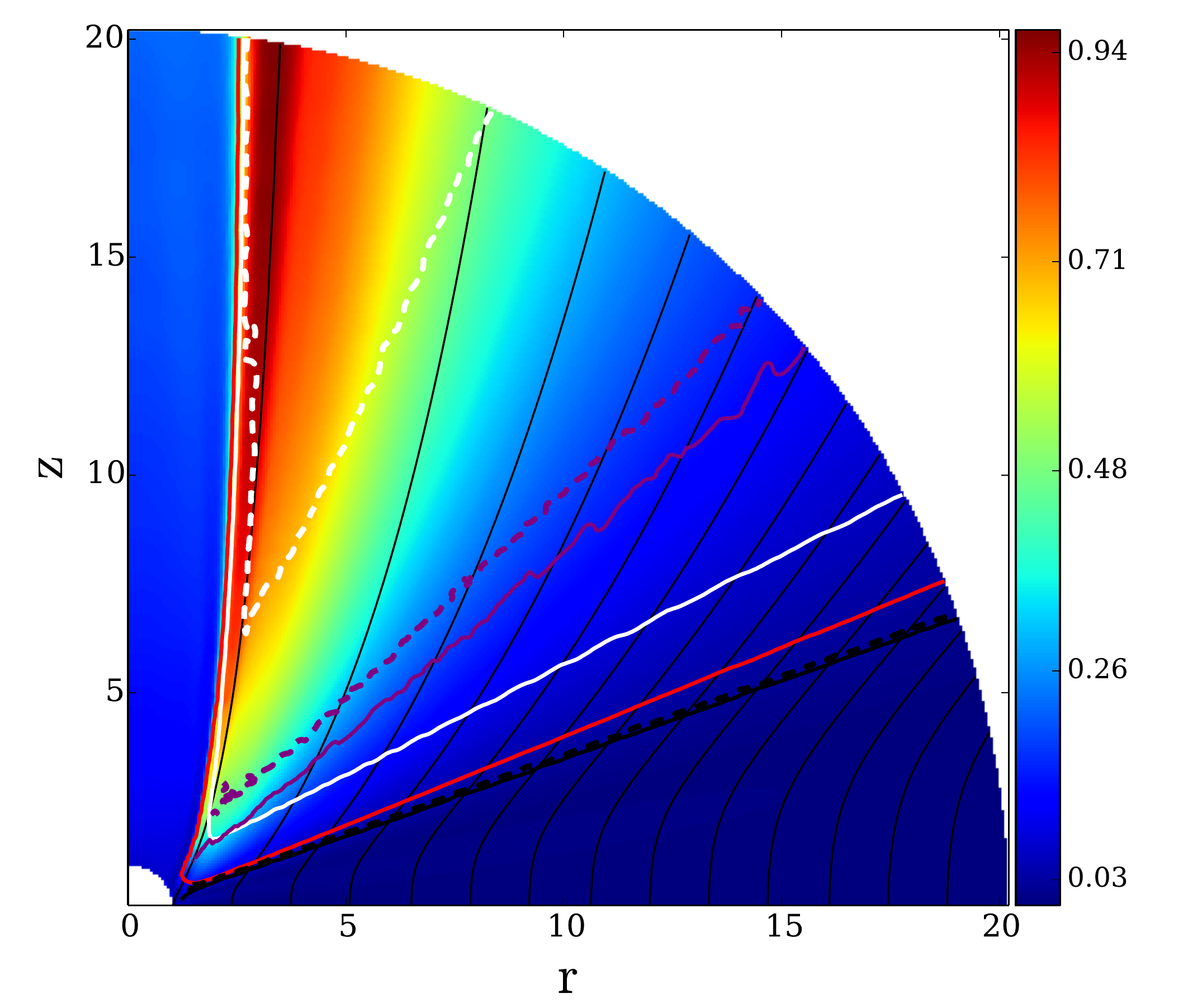}
\caption{Importance of the Lorentz force with respect to the pressure and centrifugal forces for reference simulation at T = 10,000. Shown is the poloidal speed (by colors), the poloidal magnetic field (black thin lines).
Alfv\'en (white), fast-magnetosonic (dashed white), sonic (red) surfaces.
Thick black lines denote the surface where Lorentz force is equal to pressure force components: parallel (dashed) and perpendicular (solid) to the magnetic field.
Thick purple lines denote the surface where Lorentz force is equal to centrifugal force components: parallel (dashed) and perpendicular (solid) to the magnetic field.
Thick black lines denote the ratios of Lorentz to pressure forces components: parallel (left) and perpendicular (right) to magnetic field.}
\label{fig:def_forces}
\end{figure}

\begin{figure}
\centering
\includegraphics[width=8cm]{\figurepath/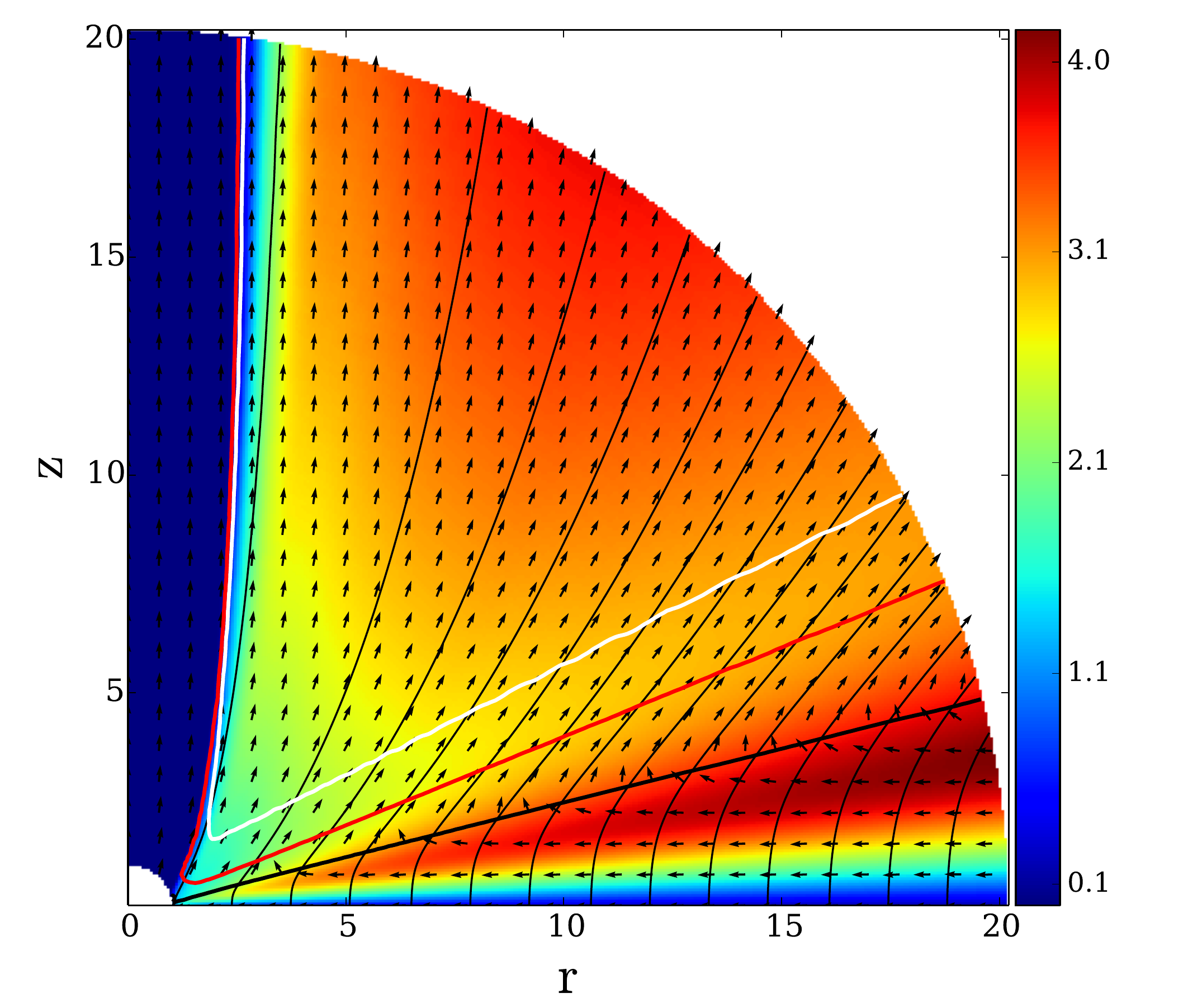}

\caption{The ratio of the toroidal to poloidal magnetic field for the reference simulation at T = 10,000. Lines represent the disk (thick black line) the sonic (red line), the Alfv\'en (white line) surfaces. Arrows show normalized velocity vectors.}
\label{fig:def_btbp}
\end{figure}


As demonstrated above (see Figure~\ref{fig:def_struct}), the magnetic torque $r F_{\phi}$ changes sign on the disk surface. 
It is negative in the disk and positive in the corona.
Thus, the magnetic field configuration established extracts angular momentum from the disk. 
The angular momentum extraction relies on the induced toroidal magnetic field component which plays a 
key role in transferring the angular momentum $\sim B_r B_{\phi}$.
Gaining angular momentum, the material that is loaded to the field lines from the accretion disk is pushed 
outwards by the centrifugal force. 

In order to illustrate the acceleration process, we show the magnitude of Lorentz force with respect to the thermal 
pressure and centrifugal forces. 
Figure~\ref{fig:def_forces} shows the contours where the perpendicular and parallel components of the Lorentz force 
are equal to the perpendicular and parallel components of pressure and centrifugal force, respectively. 
In the accretion disk both the pressure and the centrifugal forces dominate the poloidal component of the Lorentz force. 
Below the disk surface the Lorentz force (toroidal component) extracts the angular momentum from the disk. 

Since the Lorentz force increases along the outflow, it is worth to check the decomposed Lorentz force components 
$F = \nabla \times B \times B$ 
in the directions parallel and perpendicular to the magnetic field \citep{1997A&A...319..340F}. 
The ratio between toroidal and parallel components of the magnetic field,
\begin{equation} \label{eq:btbp}
\frac{F_{||} }{\Fphi} = - \frac{\Bphi}{\Bp},
\end{equation}
is shown in Figure~\ref{fig:def_btbp}.
We see that the centrifugal force is stronger in the inner area of the disk rather than in the outer parts of the disk.

At the sonic surface, the Lorentz force overcomes the pressure forces. 
From this point on the main acceleration force is the centrifugal force. 
Further along the outflow - between the Alfv\'en surface and the fast surface - the Lorentz force becomes 
the main accelerating force.

\subsection{Mass flux evolution}
We now explore the mass flux evolution namely the accretion and ejection rates.

In $\theta$-direction the control volume is enclosed by the disk surface (as defined above) $S_S$ and the 
disk midplane. 
The two other surfaces which enclose the control volume are marked by $S_1$ and $S_R$, and correspond to the vertical 
arcs at the innermost disk radius ($R = 1$) and at any other radius $R$. 
The control volume defined by these surfaces is denoted by $V(R)$.

\begin{figure*}
\centering
\includegraphics[width=5.9cm]{\figurepath/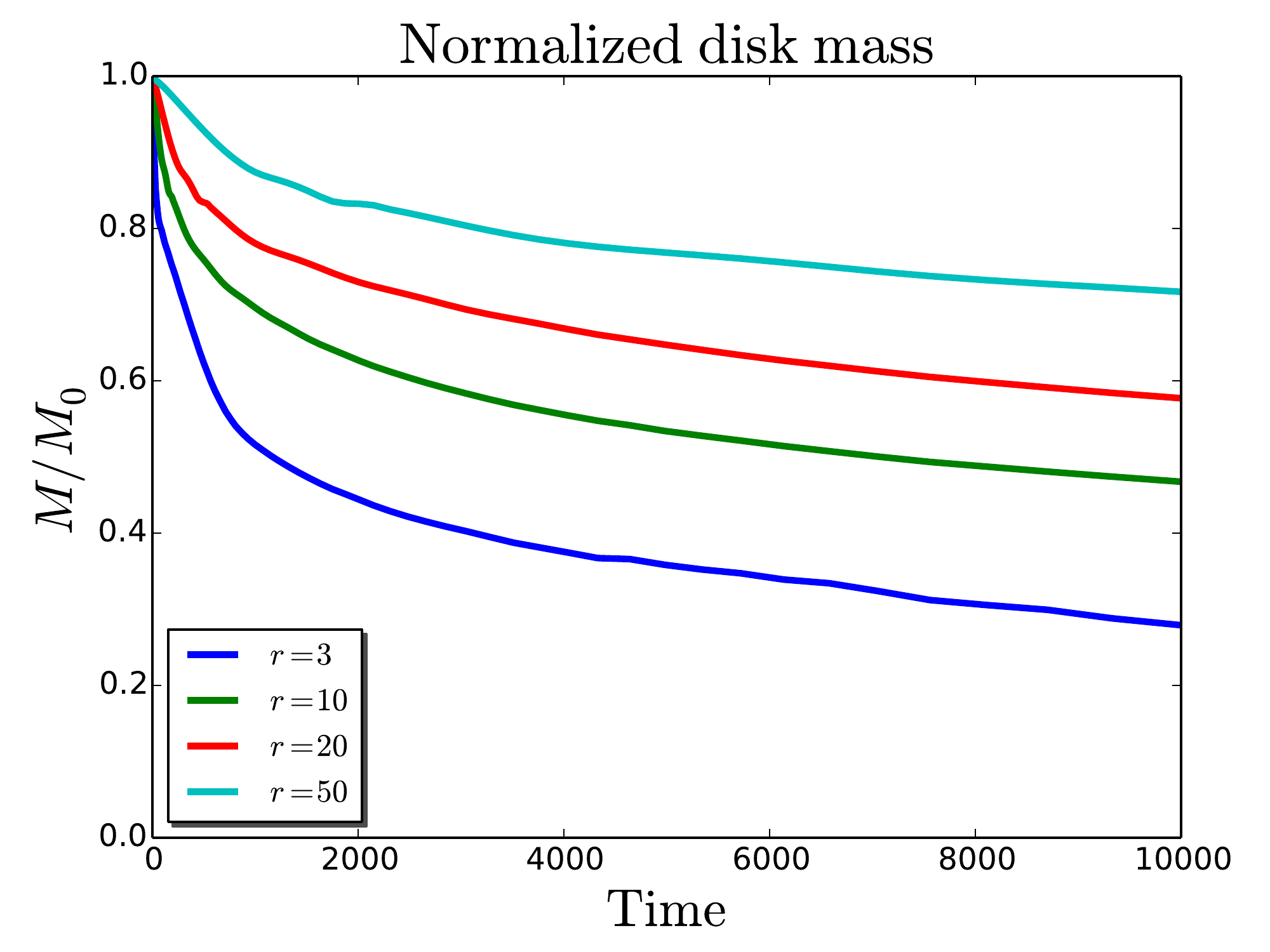}
\includegraphics[width=5.9cm]{\figurepath/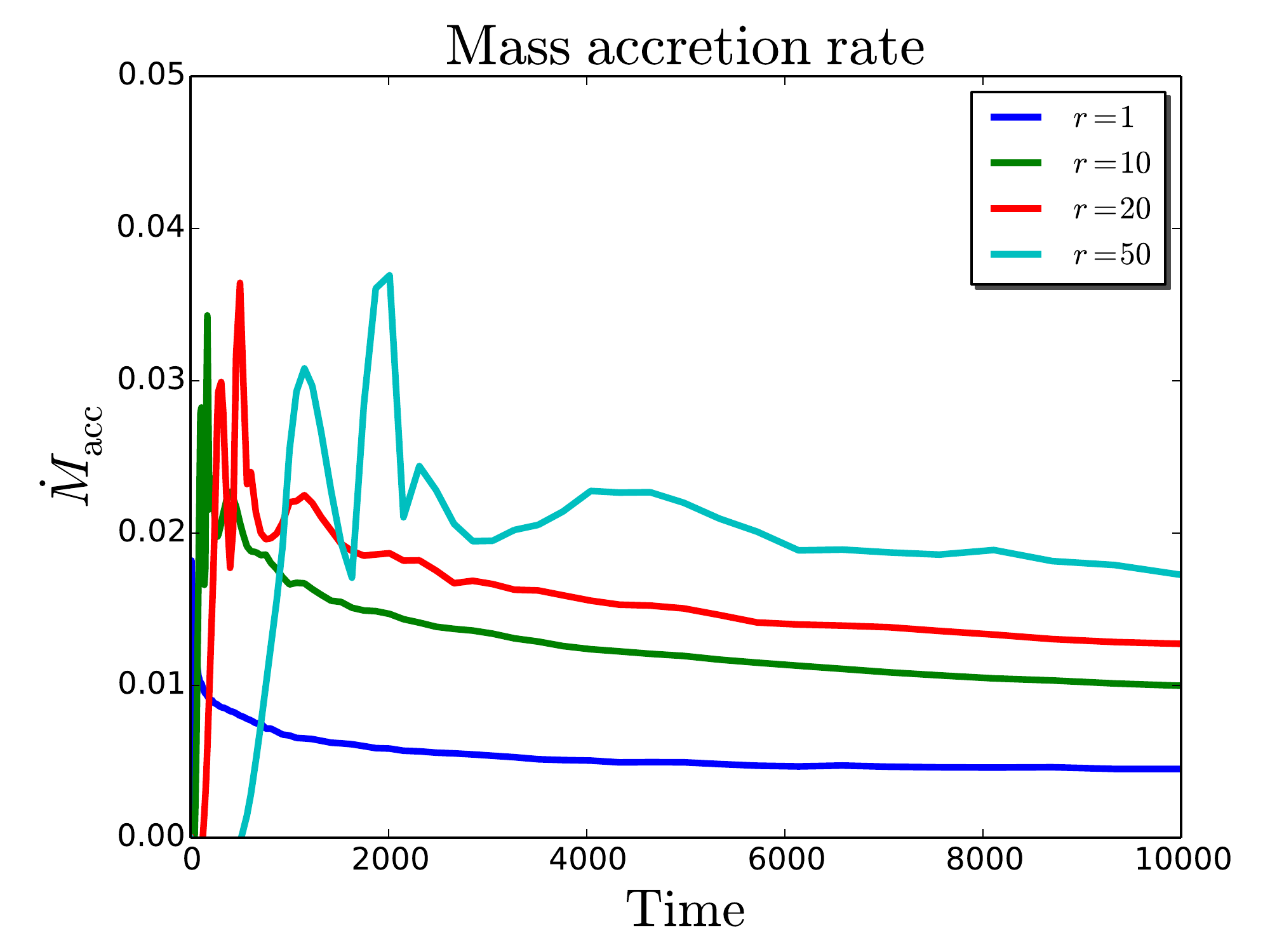}
\includegraphics[width=5.9cm]{\figurepath/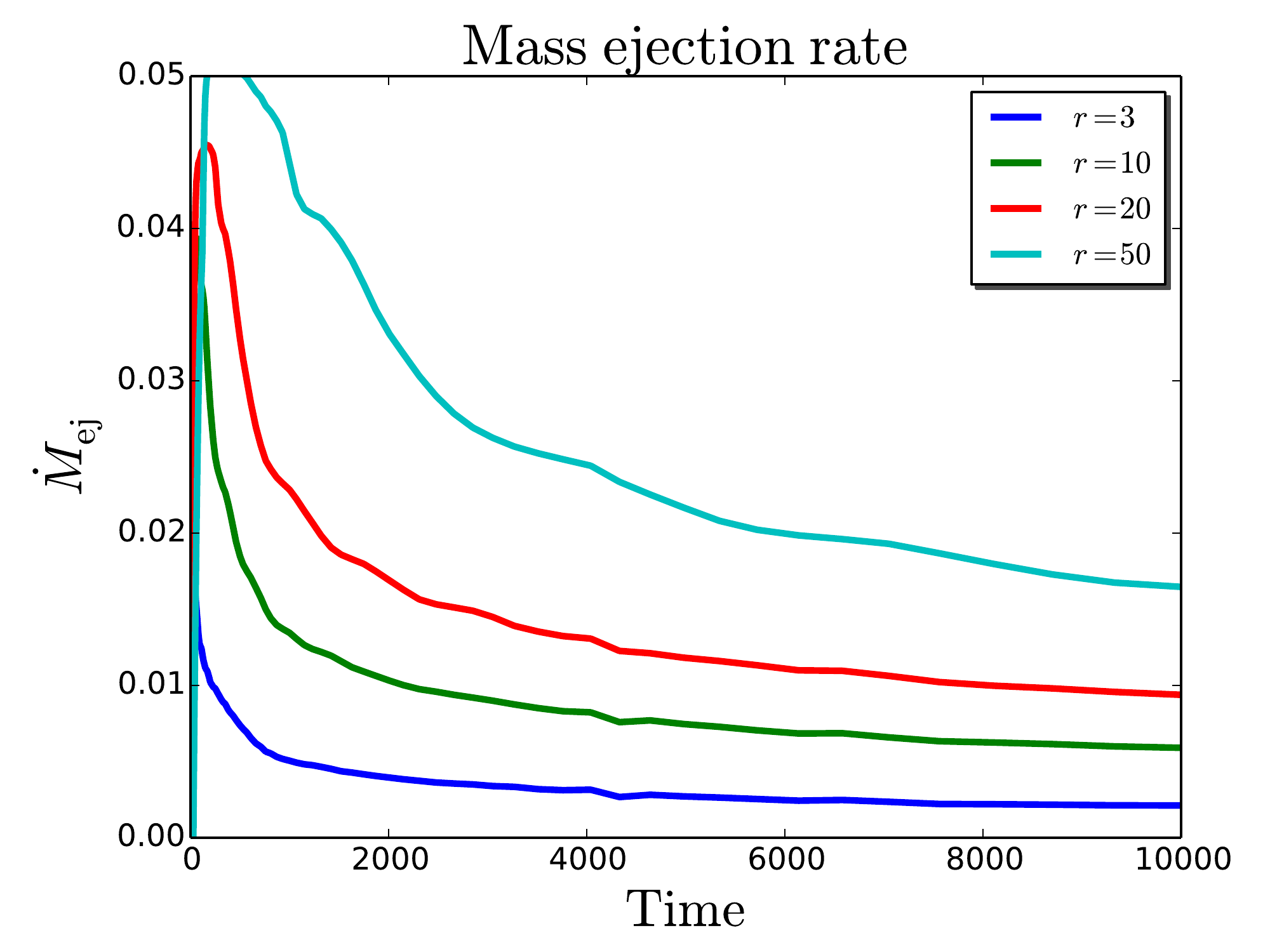}

\caption{Time evolution of the disk mass (left), the mass accretion rate (center) and 
the mass ejection rate (right) of the reference simulation at different radii.}
\label{fig:def_massflux}
\end{figure*}

Thus, the disk mass enclosed by a radius $R$ follows from
\begin{equation}
M(R) = 2 \int_{V(R)}\rho dV,
\end{equation}
while the mass accretion rate at a certain radius $R$ is
\begin{equation}
\Macc (R) = - 2 \int_{S_1}^{S_R} \rho \Vtot_p\cdot \dS,
\end{equation}
and the mass ejection rate is integrated along the disk surface,
\begin{equation}
\Mej (R) = -2 \int_{S_S} \rho \Vtot_p\cdot \dS.
\end{equation}

The mass accretion rate is defined positive if it increases the mass in the control volume. 
The mass ejection rate is defined positive if it decreases the mass of the control volume. 
The factor of two in front of the integrals takes into account the fact that only one hemisphere is treated.
Note also a minus sign in front of the integrals. 

It is common to introduce the ejection index $\xi$  which is based on the mass conservation law for a 
steady solution \citep{1995A&A...295..807F}. 
It basically measures the steepness of the radial profile of the accretion rate along the midplane. 
Setting the outer radius to $r$ and the inner radius to unity, the ejection index interrelates ejection and accretion,
\begin{equation}
\frac{\Mej}{ \Macc } = 1 - r^{-\xi}.
\end{equation}

We obtain the ejection index by a linear approximation of $\xi = -\log(1-\Mej/\Macc)/\log(r)$ within $r = [2, 10]$
The higher the ejection index, 
the higher the fraction of accreted matter being ejected within a given radius, and 
the less matter reaches the inner boundary. 
For our reference simulation $\xi \simeq 0.3$ at $T \simeq 10.000$.

Although the disk continuously loses mass (Figure~\ref{fig:def_massflux}), after dynamical times 1000-2000 
the disk mass loss is much smaller that the corresponding ejection and accretion rates. 
We therefore state that the simulation evolves through a series of quasi steady states. 
We find that a continuous disk mass loss is a typical feature of a simulation like our reference simulation. 
This is because the mass accretion from outside some outer disk radius is not able to sustain the mass which is 
lost by accretion and ejection within this radius.
We also find that the standard diffusivity model typically leads to a {\em magnetic field} distribution in the 
disk that is almost constant in time (not in space).
These two facts result in an increase of the disk magnetization in the inner disk, that in turn leads to a 
more rapid accretion in the inner disk.

Figure~\ref{fig:def_massflux} shows the time evolution of the mass accretion and ejection rates. 
Note also the general decrease of the mass fluxes over time, which is a direct consequence of the decrease 
of the disk mass. 
The behaviour and actual values of the mass fluxes are typical to the literature values. 
One should notice two distinct features of the mass fluxes.
First, the higher integration volume, the higher the mass ejection rate.
Second, in jet launching disks the mass accretion rate must increase with the radius.
These plots also indicate that the evolution of the system can be seen as a consecutive evolution 
through a series of quasi steady states.

\subsection{Magnetic field bending parameter study}
Here we discuss simulations, investigating the influence of the initial magnetic field bending parameter $m$.
We have varied $m$ from 0.2 (strongly inclined) to 0.9 (almost vertical). 

Our main result is that, although the simulations evolve slightly different initially, 
on the long-term evolution they are almost indistinguishable.

\begin{figure}
\centering
\includegraphics[width=8cm]{\figurepath/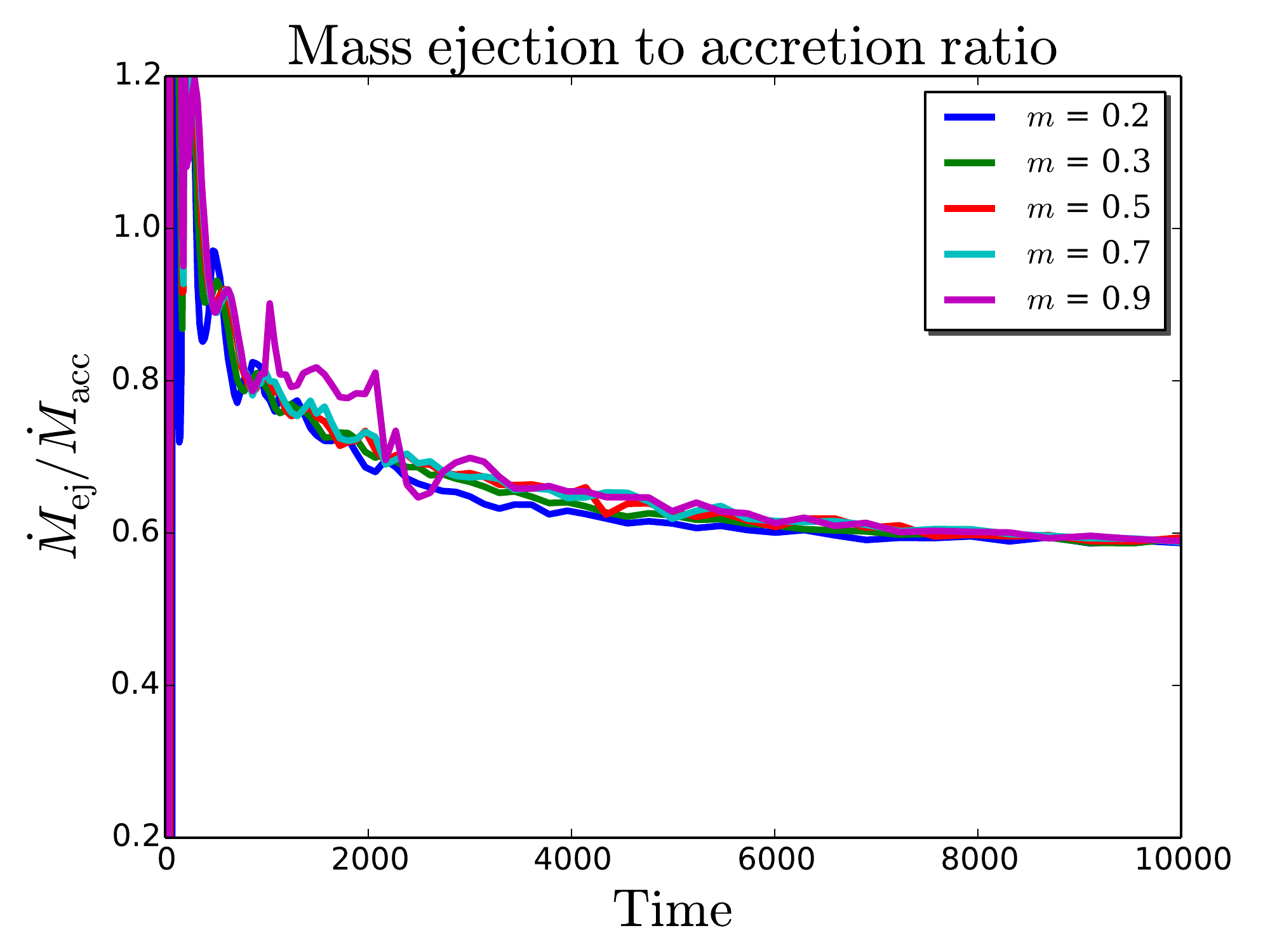}
\caption{Time evolution of the mass ejection to accretion ratio for simulations evolving from 
an initial magnetic field distribution with different initial bending parameter $m$.}
\label{fig:def_m}
\end{figure}

Figure~\ref{fig:def_m} shows the time evolution of the ejection to accretion mass flux ratio. 
The fluxes are again computed for the control volume extending to $R = 10$. 
It seems to take a few 1000 dynamical time steps for the simulation to lose the memory of 
the initial magnetic field configuration, but at $t= 10,000$ convergence has been obviously reached. 
This is also true for the fluxes of angular momentum and energy, and holds as well for the corresponding 
flux ratios.

The reason why the simulations convergence into a single - specific - configuration is 
the fact that it is mainly the diffusivity model that governs the evolution of the 
magnetic field evolution.
When we start the simulations with the same initial magnetic field strength at the midplane,
this results in exactly the same magnetic diffusivity profile.
Since we explore rather weak magnetic fields (weaker than the equipartition field),
the underlying disk structure cannot be changed substantially by the Lorentz force.
In contrary, the magnetic field distribution adjusts itself in accordance with the diffusivity model, 
which has the same vertical profile ab initio.

The convergence we observe for these simulations, starting from an initial magnetic
field with different bending, again confirms the reliability of our model in general.

\subsection{Resolution study}
Here, we briefly present example results of the resolution study.
We have performed simulations with a grid resolution of (0.5, 0.75, 1.0, 1.5, 2.0)-times
our standard resolution of 128 cells per quadrant, 
corresponding to (64, 96, 128, 192, 256) cells per quadrant, 
or approximately (8, 12, 16, 24, 32) cells per disk height $2\epsilon$. 
Note that once the resolution in $\theta$-direction and the radial extent of the disk is chosen,
the resolution in $R$-direction is uniquely determined (see Section~\ref{sec:numgrid}).

\begin{figure*}
\centering
\includegraphics[width=5.5cm]{\figurepath/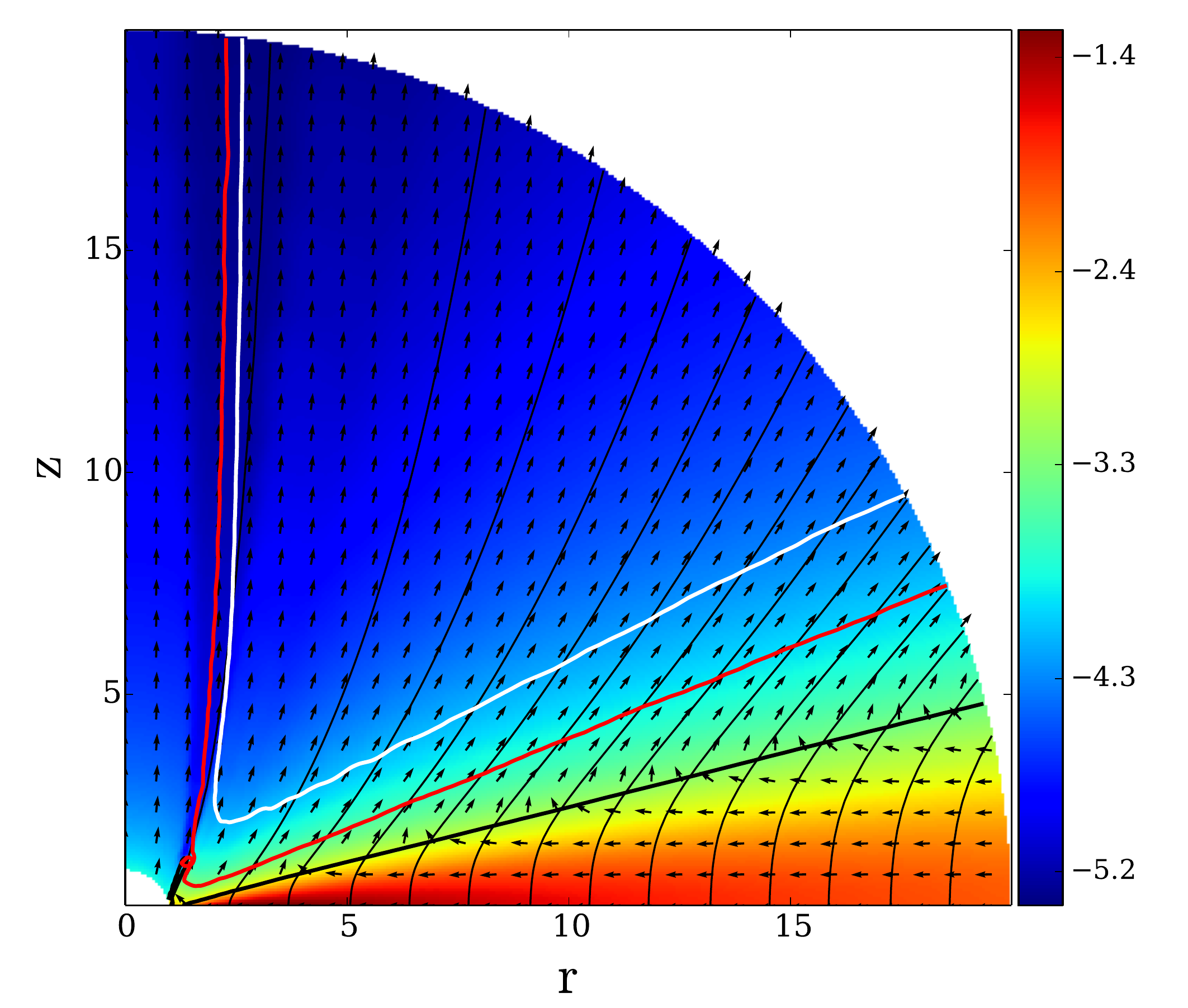}
\includegraphics[width=5.5cm]{\figurepath/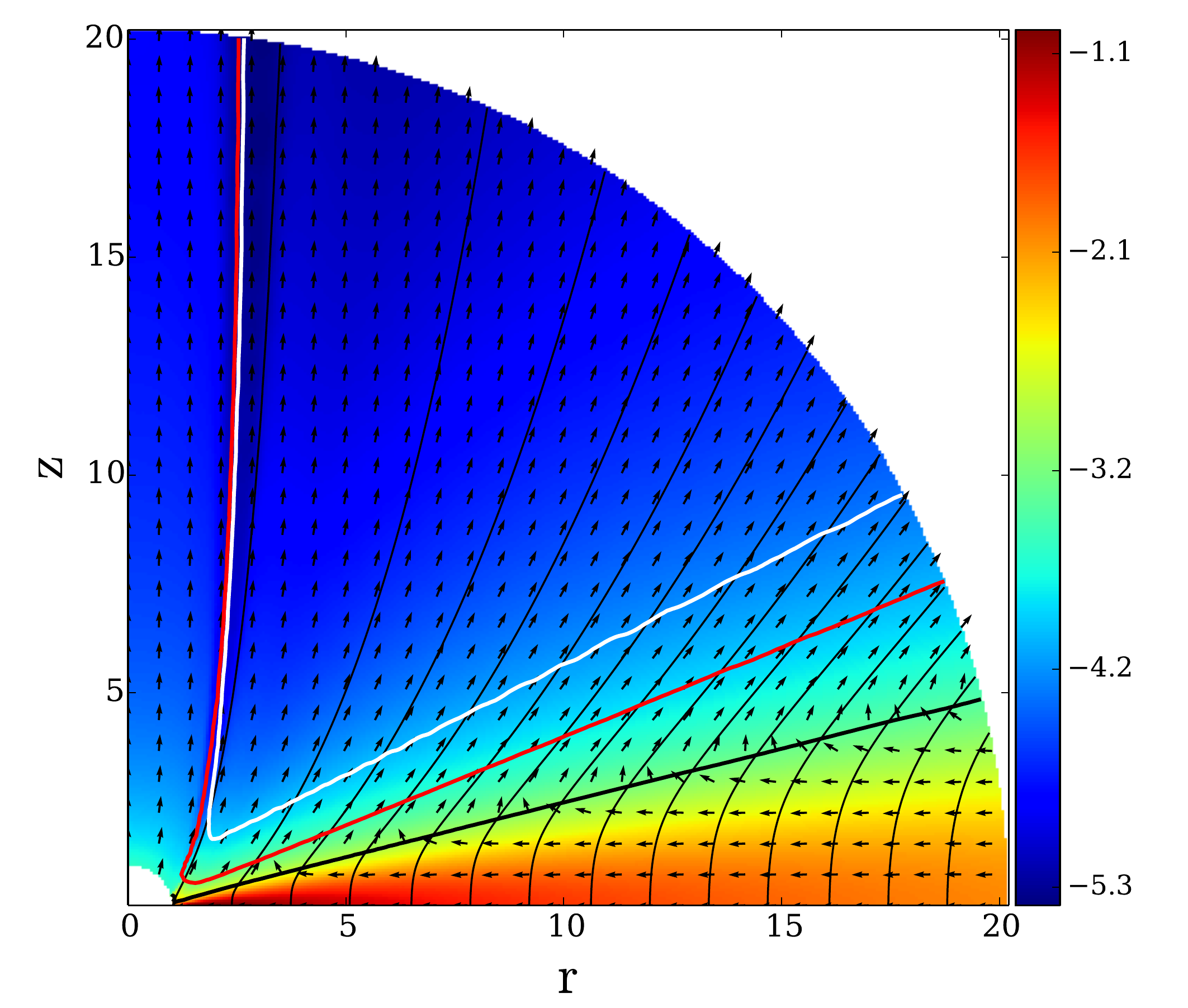}
\includegraphics[width=5.5cm]{\figurepath/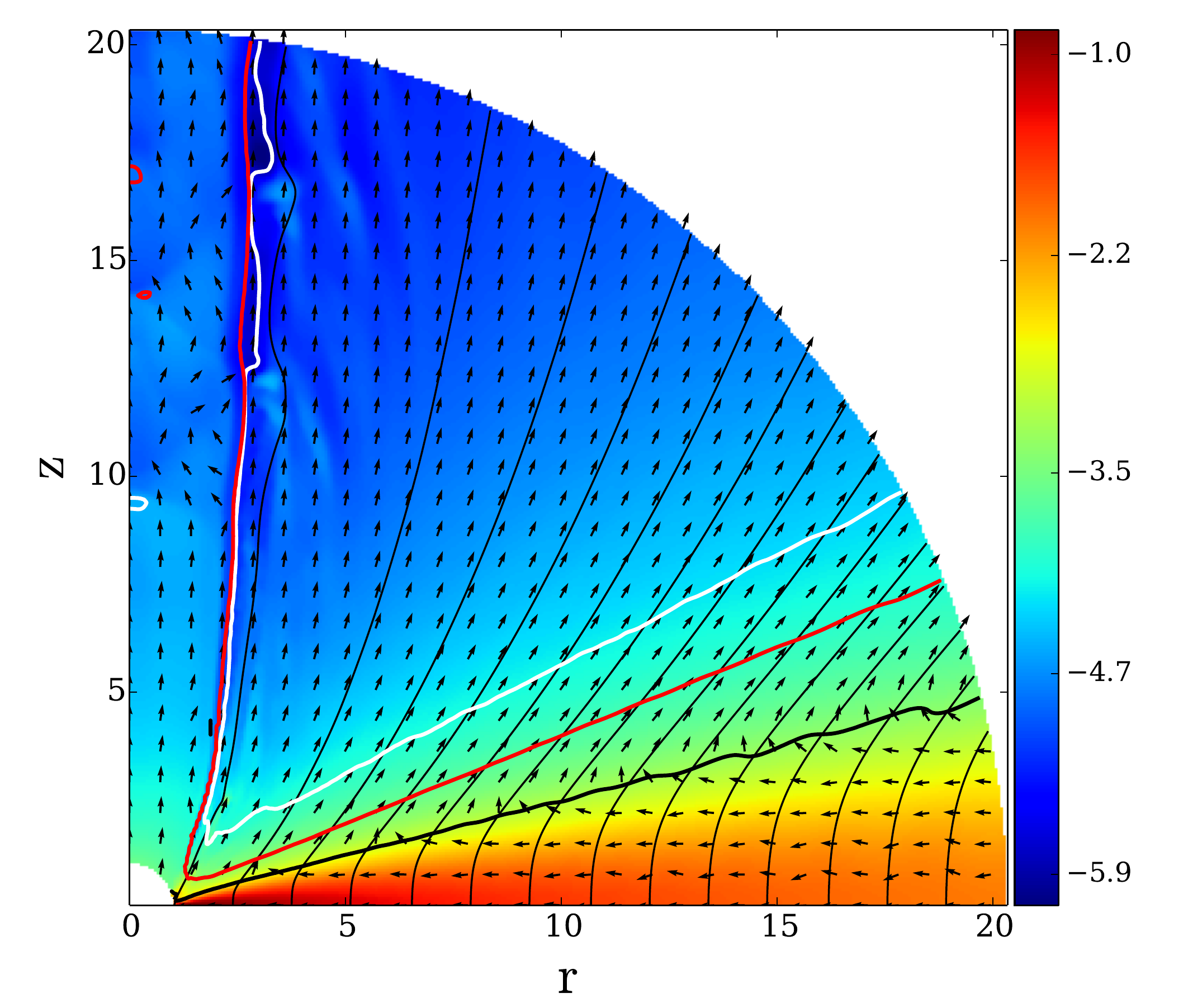}
\caption{Resolution study. 
Shown are snapshots of the density distribution at $t = 10,000$ for simulations with
different resolution. From left to right the resolution is  
(8, 16, 32)
cells per disk height ($2\epsilon$). 
Black lines mark the magnetic field lines (plotted as flux surfaces).
Lines represent the disk (thick black line) the sonic (red line), the Alfv\'en (white line) surfaces.
Arrows show normalized velocity vectors.}

\label{fig:res_all}
\end{figure*}

For the resolution study all simulations were performed up to typically 10,000 time units.
Figure~\ref{fig:res_all} shows snapshots of some of these simulations.
Essentially, we see an almost identical disk-outflow structure, indicating that numerical convergence 
has indeed been reached.

\begin{figure}
\centering
\includegraphics[width=8cm]{\figurepath/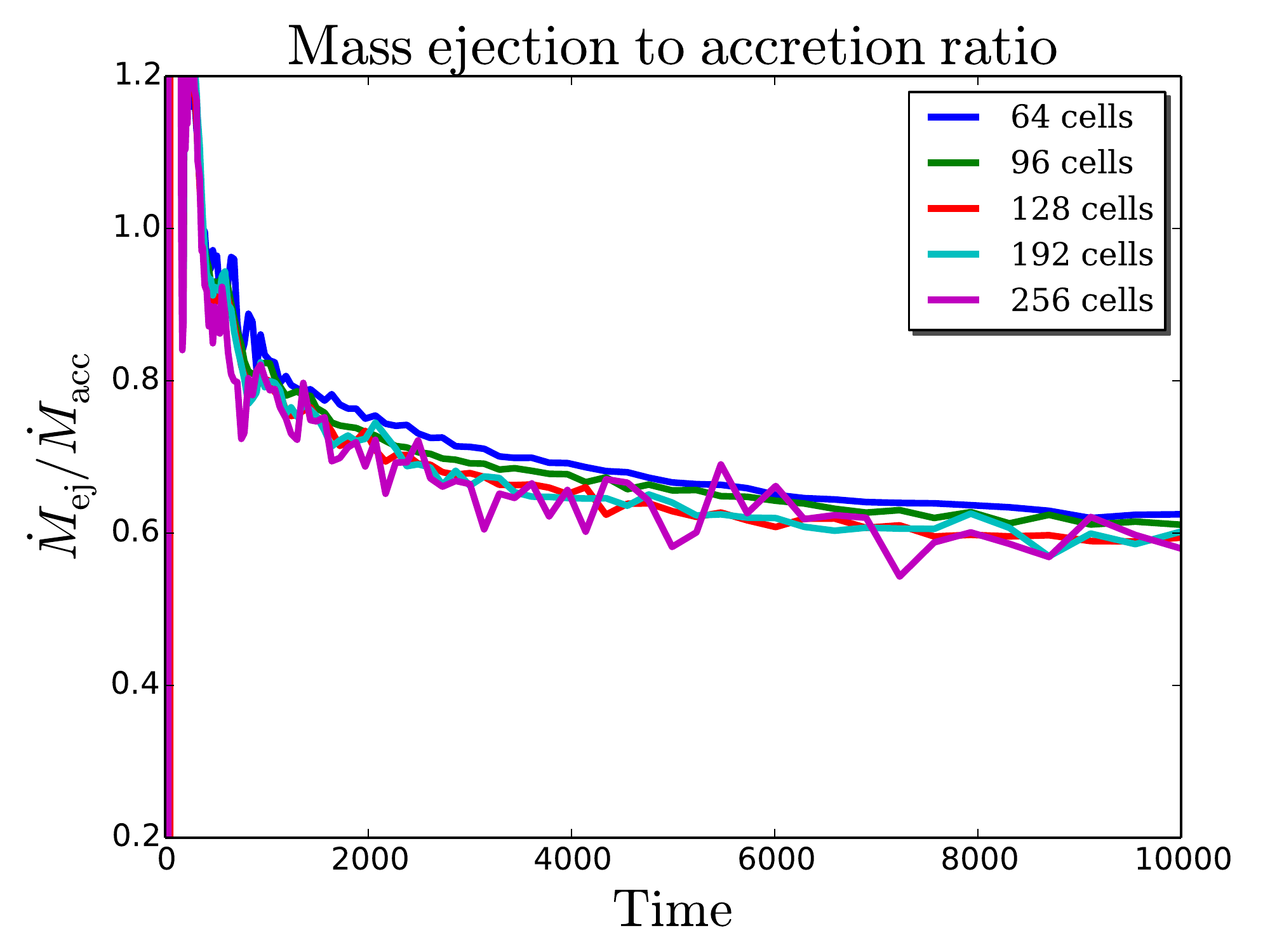}

\caption{The evolution of the ejection-to-accretion ratio at R=10 for the reference simulation with different resolution.}

\label{fig:res_tmx}
\end{figure}

Figure~\ref{fig:res_tmx} shows the time evolution of the mass ejection to accretion ratio,
again integrated throughout the control volume $R<10$,
for simulations of a different resolution.
We notice two particular issues. 
First, all curves bunch together at the mass ejection-to-accretion ratio of about $0.6$, indicating
convergence of the simulations (this is also true for other flux ratios).
Second, the simulation with the highest resolution show some intermittent behaviour 
(see Figure~\ref{fig:res_all}).
This might be related to the spatial reconstruction in non-Cartesian coordinates close to the symmetry axis
or to the ability to resolve more detailed structures.
We conclude that for the resolution chosen our simulations have converged
for the launching region.

Note that although the spherical grid is beneficial for  disk and 
outflow launching studies, mainly due to the higher resolution of the inner disk, 
its application to the jet propagation further away from the jet source is limited because of the 
lack of resolution at larger radii.

\begin{figure*}
\centering
\includegraphics[width=5.9cm]{\figurepath/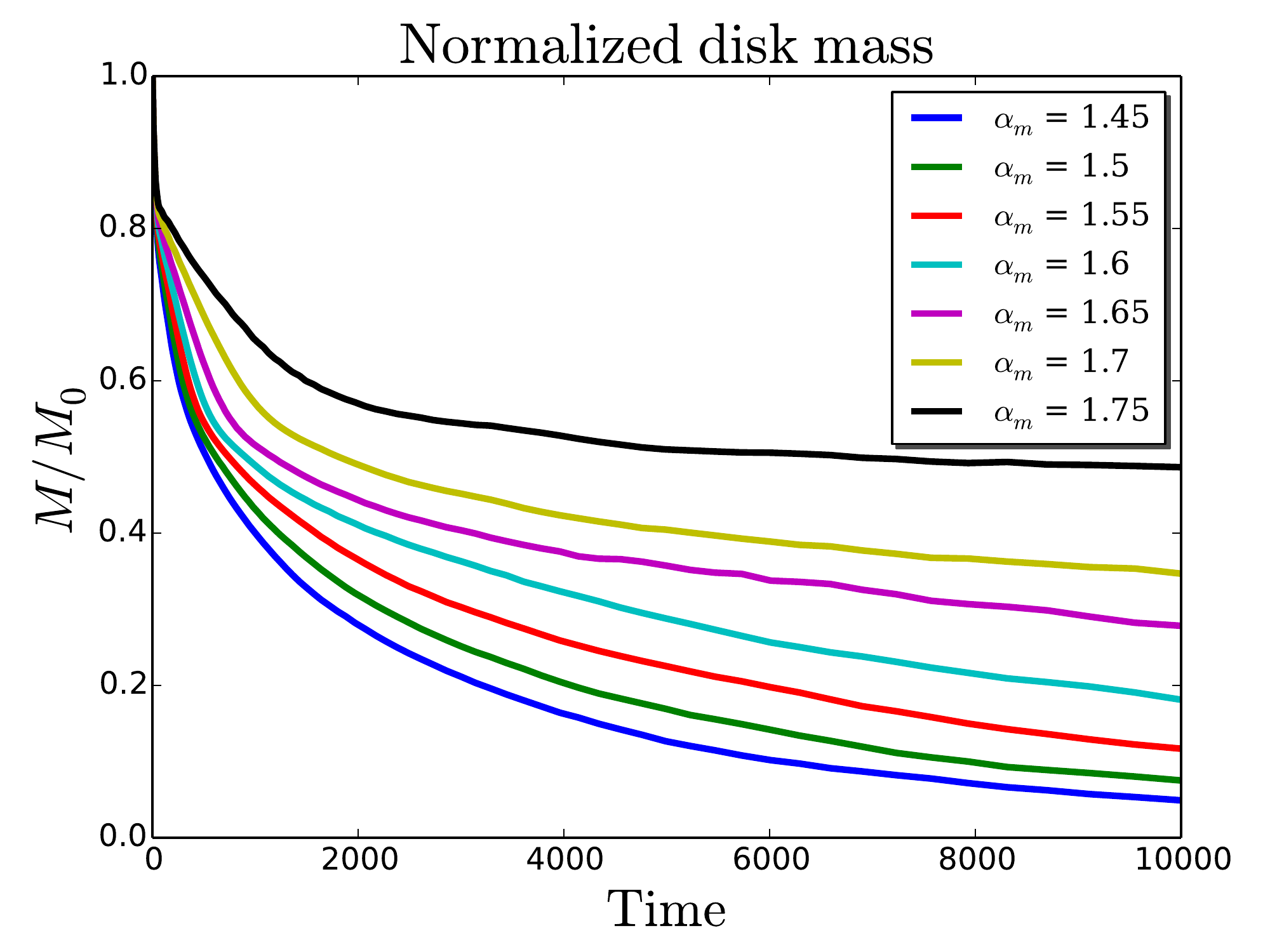}
\includegraphics[width=5.9cm]{\figurepath/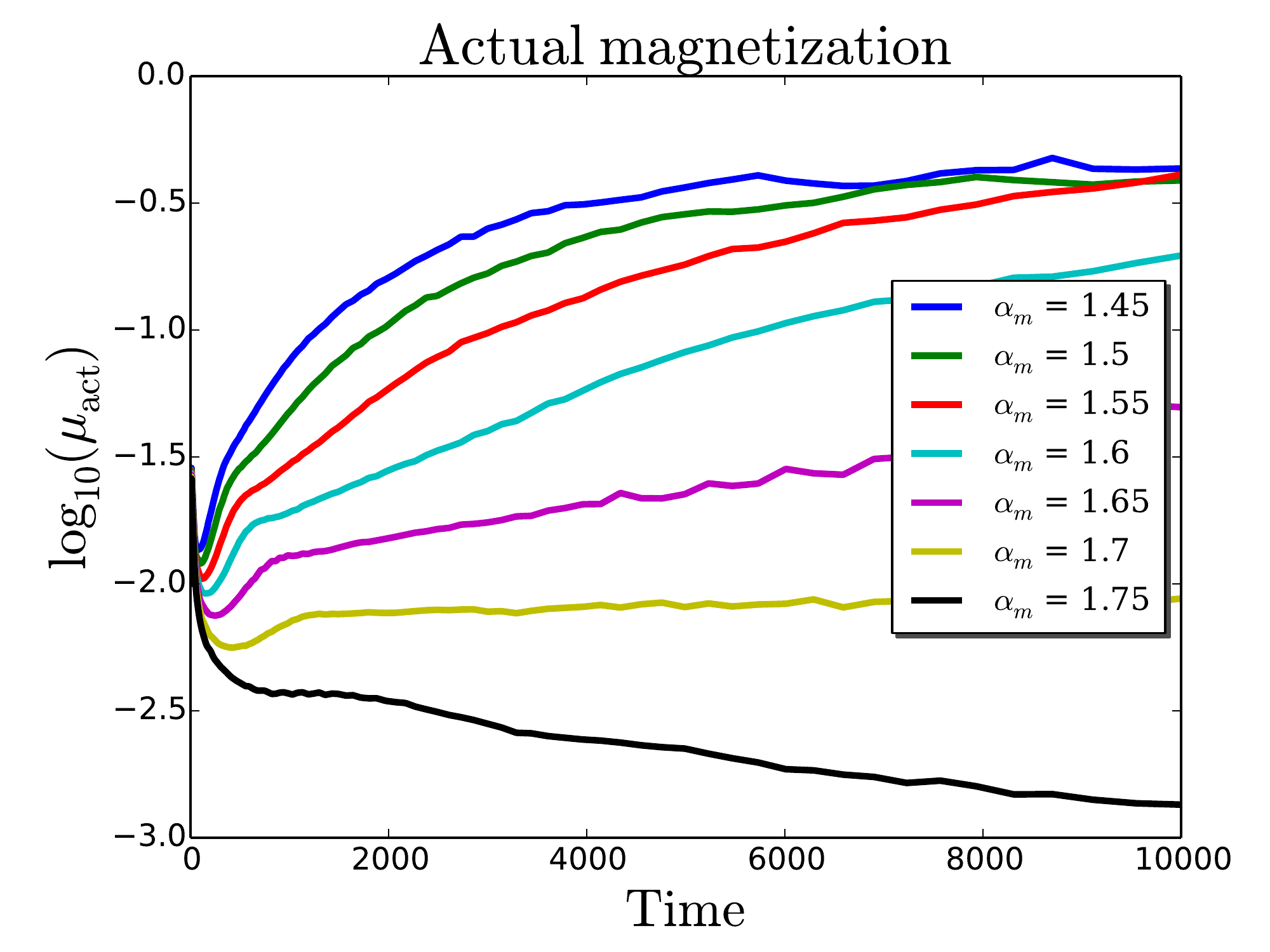}
\includegraphics[width=5.9cm]{\figurepath/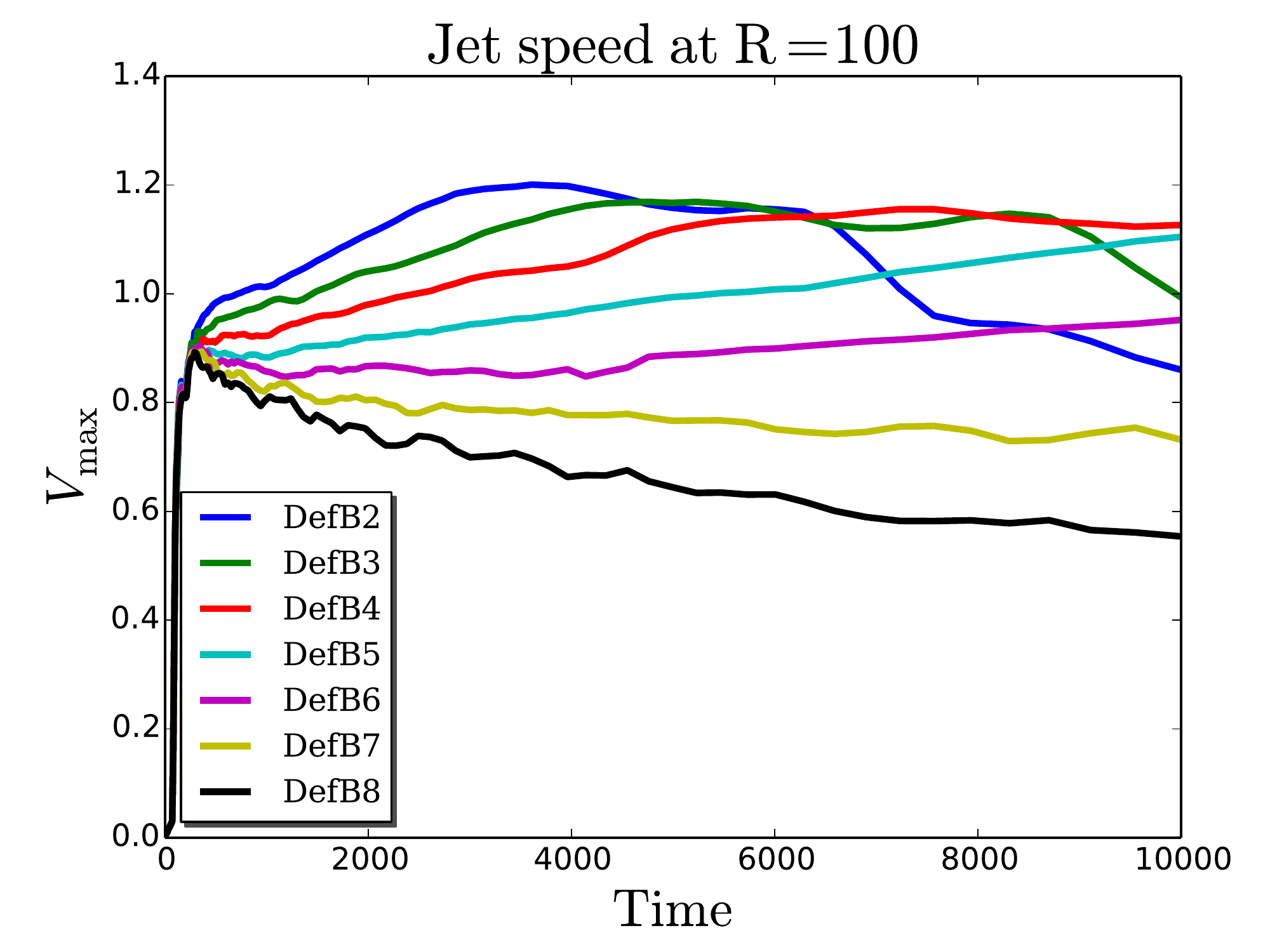}

\caption{Time evolution of the normalized disk mass (left), the actual magnetization of the disk
(center), and the jet terminal speed (right) for different diffusivity ($\am$) strength.}

\label{fig:def_xevol}
\end{figure*}

\section{Magnetization analysis}
In the following we investigate a number of physical processes of jet launching by comparing 
different simulations similar to the reference simulation.

As we have mentioned above, the evolution of our reference simulation can be seen as a
sequence of quasi-steady states. 
The slow, but constant decrease of the disk mass eventually leads to the change of the disk magnetization. 
This is more prominent in the inner part of the disk, whereas the magnetization of the outer disk does not 
change too much.
This feature brings the opportunity of studying the disk and jet quantities with respect to the {\em actual} 
disk magnetization.

In this section we present a set of simulations similar to our reference simulation, however
applying a slightly different choice of parameters.
The reference simulation was chosen such that diffusive processes are in balance with advective processes. 
Now, by choosing a slightly different diffusivity parameter $\am$, these processes now are out of the balance.
This leads either to further faster advection or diffusion of the magnetic field. 
Naturally, in case of lower $\am$, advection dominates, and, thus, magnetization grows, leading to even faster 
advection. 
In the opposite case, the disk magnetization decreases.

This approach allows us to study the evolution of the accretion-ejection in a very general way.
Each of the simulations applied has started from the same initial conditions, however, it now follows a 
different evolutionary track and finally evolves into a quite different state of the system.

Figure~\ref{fig:def_xevol} shows the time evolution of the disk mass, the average magnetization of the inner 
disk (averaged between $R = 1.1$ and $R=1.5$), and the jet
speed here defined as

\begin{equation}
{V}_{\rm jet,act} \equiv {\max}(\Vp){|}_{R=100},
\end{equation}
the maximum jet speed obtained at $R= 100$, considering only those magnetic field lines rooted in the disk.

We emphasize that the jet speed as computed
here is only an extrapolation of the terminal jet speed (at infinity).
The general behaviour of the disk is as follows.
First, the different simulations behave rather similar.
After some time they start to differ from each other.
At later stages the disks and their outflows arrive at definitely different dynamical states.
The exact times at which this happens, depends of course on the radius for which we examine of the disk
properties.
Figure~\ref{fig:def_xevol} shows this evolution in the disk and jet quantities for the inner disk.
Therefore the time when simulations start showing differences is rather small, about few hundred
dynamical time steps.

Since we start the simulation with {\em no} accretion, initially diffusion is not in balance. 
However, electric currents are induced quickly within the disk, and subsequent angular momentum transport
results in accretion.
Depending on the value of diffusivity parameter $\am$ the system results either in advection-dominated 
regime (for $\am < 1.65$), or diffusion dominated regime ($\am > 1.65$). 
In principle an equilibrium situation is possible in which these two processes are in balance. 
In case of $\MUzero = 0.01$, the diffusivity parameter for an equilibrium situation is around 
$\am = 1.65$. 
We will refer to the {\em critial} diffusivity parameter $\alpha_{\rm cr}$ as the one corresponding to the 
equilibrium state when advection and diffusion balance each other.
Generally, the lower the diffusivity, the stronger the advection and thus the resulting magnetization.

We confirm \citet{2009MNRAS.400..820T} finding that in case of a strong magnetic field with $\mu \sim 0.3$
the disk structure changes substantially - the disk becomes much thinner in the inner region of the disk.
A stronger magnetic field exerts a stronger torque on the disk, leading to a faster accretion rate. 
Thus, at some point in time the accretion velocity becomes supersonic, $ {M}_{\rm R, act} >1$.
We consider this as the limit for applying our magnetic diffusivity model.

As clearly visible from the figures discussed above, for the present setup the 
current diffusivity model (Equation~\ref{eq:stdiff}) is only marginally stable - 
all deviations from the critical diffusivity will be further amplified. 
If magnetic diffusion dominates the disk, the magnetic field becomes weaker and weaker unless at 
about $\mu \sim 0.001$ the jet outflow cannot be sustained anymore. 
On the other hand, a weaker diffusivity leads to a faster accretion that also results in a runaway process.
One way to circumvent this problem is to apply a different model for the diffusivity, 
namely $\ass(\mu)$ (see Section~\ref{sec:strong}).

As might be easily seen from Figure~\ref{fig:def_xevol}, ongoing disk mass 
depletion in most cases leads to a higher degree of disk magnetization, a process 
which happens faster for less diffusive, thus higher advective simulations. 
A change in magnetization may substantially changes the dynamics of the disk. 
The stronger magnetization, for example, leads to higher jet speed.

One might notice the deviation in the behaviour of the jet terminal speed for the low value of $\am$. 
We believe that this results from the position where we calculate the terminal speed - for this case
the jet accelerates even further out and the asymptotic velocity is not reached at $R=100$ for low $\am$ 
(or highly magnetized) case and is still in a process of transforming the magnetic energy into kinetic.
In a follow-up paper we will discuss the terminal jet speed
much more detail, demonstrating that already for a moderately
weak magnetic field, $\mu \approx 0.05$, the terminal jet speed reaches unity.

In summary, we state that it is the {\em actual magnetization} in the disk that governs ejection and 
accretion and that is directly linked to various disk-jet quantities.

\subsection{Transport of angular momentum and energy}
Here we present the analysis of the angular momentum and energy transport in our simulations.
It is common to explore the angular momentum and energy transport by means of their fluxes through a control volume. 
We define the accretion angular momentum flux $\Jacc = \JAkin + \JAmag$ as the 
sum of kinetic and magnetic parts, keeping the same control volume as for 
the mass fluxes (see Appendix \ref{app:fluxes}). 

\begin{figure}
\centering
\includegraphics[width=8cm]{\figurepath/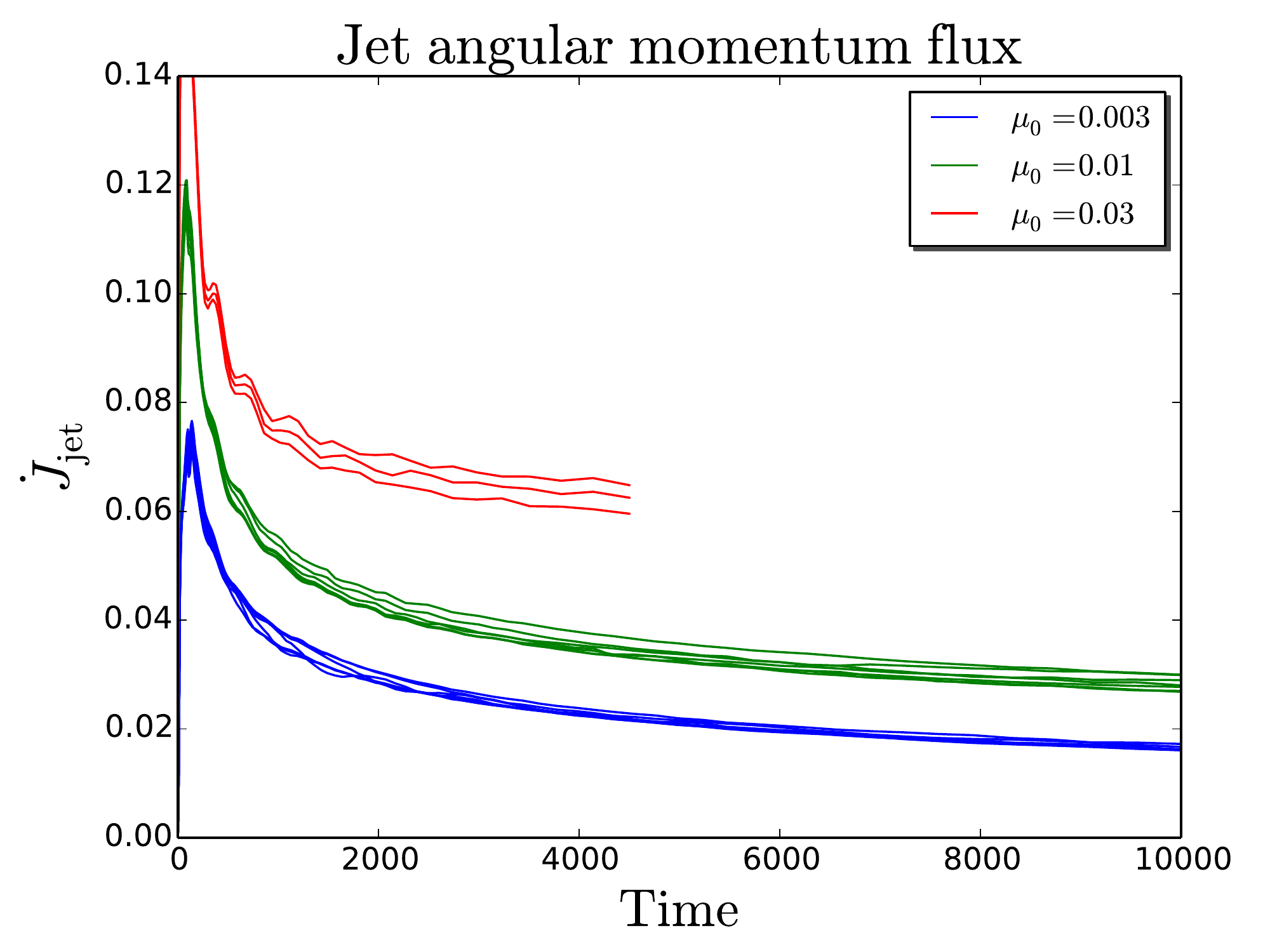}
\caption{Time evolution of the jet angular momentum flux for the reference type of simulation at radius 10. Simulations with different initial magnetic field form three distinct groups corresponding to $\MUzero = (0.003,0.01,0.03)$ (from down to up)}
\label{fig:comp_jjet}
\end{figure}


\begin{figure}
\centering
\includegraphics[width=8cm]{\figurepath/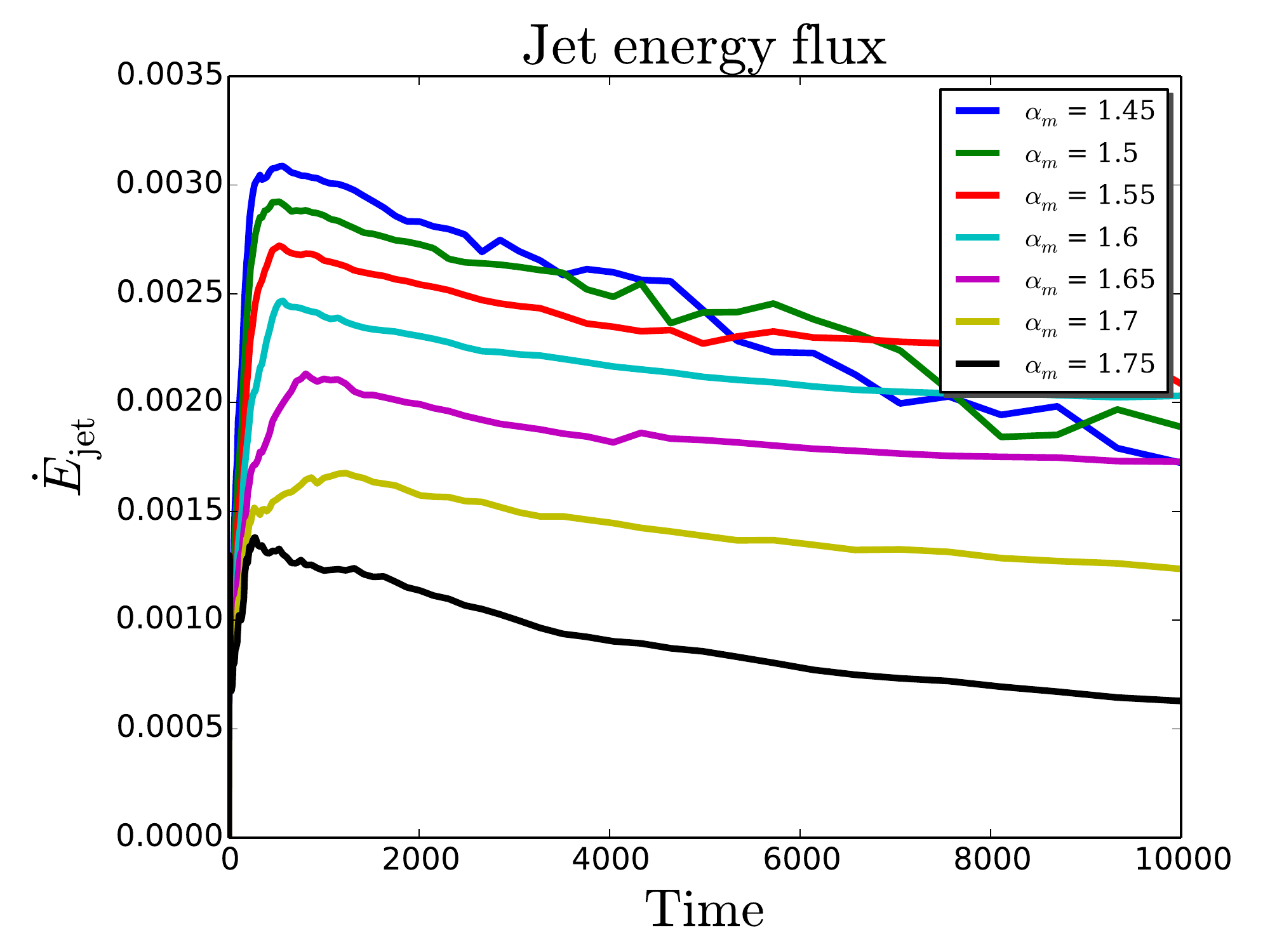}
\caption{Time evolution of the jet energy flux for the reference type of simulations at radius 10.}
\label{fig:def_B_ejet}
\end{figure}


Figure~\ref{fig:comp_jjet} shows the time evolution of the jet angular momentum 
flux for a number of simulations.
We divide them in in three different groups distinguished by their 
{\em initial} magnetization, $\mu_0 = 0.003, 0.01, 0.03$.
Different lines within each group represent simulations with different diffusivity parameter $\am$.
The jet angular momentum flux is calculated through the upper part of the control volume, thus up to R = 10.

As expected, the stronger the overall magnetic field strength is in the disk, 
the stronger torque it exerts, and thus the higher angular momentum fluxes we find.

Although the magnetic diffusivity parameter $\am$ differs within each group of lines 
(marked by different colors), the total torque measured for the corresponding 
simulations is about the same. 
This comes from the fact that the total torque is set by the global magnetic field. 
Thus, the evolution of the total torque is mainly set by the initial conditions.

We note that the simulations applying a strong initial magnetic field 
(represented by the upper bundle of curves in Figure~\ref{fig:comp_jjet}) 
have been interrupted earlier compared to usual evolution times. 
For these cases the inner part of the disk became highly magnetized.
The strong magnetic field changes the inner disk structure such that our 
model for the magnetic diffusivity cannot be applied anymore.
This is simply because the actual scale height of the disk significantly 
decreases and does no longer coincide with the initial disk surface.
The case of strong magnetic fields we consider as being beyond the scope of this paper. 
The underlying turbulence might be significantly suppressed as well.

In accordance with previous works (see e.g. \citealt{2007A&A...469..811Z}) we 
find that the ratio of angular momentum extracted by the jet to that provided 
by the disk accretion is always close to, but slightly larger than unity, 
$\dot{J}_{\rm jet} / \dot{J}_{\rm acc} $ \gax 1.
The main reason is that the accretion rate in the outer part of the disk 
is too low to compete with a strong mass loss by the disk wind at these radii.

We define the accretion energy flux (accretion power) as the sum of the 
mechanical (kinetic and gravitational), magnetic, and thermal energy fluxes,
$$\Eacc = \EAkin + \EAgrv + \EAmag + \EAthm,$$
and, similarly, the jet energy flux (jet power) by
$$\Ejet = \EJkin + \EJgrv + \EJmag + \EJthm.$$
In contrast to the angular momentum flux, most of the energy flux is 
being released from the inner part of the disk. 
This makes the energy flux very sensitive to the conditions in the inner disk. 
Indeed, Figure~\ref{fig:def_B_ejet} shows that the power liberated by 
the jet strongly depends on the diffusivity parameter, which is the main 
agent for governing the magnetic field strength. 
A weaker diffusivity parameter $\am$ leads to a higher magnetization, 
and, thus a higher jet power. The same is true for the accretion power.

We further find that the ratio of energy fluxes, namely the ratio of the
jet to accretion energy flux, is always close, but slightly lower than unity. 
This is also in accordance with \citet{2007A&A...469..811Z}.

In this section we have provided some evidence that it is the {\em actual} magnetization 
in the disk which governs the fluxes of mass, energy and angular momentum. 
In the next section we show how exactly these fluxes are connected to magnetization.

\subsection{Analysis of magnetization, diffusivity and fluxes}
In a steady state, diffusion and advection balance. 
Advection of the magnetic flux in principle increases the magnetic field strength, 
predominantly in the inner disk. 
In contrary, the diffusion smooths out the magnetic field gradient. 
Therefore, the diffusivity model applied is a key ingredient for
these processes, directly influencing the disk structure and evolution.

In the following analysis we will not focus on the profile or the magnitude of the magnetic 
diffusivity, but concentrate on the resulting magnetization and its time evolution. 
As discussed above, by changing the magnetic diffusivity parameter $\am$ we are able to explore how the 
actual disk magnetization influences various properties of the disk-jet system.

For each parameter run performed, we measure the {\em actual} physical variables 
in the disk-jet system, such as the {\em time-dependent} mean magnetization, accretion 
fluxes, jet fluxes, or the accretion Mach number. Naturally, the {\em actual} 
value of a certain property has evolved from the initial value during the simulation.
With mean value we denote the values averaged over a small area of the inner disk,
\begin{equation}
X \equiv <X (r,z=0)>.
\end{equation}
All mean quantities discussed are averaged over the inner disk midplane from $R = 1.1$ to $ R = 1.5$. 
The choice of the averaging area is motivated as follows. 
First, in order to avoid any influence of the inner accretion boundary we have moved the
inner integration radius about 10 grid cells away from it. 
Second, we are interested to examine the inner part of the disk, since it is the region where the 
magnetization is changing predominantly and it is the launching area for the most energetic
part of the jet. 
Third, we want to avoid large magnetic gradients affecting our averaging area. 
Although most of jet energy is launched from region broader than this small area, the area is seen as
representative.
We emphasize that the profiles of the jet power along the disk surface are similar for all simulations. 
In all cases we investigated, we find that the {\em general} behaviour of the 
physical outflow or disk quantities with respect to underlying disk magnetization does 
not depend on the area where the averaging is done (neither on the location nor on the size). 

Keeping all other parameters the same, we have carried out simulations varying 
the initial magnetization $\MUzero$ and the strength of the magnetic diffusivity $\am$.
For all our simulations, starting with different initial magnetization, 
$\MUzero=0.003$, $\MUzero=0.01$, $\MUzero=0.03$, we find that the 
interrelation between the different jet or disk quantities and the disk magnetization, 
is essentially the same.
We therefore present only one group of simulations, namely that with $\MUzero=0.01$. 
Although the initial magnetic field strength differs, 
we can already suspect at this point that it is the {\em actual} rather than the 
initial strength of the magnetic field in the disk that governs the disk accretion and ejection of the 
jet, and, thus, playing the major role in the launching process.
This has not yet been discussed in the literature so far, as 
most publication parametrize their simulations by the initial parameters.
An exception might be \citet{2012ApJ...757...65S}, who pointed out substantial changes in the disk 
plasma beta during the time evolution.
While the initial magnetization does not play a leading role for the launching process, it is,
as was previously shown, responsible for the magnitude of overall torque exerted on the disk.

An interesting representation of the evolution of the main disk-jet quantities are
($\mu, X$)-plots, where X stands for the examining variable. 
Note, that in these plots the time evolution is hidden.

\subsubsection{Accretion Mach number}
As it was shown by \citet{2011ppcd.book..283K}, there is a link between 
the mean accretion Mach number and the disk magnetization. 
In our terms this relation can be expressed as
\begin{equation}
{M}_{\rm R, act} = \frac{2}{\sqrt{\gamma}} q \MUact,
\end{equation}
where $q$ is the magnetic shear,
\begin{equation}
q = \frac{H\Jr}{\Bp} = -\frac{B_\phi^{\rm +}}{\Bp},
\end{equation}
and where the plus sign denotes a variable estimated on the disk surface. 
Note that in our case there is no viscous contribution and the 
factor $1/\sqrt{\gamma}$ appears in the relation since the accretion 
Mach number is calculated using adiabatic sound speed.

\begin{figure}
\centering
\includegraphics[width=8cm]{\figurepath/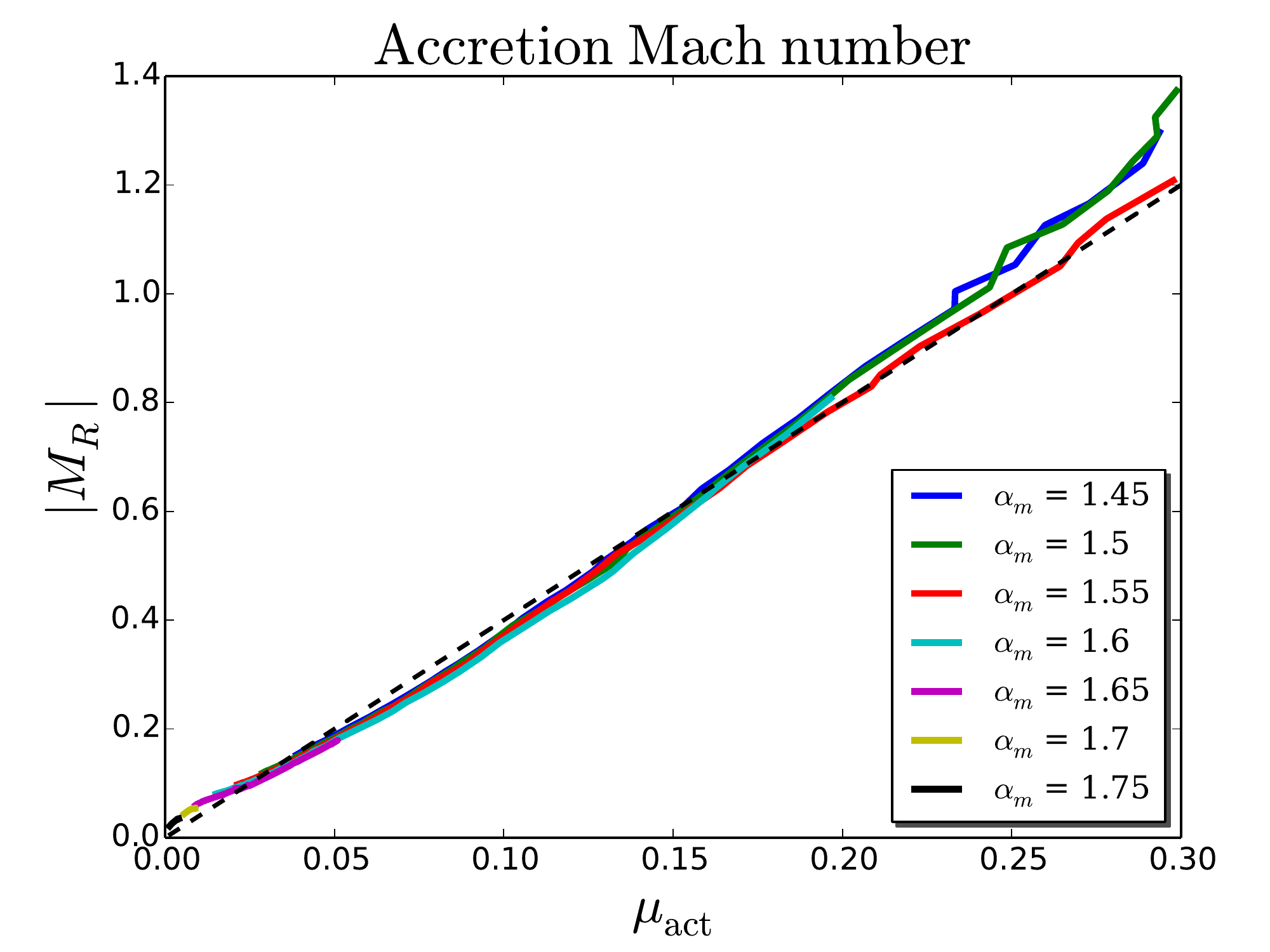}
\caption{The relation of the accretion Mach number to actual magnetization for the initial
magnetization $\mu_0 = 0.01$. 
The linear approximation $M_R = 4 \mu$ is shown by the dashed black line.
}
\label{fig:def_B_mumr}
\end{figure}

Figure~\ref{fig:def_B_mumr} shows that setting $q$ to constant 
$q =2\sqrt{\gamma}$ (thus $M_R = 4 \mu$) is a good first approximation, 
especially for the strong magnetization cases. 
As we will see in the next section, in case of a weak magnetic field 
the magnetic shear $q$ behaves far from being constant.
The closer examination of the magnetic shear $q$ reveals a presence 
of two different jet launching regimes.

\subsubsection{Magnetic shear}
A tight relation exists between the magnetic shear $q$ and the 
ratio between the toroidal and poloidal magnetic field components. 
We define the magnetic shear $q$ by the radial electric current at the disk midplane, 
since it does not require to apply the notation of the disk surface, 
which, in case of a strong magnetic field, might change by up to 40 percent.
Note, that the magnetic shear is the first derivative of
the accretion Mach number with respect to the magnetization.
Therefore it shows the growth rate of the local Mach number or steepness of the curve.

\begin{figure}
\centering
\includegraphics[width=8cm]{\figurepath/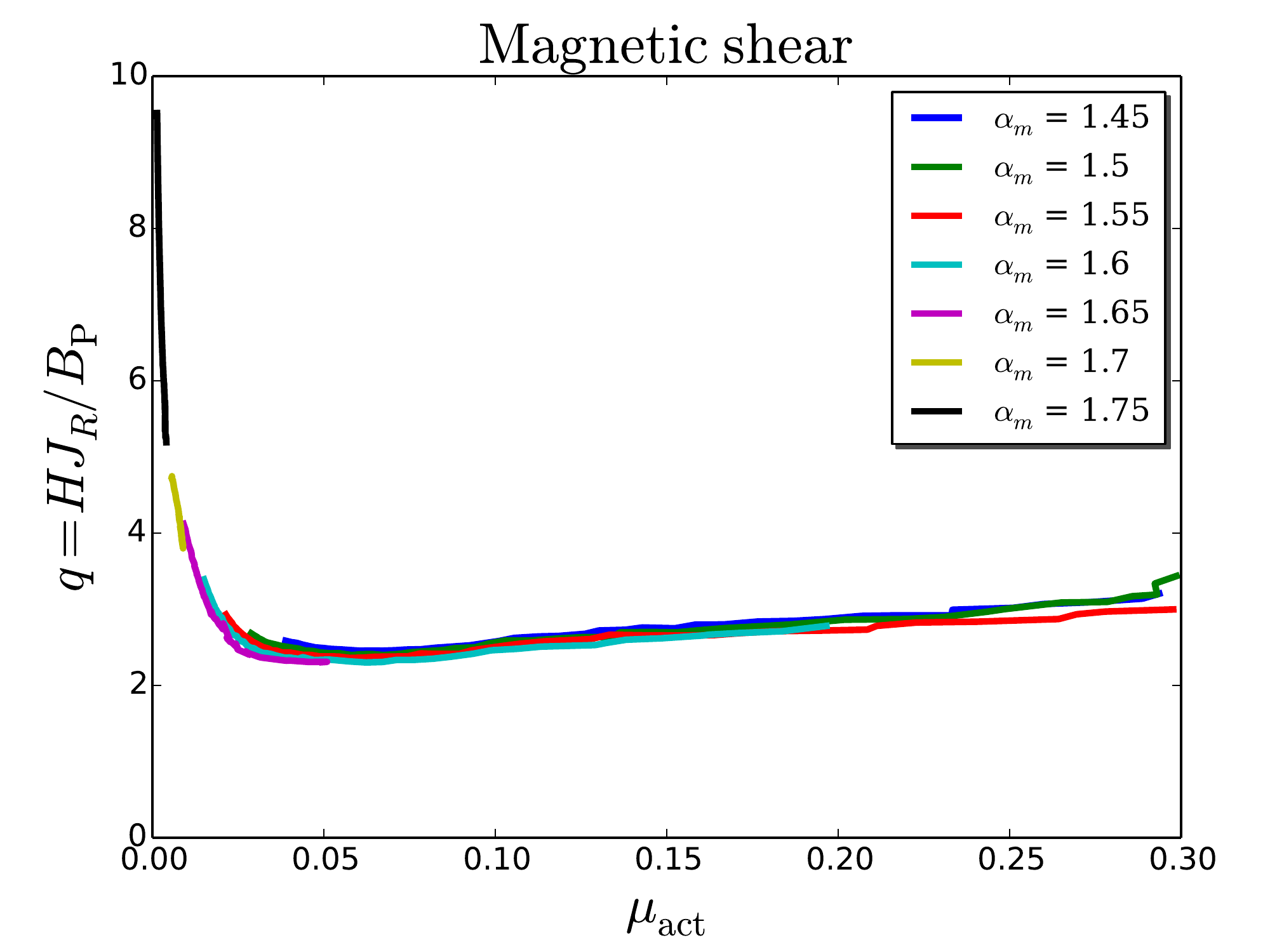}
\caption{The relation of the magnetic shear with respect to actual magnetization}
\label{fig:def_mujbx}
\end{figure}

Figure~\ref{fig:def_mujbx} shows the magnetic shear with respect to the underlying inner disk 
magnetization.
We see that the magnetic shear behaves in two different ways - in case of 
low magnetization, $\mu \leq 0.03-0.05$, the magnetic shear is substantially 
higher in comparison to the case of high magnetization, $\mu \geq 0.03-0.05$. 
The explanation is straightforward: there is a turning point concerning the generation of the 
toroidal magnetic field versus flux losses through the disk surface (by the outflow).
To understand this one needs to set apart the generation processes of the 
toroidal magnetic field from the loss processes.
The rate of the generation of the magnetic field in Keplerian disks is primarily set by 
the structure of the magnetic field and is rather constant in case of a quasi steady state
On the other hand, the outflow speed (through the disk surface) is highly dependent 
on the actual disk magnetization.
In case of a weak magnetization the outflow speed is rather small, 
which makes it possible to sustain a stronger magnetic shear.
A strong disk magnetization results in a fast outflow, 
thus setting the maximum limit for the magnetic shear.

\subsubsection{Mass and energy flux}
The magnetic shear has a great impact on the outflow launching \citep{1995A&A...295..807F}.
We confirm this finding by presenting the mass and energy ejection and accretion fluxes.

\begin{figure}
\centering
\includegraphics[width=8cm]{\figurepath/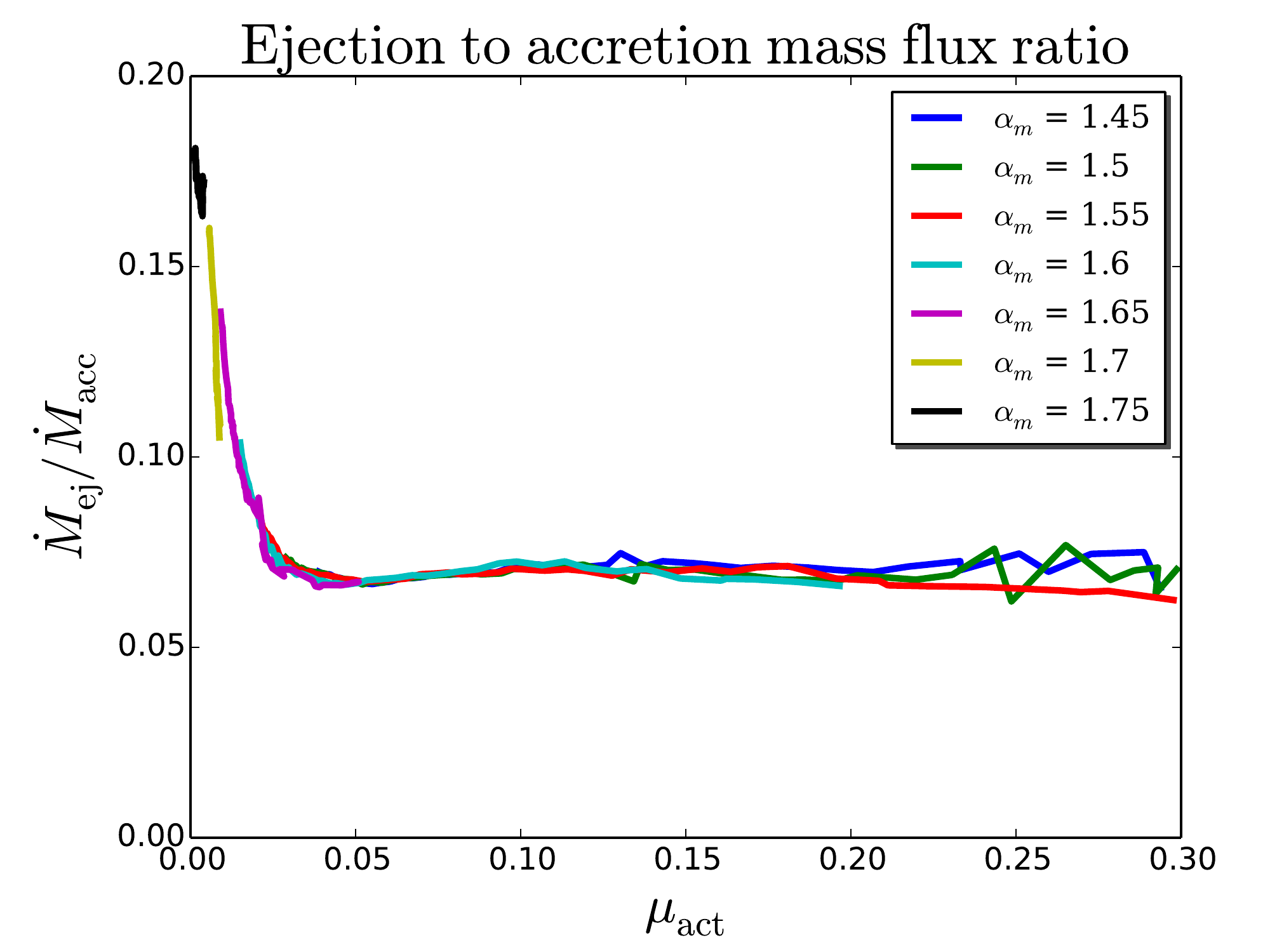}
\caption{The relation of the mass ejection-to-accretion ratio with respect to actual magnetization}
\label{fig:def_mumx}
\end{figure}

Figure~\ref{fig:def_mumx} shows the ratio of the mass ejection rate, $ \Mej(1.5) - \Mej(1.1)$, to the accretion rate, 
both averaged over the same area. 
Obviously, the ejection efficiency is higher for weaker magnetized disks. 

This is easy to understand considering Equation~\ref{eq:btbp}. 
In case of a weak magnetic field, the strong magnetic shear 
(the high toroidal to poloidal magnetic field ratio) leads 
to faster poloidal acceleration, caused by the force component 
parallel to the magnetic field. This force also extracts the matter from the disk.
In case of a strong disk magnetization, the acceleration of a matter 
is primarily supported by the centrifugal force.

We note that studying the ejection index (calculated within an area $R=2 ... 10$) 
with respect to the mean disk magnetization leads to very similar results, 
that is a saturation to values of about 0.35-0.40 in case of a high magnetization 
and a significant increase in case of a low magnetization.

\begin{figure}
\centering
\includegraphics[width=8cm]{\figurepath/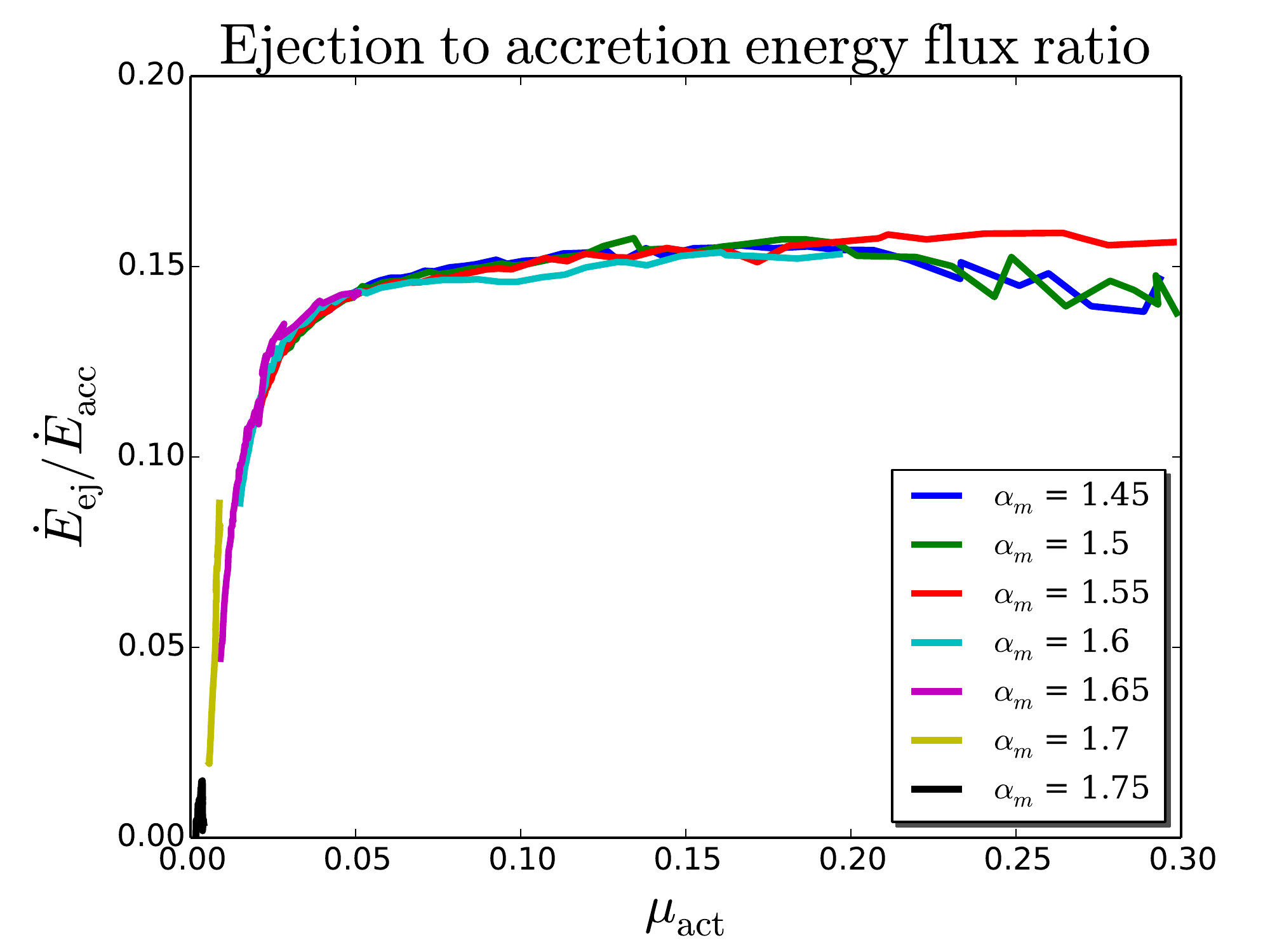}
\caption{The relation of the energy ejection-to-accretion with respect to actual magnetization}
\label{fig:def_muex}
\end{figure}

Figure~\ref{fig:def_muex} shows the ratio of the energy ejection density to the 
average accretion energy, computed in the same way as for the mass fluxes.
Compared to the mass fluxes, the energies show the opposite behaviour - 
the ejection to accretion power is increasing function with magnetization. 
This is a highly important relation, since it relates two observables. 
Note that following our findings, there is not a fixed value for the ratio between jet and accretion power.
This should be considered when comparing observational results to the theory.

Essentially, this result shows the general importance of the magnetic 
energy flux compared to the mechanical energy.
The mechanical energy flux is always negative, while the magnetic energy flux is positive.
In case of a strong magnetic field our results are similar to \citep{2007A&A...469..811Z}, 
namely that magnetic energy flux dominating the mechanical flux. 
We also see a saturation of the flux ratio in case of a moderately strong magnetization.
We find that in a weak magnetization case the energy flux ratio can be very small.

We also find that the ejection Mach number, Equation~\ref{eq:ejmach}, 
increases almost linearly with magnetization. 

Essentially, the  {\em general} behaviour of the mass and energy flux ratios does  depend 
 on the averaging area or its position.

The accretion power (see Appendix, Equation~\ref{eq:accpow}) is mainly 
determined by the accretion rate at the inner disk radius.
Assuming a typical scale height for the mass accretion as $\epsilon r$, 
the accretion power can be estimated considering
the magnetic shear and the actual magnetization of the disk,
\begin{equation}
{\Eacc} \simeq 0.06\,q\,\MUact \dot{E}_0,
\end{equation}
where $\dot{E}_0$ is the unit power (see Appendix~\ref{app:units}).

In case of strongly magnetized disks one can assume that the magnetic shear 
is approximately constant $q \approx 3$, 
and this relation transforms into ${\Eacc} \simeq 0.2 \MUact \dot{E}_0$.
Note that this result connects two essential quantities - the accretion 
power which manifests itself by the accretion luminosity to the disk 
magnetization, which is intrinsically hidden from the observations.

\section{A stable long-term evolution}\label{sec:strong}
In this section we discuss the commonly used diffusivity model and the reasons 
why we think that it fails in case of very long-term simulations, 
in particular when treating weakly magnetized disks. 
In order to overcome this problem - the accretion instability - we propose another magnetic diffusivity model.
This new model enables us to simulate the evolution of the disk-jet system for much longer times.

\subsection{Constraints on the diffusivity parameters}
The simple idea that the induction of the magnetic flux in steady state is compensated by the flux losses, 
both by diffusion and magnetic flux escape through the disk surface,
becomes hardly applicable in case of a weak magnetic field.
As discussed previously (see Section~\ref{sec:andiff}), 
in order to keep the magnetic field distribution properly bent, the magnetic 
diffusivity parameter $\am$ must be linked to the anisotropy parameter $\chi$.
Equation~\ref{eq:curv} was obtained considering the standard magnetic diffusivity model.
A more general relation can be derived assuming a curvature of the magnetic field of about 0.5, that is the 
mean curvature of the initial field distribution (see Equation~\ref{eq:Bpot}),
\begin{equation}
\left(\ass \chi - \Mth \right)\ass \leq \mu.
\end{equation}
Solving this inequality for $\ass$ and assuming $\Mth \propto \beta \mu$, we find
\begin{equation}\label{eq:asslim}
\ass \leq \alpha_0 = \frac{\beta}{2}  \mu_\chi + \sqrt{ \left(\frac{\beta}{2}\right)^2 \mu_\chi^2 + \mu_\chi },
\end{equation}
where $\mu_\chi = \mu/\chi$, and $\beta \approx 6$ in our simulations. 
This relation shows that in order to keep the disk magnetic field properly bent, 
the $\ass$ should behave differently in the two limits of magnetization.
A linear relation to the magnetization in case of a strong magnetic field, 
and proportional to the square root of magnetization in case of a weak field. 
In case of a strong deviation from this relation, the magnetic field structure will 
be substantially affected, resulting in a high field inclination (for $\ass \ll \alpha_0$), 
or a strong outward bending (for $\ass \gg \alpha_0$).

\begin{figure}
\centering
\includegraphics[width=8cm]{\figurepath/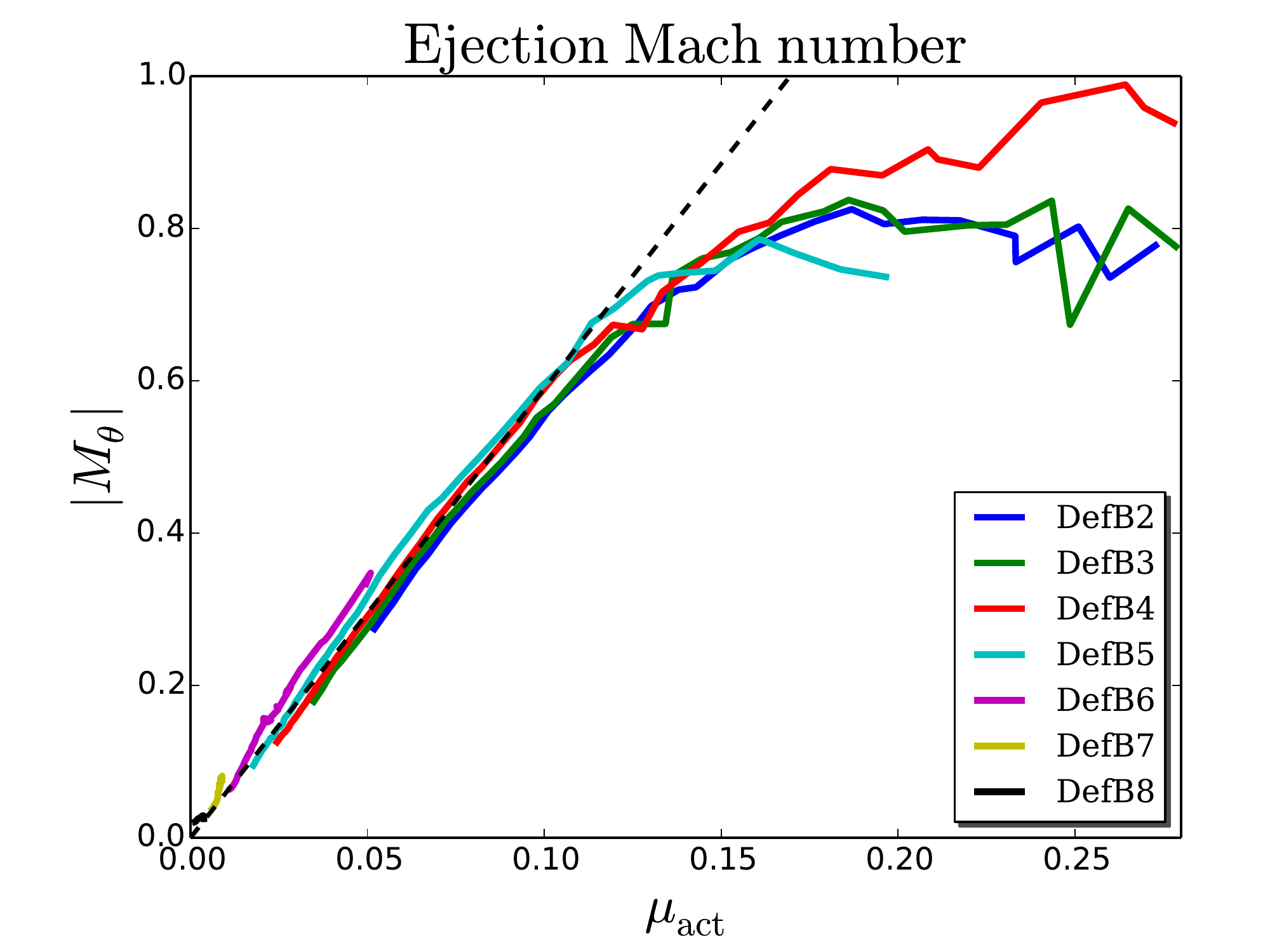}
\caption{The relation of the ejection Mach number with respect to actual magnetization. The slope of dashed line $\beta=5.9$.}
\label{fig:def_mumth}
\end{figure}

For equation \ref{eq:asslim} we have implicitly assumed a linear relation between the ejection Mach number 
(at the disk surface) and the magnetization.
In fact our simulations approve such an interrelation.
Figure~\ref{fig:def_mumx} shows that for a moderately strong magnetic field, 
there is a linear relation between the ejection Mach number and
underlying disk magnetization.
This behaviour is also consistent with ejection to accretion mass flux ratio we haqve discussed above.

As a consequence of Equation~\ref{eq:jrjphi}, the $\ass$ plays a direct role 
in determining the strength of the poloidal electric current with respect to the toroidal current.
We find in our simulations that in order to sustain jets, the ratio of the poloidal to the toroidal current should be
sufficiently high (about 15).

The difficulty in performing simulations of weakly magnetized outflows is that 
the specific torques may {\em increase} towards the disk surface area, just 
because of the low densities over there.
This will lead to a layered accretion along the disk surface, and, thus, much lower 
accretion along the midplane. 
Although this might be a relevant process in reality, our numerical setup is not 
suited for the treatment of such a configuration. 

The standard, commonly used magnetic diffusivity model is parametrized by two constants, $\am$ and $\chi$. 
In general, choosing a high anisotropy parameter $\chi$ implies a low diffusivity parameter $\am$. 
Together, this will lead to a decrease in the poloidal electric current and an increase in the toroidal current 
(see Equation~\ref{eq:jrjphi}).
Thus, the resulting torque will not be sufficient to brake the dense matter along the midplane, and will lead to
layered accretion.
An anisotropy parameter lower than unity has proven to lead to a smoother time evolution, 
since it allows for stronger poloidal currents at the midplane, and, thus, 
a mass accretion which is developed over the full disk height.
Another option to have $\chi> 1$ would be to modify the vertical diffusivity profile such 
that it reaches the maximum not at the disk midplane, but at the disk surface. 
This would also help to develop a strong electric current in the disk midplane. 

\subsection{The accretion instability}
Here, we discuss another problem, which the standard diffusivity model is prone to.
As previously shown, most of the simulations suffer from the mass loss from the disk, 
leading to the increase of the magnetization. 
However, the increase of the magnetization further amplifies the mass loss. 
This is known as the accretion instability, first studied by \citet{1994MNRAS.268.1010L},
and later confirmed for more general cases \citep{2009MNRAS.392..271C}.
If, on the other hand, the diffusivity is too high (chosen by a high $\am$ parameter), 
inevitable diffusion of the magnetic field will lead to the situation that 
a jet cannot be sustained anymore.

We like to emphasize that the main reason for the increase of magnetization 
is the mass loss, but not the actual magnetic field amplification. 
This is a direct consequence of the accretion instability, namely, a lack of sufficient feedback that could 
bring the accretion system back to a stable state.
In other words, in order to run long-term simulations one needs to apply a diffusivity model that provides 
a stronger feedback to the diffusivity profile than the standard choice $\propto \sqrt{2\mu}$.

\subsection{A proper magnetization profile}
The direct consequence of the accretion instability is that the magnetization 
increases towards the center. It is easy to show that the behaviour of 
the magnetization has to be opposite.
In accretion disks producing outflows, the mass accretion rate must naturally increase with radius.
Assuming a radial self-similarity of the disk and taking into account that the accretion Mach 
number is linearly related to the magnetization and that
$\rho \propto \Cs^3$, one derives
\begin{equation}\label{eq:muxics}
\beta_{\mu} = \xi -2 - 4\beta_{\Cs},
\end{equation}
where $\xi$ is the ejection index and $\beta_X$ represent the power law index of a physical quantity $X$.
Considering that magnetized disks are very efficient in producing outflows, $\xi \approx 0.2-0.4$, one 
may expect $\beta_{\mu}$ to be positive (if $\beta_{\Cs} \approx -1/2$), thus, an increasing function with radius.
However, the disk structure itself can be re-arranged such that $|{\beta_{\Cs}}| \le 1/2$, 
that eventually will satisfy the relation \ref{eq:muxics}. 
This is indeed what we find.

One should, however, keep in mind that this equation is a rough
estimate and might be subject to the different disk physics involved.
If the magnetic torque is not the only supporter of the
accretion as in case of viscous simulations of 
\citet{2010A&A...512A..82M} the above presented relation
might be relaxed.
However, the similar analysis can be performed to set the limit
of the magnetization with respect to other quantities.

\subsection{A modified diffusivity model}
In this section we present a diffusivity model, which does not suffer 
from the accretion instability. 
Although the standard diffusivity model gave us an opportunity to probe 
a wider parameter space, it is not applicable for very long-term studies. 

We emphasize that the transition from the a direct simulation of turbulence 
to the mean field approach, which lacks the small scales by design, is indeed subtle.
So far, in the literature, the jet launching problem is addressed without 
considering the origin of the magnetic field by a dynamo.
Therefore, the only way of amplifying the magnetic field is by 
advection (or stretching in case of a toroidal field). 
There is also no intrinsic angular momentum transport by the turbulence itself. 
The only term we need to model when applying a small scale turbulence, is the effective magnetic diffusivity.
This might have surprising consequences.
In order to suppress the turbulence, one should rather {\em amplify} the effective diffusivity - leading 
to a stronger decay of the magnetic field and resembling the quenching of 
diffusivity (or dynamo) - rather than decrease it - leading to stronger advection and 
thus an amplification of the magnetic field.
The main motivation of our new model for the diffusivity - is to consider stronger 
feedback by the disk magnetization. We overcome the accretion instability by 
assuming a stronger dependence of $\ass$ on the magnetization,
\begin{equation}\label{eq:strongdiff}
{\ass} = \am \sqrt{2\mu_0} \left(\frac{\mu}{\mu_0}\right)^2,
\end{equation}
where we choose $\am = 1.55$ and $\mu_0 = 0.01$. 
Here we keep the previous overall form and constants, indicating that both 
models are the same at the magnetization $\mu_0$.
A quadratic dependence on $\mu$ was chosen in order to amplify the feedback.
Choosing a lower power than two might have revealed other complications, for example 
a feedback too weak to work fast enough, 
and keep $\ass$ under the constraint of Equation~\ref{eq:asslim}. 
We will refer to this diffusivity model as {\em strong diffusivity model}.

\subsection{The long-term disk-outflow evolution}
Applying our strong diffusivity model enables us to overcome the accretion instability.
As a result, we were able to perform our simulations for {\em much} longer times, 
reaching evolutionary time steps of $t > 150,000$ which corresponds to 
approximately 25,000 revolutions at the inner disk orbit. 

\begin{figure*}
\centering
\includegraphics[width=8cm]{\figurepath/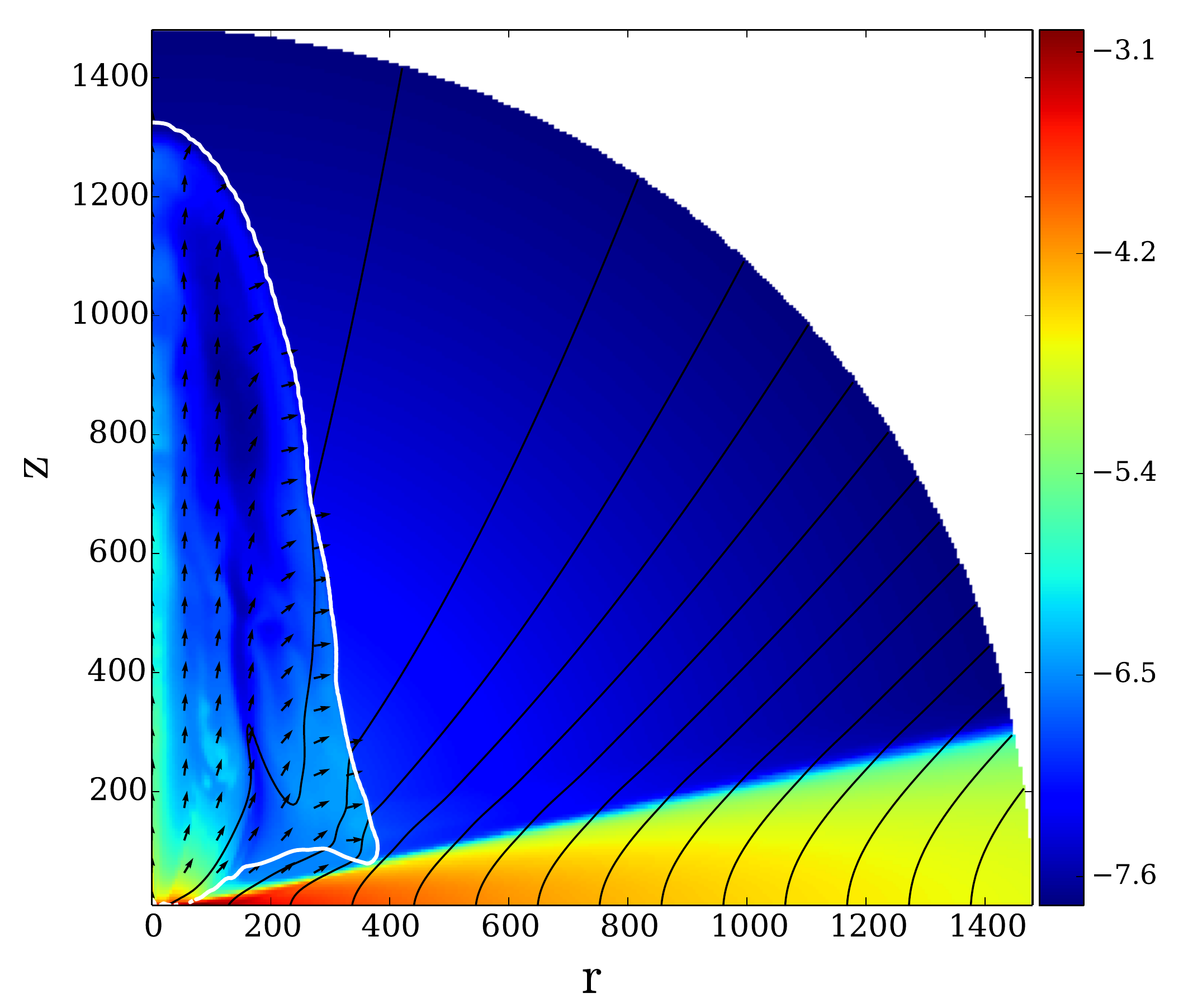}
\includegraphics[width=8cm]{\figurepath/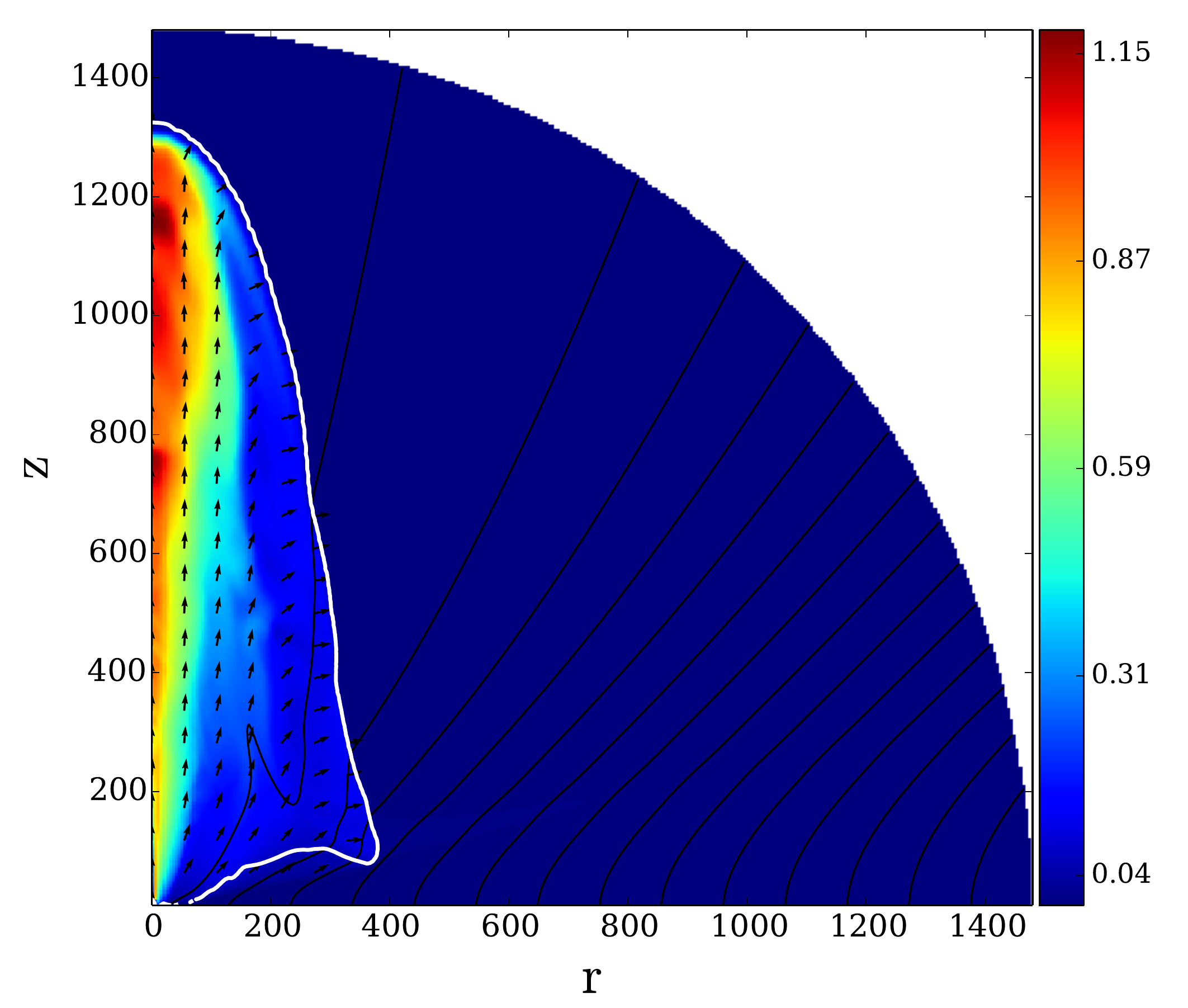}
\caption{Initial evolution of the strong diffusivity setup at T = 1500. Colors represent the logarithm of density (left) and speed (right), black lines denote the magnetic field, arrows the normalized velocity, and white line shows the Alfv\'en surface. Arrows show normalized velocity vectors.}
\label{fig:mu2_cocoon}
\end{figure*}

\begin{figure*}
\centering
\includegraphics[width=8cm]{\figurepath/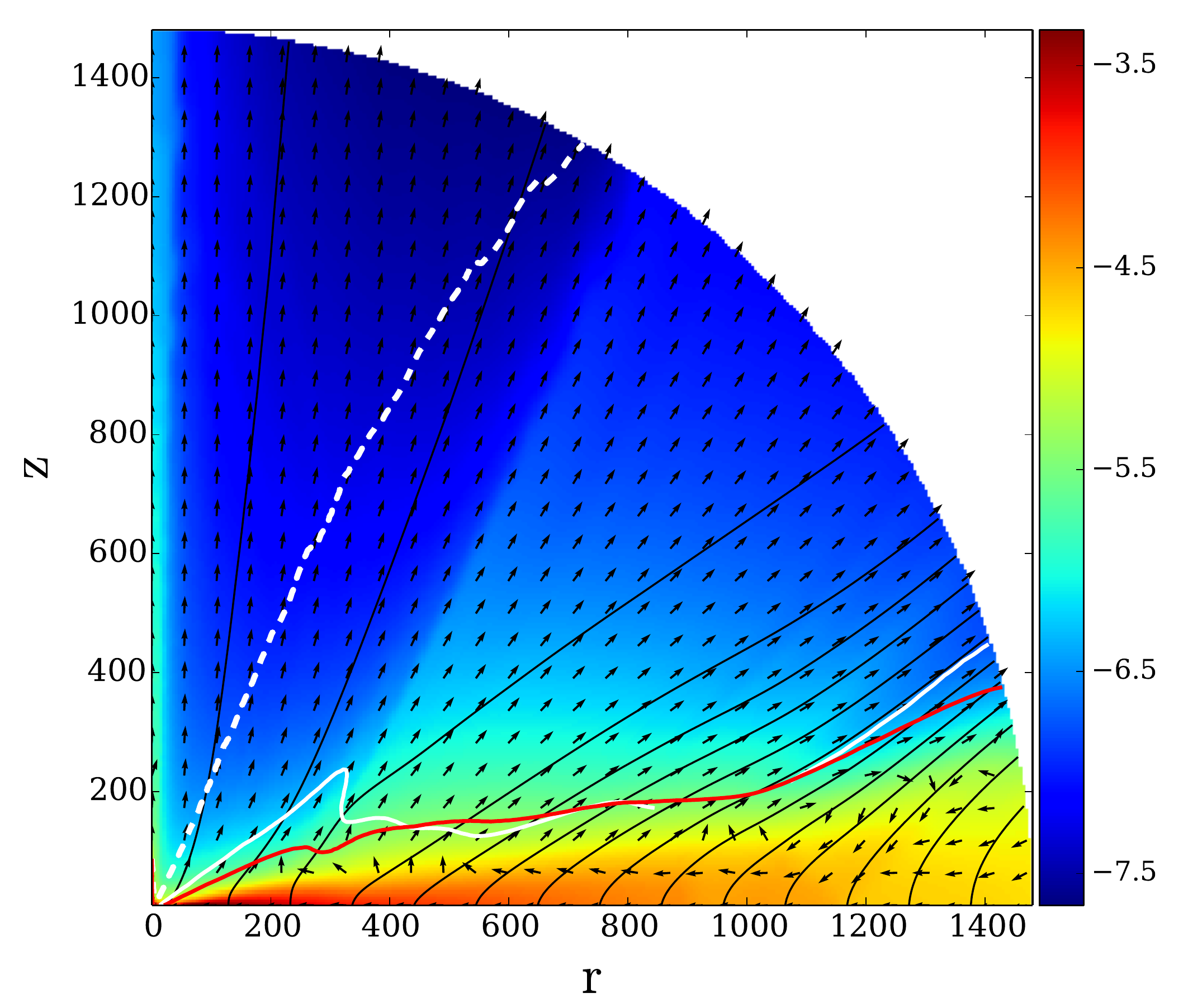}
\includegraphics[width=8cm]{\figurepath/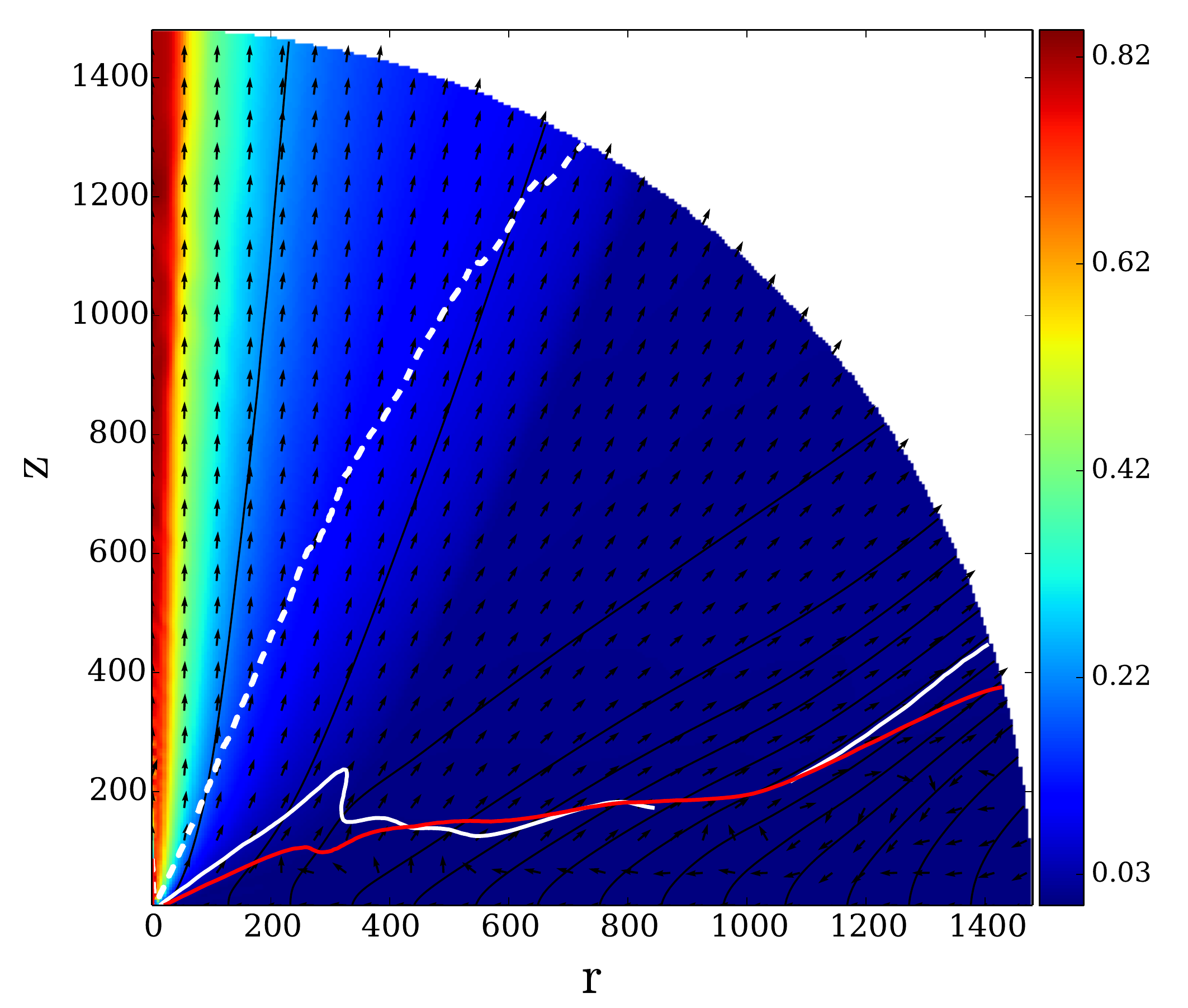}

\caption{Time snapshot of the strong diffusivity setup at T = 150,000. 
Shown is the density (colors, in logarithmic scale), 
the poloidal magnetic flux (thin black lines),
the sonic (red line), Alfv\'en (white line), and fast magnetosonic (white dashed line) 
surfaces. Arrows show normalized velocity vectors.}
\label{fig:mu2_big}
\end{figure*}

Figure~\ref{fig:mu2_cocoon} shows the typical computation domain and the initial dynamics of the system. 
As usual, the evolution starts with the propagation of the toroidal Alfv\'en wave, 
resulting in a propagating cocoon. 
At this point, the innermost area ($R \leq 10$) has reached a quasi steady state, while the outer 
part has not even slightly moved from the initial state.
Figure \ref{fig:mu2_big} presents the same simulation at a later state when a strong 
outflow has been developed.
Notice the difference between the inner part, $r\le 200$ and the outer part, $r\ge 200$.
The inner part has already relaxed to a steady state, while the outer region shows 
rapid accretion and ejection patterns. This is a direct consequence of the new diffusivity model. 
The logic behind implementing the enhanced feedback is valid only when 
the accumulation of the flux is possible. 
The initial imbalance between advection and diffusion in the outer part 
leads to a rapid advection of the magnetic flux to the inner disk.
As a result, the rapid accretion further leads to higher 
inclination angles of the magnetic field (smaller angle with respect to the disk surface).
This results in a higher efficiency for the toroidal magnetic field induction, thus 
leading to even more rapid accretion and ejection.

\begin{figure*}
\centering
\includegraphics[width=18cm]{\figurepath/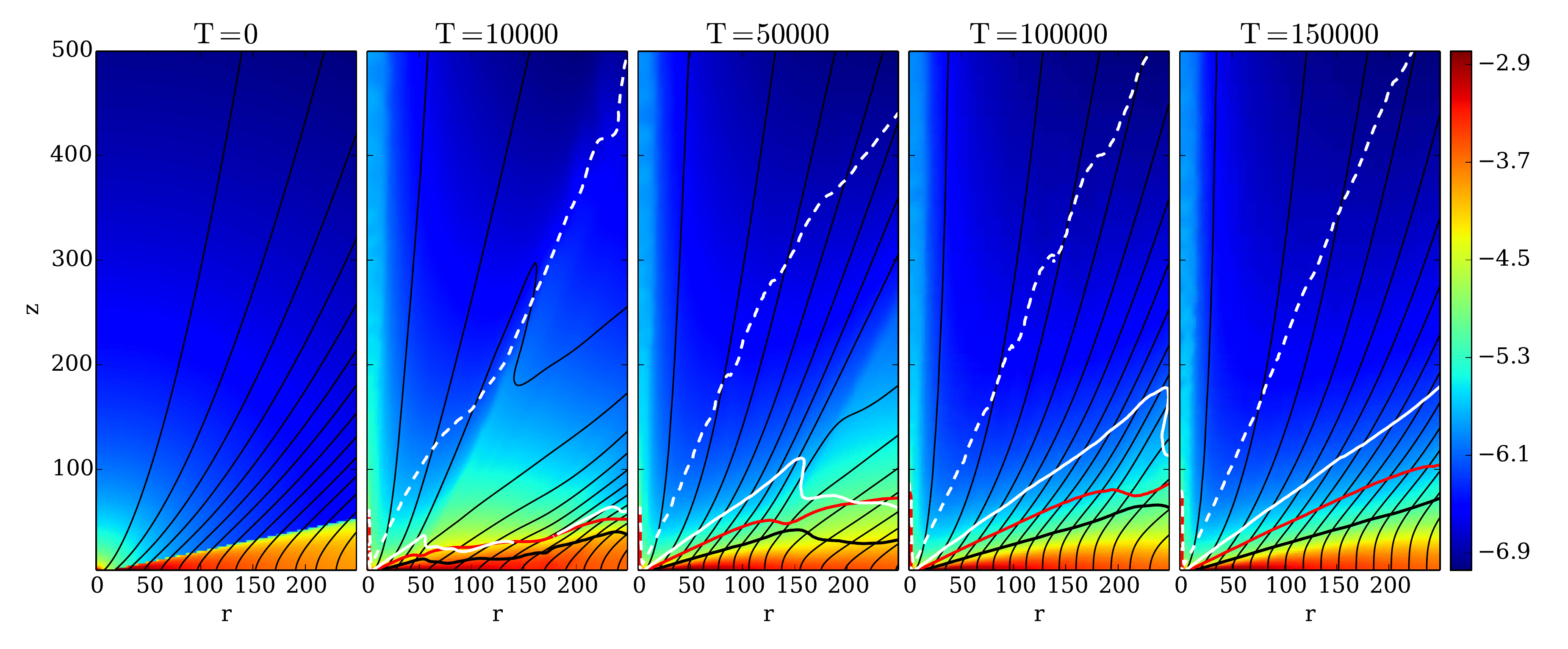}
\caption{Time evolution of the disk-jet structure for the strong diffusivity simulation.
Shown is the evolution of density (colors, in logarithmic scale), 
the poloidal magnetic flux (thin black lines), the disk surface (thick black line) 
the sonic (red line), Alfv\'en (white line), and fast magnetosonic (white dashed line) 
surfaces. }

\label{fig:mu2_tevol}
\end{figure*}

Figure \ref{fig:mu2_tevol} shows a long term evolution on a small sub-grid of 
the simulation with our strong diffusivity model.
As can be seen, until time $t=10,000$ only a small fraction of the disk has 
dynamically evolved (up to $R \approx 50$), while at later times also the outer parts 
of the disk do reach a new dynamic state.
A steady outflow establishes from the whole disk surface (shown on this sub-grid), and 
reaches super-fast magnetosonic speed. 
Note that the positions of the critical MHD surfaces are constant in time, which
is a further signature of a steady state.

The outflow reaches maximum velocities typically of the order of 100 $\Vx$ for YSO, 
or 70,000 $\Vx$ in case of AGN. 
Concerning observationally relevant scales, our simulations compare to the following numbers.
Our numerical grid is comparable to 150\,AU for YSO, and 0.14\,pc in case of AGN
Physically more meaningful is the grid size where our simulation has reached 
a steady state. 
That is a size comparable to 25 AU for YSO, and 5000 AU in case of AGN, but can be extended 
by running the simulations longer.
The dynamical time scale of 150,000 time units (or 25,000 disk orbits) corresponds to 
about 550 years in case of YSO and about 200 years in case of AGN.
Typical accretion rates of our simulations are $3\times 10^{-7}\Mx$ for YSO, and $0.1 \Mx$ in case 
of AGN, but one has to keep in mind that these values depend not only on the intrinsic scaling of central mass 
and inner disk radius, but also on the scaling of density.
Therefore, we herewith present the most extended and longest 
MHD simulations of jet launching obtained so far - connecting 
the jet launching area close to the central object with the asymptotic 
domain which is accessible by observations.

Although a spherical grid is computationally very efficient and may allow to extend the computational domain to
almost any radius, in reality its application for the jet launching simulations is somewhat limited.
There are two reasons for that.
First, it takes obviously much longer time for the outer disk areas to 
evolve into a new dynamical steady state. 
Thus, outer disk will remain close to the initial state of the simulation for quite some time. 
Second, and a more severe drawback is the lack of resolution for the asymptotic jet.
For example, for distances $R > 500 \Ri$ along the rotational axis, 
the jet radius of about  $r_{\rm jet} \approx 25$ 
can be resolved only by about 5 grid cells (applying our typical resolution). 
We therefore restrict our computational domain for such grid size to about 
$R_{\rm out} \simeq 1000 - 2000 \Ri$.
The above mentioned resolution issue is in fact one of the advantages 
for using cylindrical coordinates for jet formation simulations.

\subsection{Results of the strong diffusivity model}
\begin{figure}
\centering
\includegraphics[width=8cm]{\figurepath/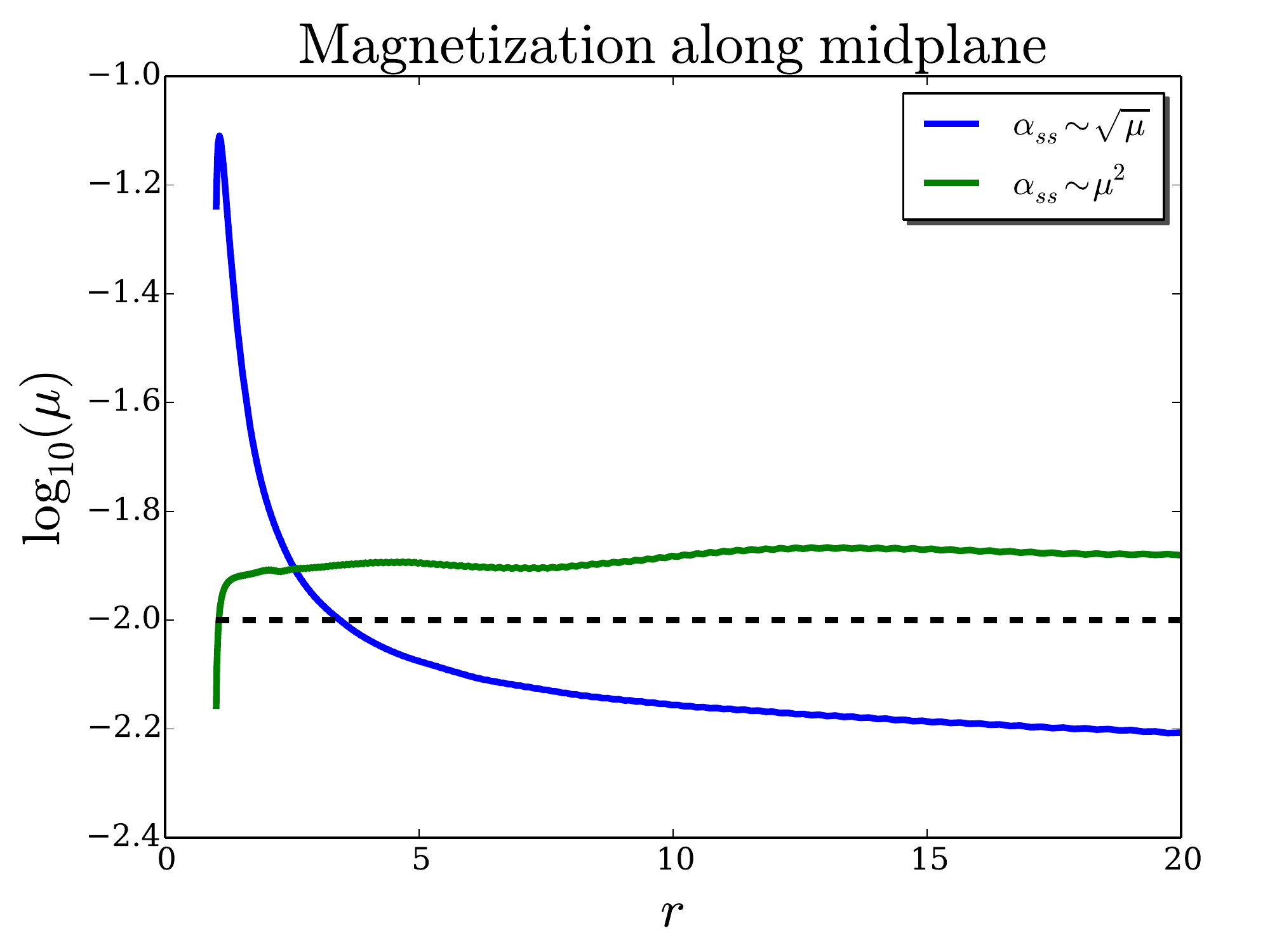}
\caption{Magnetization distribution throughout the disk for reference and 
strong diffusivity model at T=10,000. The dashed line marks the initial magnetization.}

\label{fig:comp_mu}
\end{figure}

By design, the purpose of our strong diffusivity model was to avoiding the accretion instability. 
As a consequence of this application, magnetization profile does not 
decrease with radius (see Figure~\ref{fig:comp_mu})
Although both simulations (with standard and strong diffusivity model)
start from the same initial disk magnetization, the disk evolution results in a quite 
different magnetization distribution.
The standard diffusivity model (our reference simulation from above) results in a magnetization profile decreasing with radius.
In contrast, for the strong diffusivity model a rather flat magnetization profile emerges.
In non-viscous simulations, assuming radial self-similarity, a flat (or not decreasing with radius) 
profile is essential for sustaining a continuous accretion flow at any given radius.

As soon as a steady state is reached, the evolutionary track for this simulation is represented
by a simple dot in all ($\mu, X$)-diagrams (at least from 1000 to 150,000 time units). 
The mean inner disk magnetization is $\mu \approx 0.012$. 
We find that this simulation fits to every relation presented above, such as mass and energy flux ratios, 
or magnetic shear, that were derived applied a standard diffusivity model.
In other words, the aforementioned dots belong to the curves drawn in the ($\mu, X$)-diagrams.

\begin{figure}
\centering
\includegraphics[width=8cm]{\figurepath/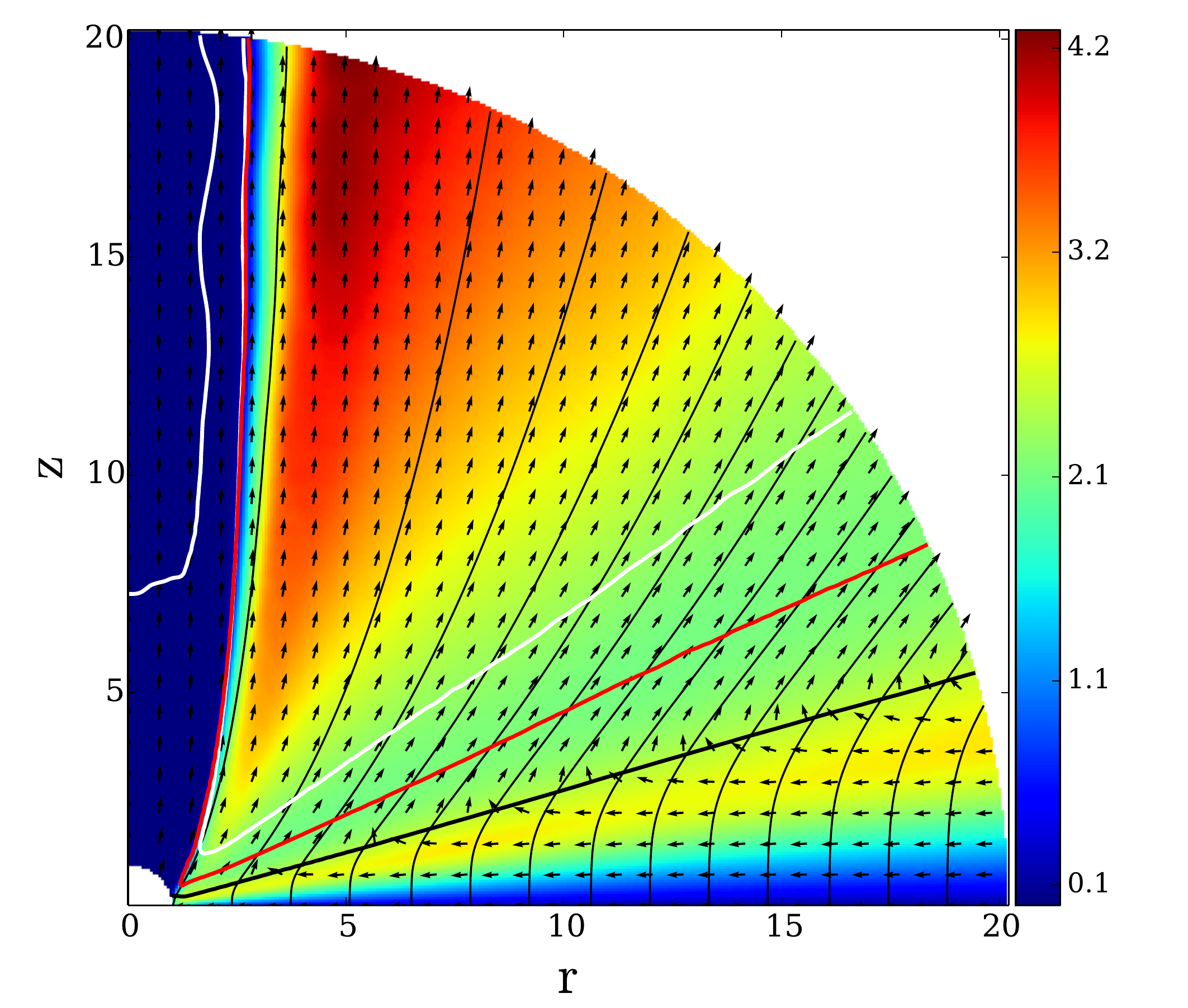}
\caption{The ratio of the toroidal to poloidal magnetic field for the strong diffusivity model at T = 10,000. 
Lines represent the disk (thick black line) the sonic (red line), the Alfv\'en (white line) surfaces.
Arrows show normalized velocity vectors.}

\label{fig:mu2_btbp}
\end{figure}

There are several distinct features one can derive from the resulting magnetic field structure.
Figure~\ref{fig:mu2_btbp} shows the toroidal to poloidal magnetic field component ratio. 
Taking into account that the disk magnetization (calculated from poloidal component only) is uniform, 
three different regions can be distinguished.
The first region is between the midplane and the disk surface where the 
toroidal magnetic field reaches its maximum and the torques change sign.
The second region is between the disk surface and the Alfv\'en surface 
where the ratio of the field components is quite constant.
The third region is beyond the Alfv\'en surface when the poloidal component 
of the magnetic field becoming weak enough to keep a rigid magnetic field 
structure and toroidal component starts to dominate the poloidal one.

\begin{figure}
\centering
\includegraphics[width=8cm]{\figurepath/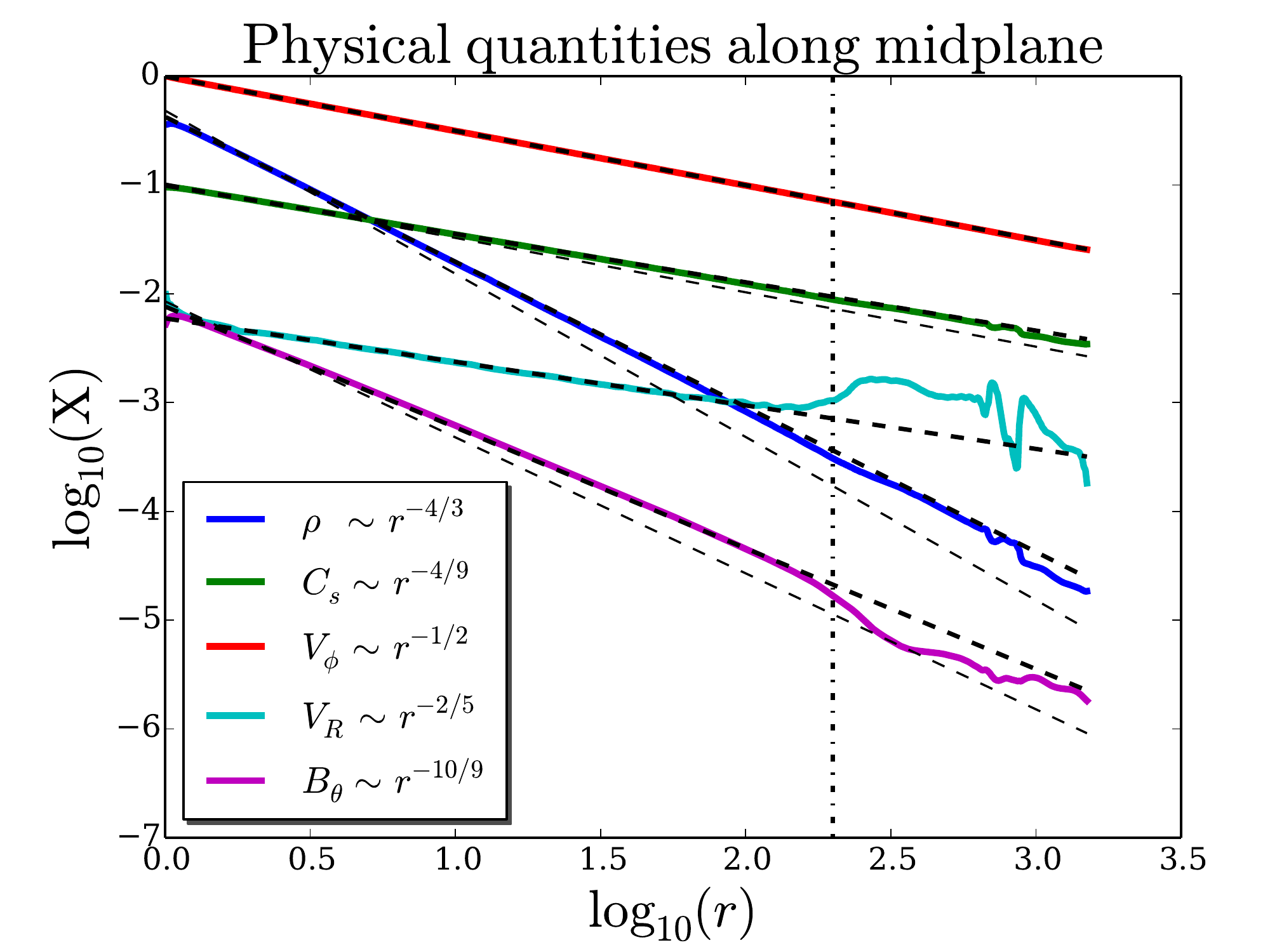}
\caption{Physical quantities along the midplane for the strong diffusivity model. Colors show different variable profile, thick dashed lines correspond to certain power law, the mismatched thin dashed lines correspond to initial distributions of variables. Vertical dashed line marks $r = 200$. }

\label{fig:mu2_profiles}
\end{figure}

\subsection{Dynamical profiles of a steady state accretion disk}
In this subsection we further explore the disk structure in a steady state. 
In Figure~\ref{fig:mu2_profiles} we present the radial profiles of certain magnetohydrodynamical 
variables along the midplane.
We show the profiles derived from our numerical simulations with thier approximations by power laws, 
and compare them to the initial distribution.
These radial profiles are obtained along the disk midplane, however, 
they also hold at least for one disk semi-height.
The thetoidal profiles that normalized to the corresponding midplane value (not shown here) 
almost coincide with each other, indicating that the assumption 
of a self-similar disk is in fact reasonable, though different power
indexes should be used.

In particular, Figure~\ref{fig:mu2_profiles} shows how the disk structure 
evolves from a certain initial power law distribution into another power law profile.
We see distinct power law profiles for radii up to  $R \leq 250$. 
This corresponds to the area where the disk evolution has reached a steady state. 
For very small radii $R \leq 1.1$ we also see a deviation from a power law profile, that we consider as a  boundary effect.

At time $t= 150,000$, we find the following numerical values for the power law coefficients 
$\beta_X$ for the different variables at the midplane $X(r, \theta = \pi/2) \sim r^{\beta_X}$.
The disk rotation remains Keplerian throughout the whole time evolution, thus $\beta_{\Vphi} = -1/2$.
The radial profiles for density and gas pressure slightly change from their initial distribution.
The density power low index changes from  $\beta_\rho = -3/2$ to $\beta_\rho = -4/3$, while
for the pressure it changes from $\beta_{\rm P}= -5/2$ to $\beta_{\rm P}= -20/9$.
For the accretion velocity we find a profile of $\beta_{\Vr} = -2/5$, and $\beta_{\Bth} = -10/9$ for the magnetic field.
As a consequence, the profile for the magnetization remains about constant $\beta_{\mu} \sim 2\beta_{\Bth} / \beta_{\rm P} = (-20/9) / (20/9)$.
The accretion velocity remains subsonic over the whole disk with an accretion Mach number of $V_R / \Cs \simeq 0.1$.

Following \citet{1995A&A...295..807F} and considering the mass accretion 
$\Macc \sim R^2 \rho \Vr$ it is easy to get the ejection index $\xi = 0.26$. 
This is in accordance with previous work \citet{2012ApJ...757...65S}.

\subsection{Discussion of the new diffusivity model}
Our strong diffusivity model, Equation~\ref{eq:strongdiff}, does not {\em necessarily} 
lead to a flat magnetization profile.
The model does not directly force the magnetization to be uniformly distributed - 
in opposite, the profile is expected to be outwardly increasing. 
As we previously showed the magnetization profile is linked to the ejection index 
and it has to be positive if the sound speed stays as initially distributed. 
What we found is that the disk hydrodynamics changes such that the magnetization of the 
disk remains flat, thus satisfying Equation~\ref{eq:muxics}.

Another way of reasoning is the following.
In case of a flat magnetization profile the result of the simulation is not 
sensitive to the diffusivity model anymore. 
In other words, it is possible to switch back from the new diffusivity model to the standard model 
when the radial magnetization profile is uniform (along the midplane). 
However, the diffusivity parameter $\am$ has to be correspondingly re-normalized. 
The only substantial deviation from a uniform profile is in the innermost disk, 
which might be influenced by the boundary condition.
In fact, we performed such a test simulation, switching back from strong diffusivity 
to standard diffusivity model. 
As expected, the accretion instability starts to manifest itself similarly as before, 
leading to the typical magnetization profile (increasing towards the center). 
However, it takes much longer time to substantially affect the outer parts of the disk.

The steady state solution achieved when using our strong diffusivity model, perfectly
fits to the results obtained by the standard model (shown previously as a dot on the plots).
This actually approves our understanding that the main agent in driving 
outflows is the {\em actual} magnetization, and that the magnetic diffusivity
is only the {\em mediator} through which the magnetic field structure is being governed. 
A self-consistent treatment of turbulence is not feasible yet in the context of outflow launching.
Therefore, what should be considered at first place, is the resulting magnetic field strength and distribution, 
but not the diffusivity model itself.

\section{Conclusions}
We have presented results of MHD simulations investigating the launching of jets and outflows
from a magnetically diffusive accretion disk.
The time evolution of the disk structure is self-consistently taken into account.
The simulations are performed in axisymmetry applying the MHD code PLUTO 4.0.
In contrast to previous work we have applied a {\em spherical} coordinate system and numerical
grid, which implies substantial benefits concerning the numerical resolution and the stability
(in time evolution) of the simulations.

In particular, we have obtained the following results.

(1) Our numerical setup in spherical coordinates for disk-jet related problems is very robust.
The use of spherical geometry in the context of the outflow launching cannot be underestimated. 
It allows to study the launching of outflows for very long time (more than 150,000 time units) 
on highly resolved (up to 24 cells per disk height) and at the same time
very large (1500 $\ri$) domains.
On the other hand, a spherical grid somewhat limits the study of jet propagation, since the 
resolution far from the origin becomes low. 
The rather low resolution in the outer disk region, where the dynamical timescales are long, 
helps to smooth out small-scale disturbances, thus helping in establishing a steady state.

(2) Our study has approved a robust disk-outflow structure, 
however for the highest resolution the evolution is prone to have some more fluctuations.
The ability to evolve the disk for very long time disentangles the complex 
interrelations between the essential quantities for the jet launching.
Those are the disk actual magnetization ( at a certain time and averaged for a certain location),
the mass, energy and angular momentum fluxes, and the jet terminal velocity.

(3) Our main result is that it is the {\em actual} rather than the initial disk magnetization that plays 
a key role for the jet formation and directly affects the accretion power.
The value of the {\em initial} magnetization can fail to properly characterize the disk-jet 
properties, but sets the overall jet torque and the disk's magnetic reservoir.
This becomes obvious for very weakly magnetized disks ($\mu_0 = 0.003$). 
In this case, when choosing a low magnetic diffusivity, the magnetic flux can 
still be accumulated to a high magnetization in the inner disk.
We find that the actual magnetization necessary for sustaining a stable jet is 
of order of $10^{-3}$, in accordance with \citet{2010A&A...512A..82M}.

(4) We showed that the ejection Mach number in case
of moderately strong magnetization ($\mu < 0.15$) is linearly
related with respect to the disk magnetization. 
This is indeed consistent with the linear to magnetization
mass ejection to accretion relation.
The mass ejection index (the ratio between ejection and accretion) is about 0.3 and thus similar
to the literature values.

(5) We found that in case of uniform magnetization, the MHD disk quantities show a self-similar 
structure, i.e. resulting in approximately the same vertical profile, and a radial power law distribution.
In case of the strong diffusivity model we have presented the corresponding power law indices for all MHD quantities,
although we believe that these power law indices may directly depend on the actual strength of the magnetization.
This would be a natural consequence of the ejection index being a function of magnetization as well.

(6) We showed that there are two principally different regimes for outflow launching,
complementary to each other.
In case of weak magnetic fields (below $\mu \approx 0.03$) we see signatures of a 
strong magnetic shear, which results in less powerful, but more efficient (higher ejection index) outflows. 
In case of a higher magnetization, the magnetic shear, the ejection efficiency and the energy ejection to accretion flux ratio do not strongly depend on the magnetization

(7) We found the upper theoretical limit for the parameter specifying the 
anisotropy $\chi$ in the magnetic diffusivity in case of the standard diffusivity model, 
essentially depending on the actual magnetization in the disk.
In the limit of low magnetization the anisotropy parameter must satisfy $\chi \leq 1/\am^2$.

(8) We showed that in
non-viscous steady state, assuming radial self-similarity, the magnetization profile should be 
non-decreasing function of radius. 
In steady state, jet launching disks must have a radially increasing profile of 
the mass accretion rate. 
This is a requirement of a positive ejection index. 
Taking into account that i) the accretion Mach number is proportional to the magnetization and 
ii) assuming that the radial profile of the sound speed does not strongly deviate from Keplerian, 
we showed that the index of the magnetization profile is non-negative.
 
This is paper I in a series of papers, that studies the long term evolution of outflow-generating disks.
In two follow-up papers we will present the 
 i) connection of the jet properties (the potential observables) with the underlying disk quantities, and we will
ii) extend the present setup to simulations including a disk magnetic field that is self-generated by a mean-field dynamo.

\acknowledgements
We thank Andrea Mignone and the PLUTO team for the possibility to use their code.
We appreciate many helpful discussions with Andrea Mignone and 
also competent suggestions by the unknown referee.
The simulations were performed on the THEO cluster of Max Planck Institute for Astronomy.
This work was partly financed by the SFB 881 of the German science foundation DFG.

\appendix

\section{Units and Normalization}
\label{app:units}
Here we write up the typical normalization to be used to apply our code units to different
astrophysical jet-launching objects, such as young stellar objects (YSO), and active galactic nuclei (AGN).
The main normalization units are the Keplerian speed at the inner disk radius,
%

$$
V_{\rm K0} =                94\,{\rm km\,s^{-1}}\, \Myso^{1/2} \Ryso^{-1/2} 
           = 6.7 \times 10^{4}\,{\rm km\,s^{-1}}\Ragn^{-1/2}
           \hfill
$$
%
the time unit that is expressed in units of $T_{\rm 0} \equiv R_0 / V_{\rm K0}$,
%
$$
T_{\rm 0} = 1.7\,{\rm days}\,\Myso^{-1/2} \Ryso^{3/2} = 0.5 \,{\rm days} \Ragn^{3/2}
\hfill
$$

The mass accretion rate is a parameter which is in principle accessible by observation. 
Thus, the normalization of density $\rho_0$ can be chosen by setting suitable accretion 
rates $\dot{M}_0 = \Ri^2 \rho_{\rm 0} V_{\rm K0}$. 
Assuming 
$\Macc \simeq 10^{-7} \,\Mx$, 
$\Macc \simeq 10 \,\Mx$,
and taking into account that the typical accretion rates our simulations provide are of 
order of $\Macc \simeq 0.01$ (in code units), one gets
%
%
$$ 
\dot{M}_0 = 3\times 10^{-5} \Mx \DNyso \Myso^{1/2} \Ryso^{3/2} = 10\,\Mx \DNagn \Magn^{1/2} \Ragn^{3/2}
\hfill
$$

The torque and power are given in units of
$\dot{J}_0 = \Ri^3 \rho_{\rm 0} V_{\rm K0}^2$ and $\dot{E}_0 = \Ri^2 \rho_{\rm 0} V_{\rm K0}^3$ 
respectively,
%
%
$$
\dot{J}_0 = 3.0\times 10^{36} \Jx \DNyso \Myso \Ryso^2 = 1.2\times 10^{51} \Jx \DNagn \Magn^3 \Ragn^2
$$
$$
\dot{E}_0 
= 1.9\times 10^{35} \Ex \DNyso \Myso^{3/2} \Ryso^{1/2} = 2.6\times 10^{46} \Ex \DNagn \Magn^2 \Ragn^{1/2}
$$

The magnetic field is normalized to its values at the midplane, $B_{\rm 0}= \sqrt{8 \pi P_{\rm 0}\mu }$,
\begin{eqnarray}
B_{\rm 0} 
& = & 15 {\rm G } \MUx^{1/2} \EPSx \DNyso^{1/2} \Myso^{1/2} \Ryso^{-1/2} \hfill
\nonumber \\
& = & 1 {\rm kG } \MUx^{1/2} \EPSx \DNagn^{1/2} \Magn^{1/2} \Ragn^{-1/2} \hfill
\nonumber 
\end{eqnarray}

\section{Control volumes and the fluxes in disk and jet}
\label{app:fluxes}
For the fluxes of energy and angular momentum, we keep the same notation as for the mass flux,
\begin{equation}
\JAkin = - 2 \int_{S_1}^{S_R} r\rho \Vphi \Vtot_p\cdot \dS, \quad\quad
\JAmag =  2 \int_{S_1}^{S_R} r \Bphi \Btot_p\cdot \dS.
\end{equation}

We define the jet angular momentum flux $\Jjet = \JJkin + \JJmag$ with
\begin{equation}
\JJkin = - 2 \int_{S_S} r\rho \Vphi \Vtot_p\cdot \dS, \quad\quad
\JJmag =   2 \int_{S_S} r \Bphi \Btot \cdot \dS.
\end{equation}

Similarly, we define the accretion power $\Eacc = \EAmec + \EAmag + \EAthm$ as the sum of the mechanic, magnetic, and 
thermal energies,
\begin{equation} \label{eq:accpow}
\EAmec = - 2 \int_{S_1}^{S_R} \left(\frac{\Vtot^2}{2}+\Grav \right) \rho \Vtot_p\cdot \dS,\quad
\EAmag = - 2 \int_{S_1}^{S_R} \bit{E}\times\Btot\cdot \dS, \quad
\EAthm = - 2 \int_{S_1}^{S_R} \frac{\gamma}{\gamma-1} P \Vtot_p\cdot \dS,
\end{equation}
and the jet power $\Ejet = \EJkin + \EJgrv + \EJmag + \EJthm$ with
\begin{equation}
\EJkin = - 2 \int_{S_S} \frac{\Vtot^2}{2} \rho\Vtot_p\cdot \dS, \quad
\EJgrv = - 2 \int_{S_S} \Grav \rho\Vtot_p\cdot \dS, \quad 
\end{equation}
\begin{equation}
\EJmag = - 2 \int_{S_S} \bit{E}\times\Btot \cdot \dS,\quad
\EJthm = - 2 \int_{S_S} \frac{\gamma}{\gamma-1} P\Vtot_p\cdot \dS.
\end{equation}
%


\bibliographystyle{apj}


\begin{thebibliography}{44}
\expandafter\ifx\csname natexlab\endcsname\relax\def\natexlab#1{#1}\fi

\bibitem[{{Bai} \& {Stone}(2013)}]{2013ApJ...767...30B}
{Bai}, X.-N. \& {Stone}, J.~M. 2013, \apj, 767, 30

\bibitem[{{Balbus} \& {Hawley}(1991)}]{1991ApJ...376..214B}
{Balbus}, S.~A. \& {Hawley}, J.~F. 1991, \apj, 376, 214

\bibitem[{{Banerjee} {et~al.}(2007){Banerjee}, {Klessen}, \&
  {Fendt}}]{2007ApJ...668.1028B}
{Banerjee}, R., {Klessen}, R.~S., \& {Fendt}, C. 2007, \apj, 668, 1028

\bibitem[{{Beckwith} {et~al.}(2011){Beckwith}, {Armitage}, \&
  {Simon}}]{2011MNRAS.416..361B}
{Beckwith}, K., {Armitage}, P.~J., \& {Simon}, J.~B. 2011, \mnras, 416, 361

\bibitem[{{Blandford} \& {Payne}(1982)}]{1982MNRAS.199..883B}
{Blandford}, R.~D. \& {Payne}, D.~G. 1982, \mnras, 199, 883

\bibitem[{{Cabrit} {et~al.}(1990){Cabrit}, {Edwards}, {Strom}, \&
  {Strom}}]{1990ApJ...354..687C}
{Cabrit}, S., {Edwards}, S., {Strom}, S.~E., \& {Strom}, K.~M. 1990, \apj, 354,
  687

\bibitem[{{Camenzind}(1990)}]{1990RvMA....3..234C}
{Camenzind}, M. 1990, in Reviews in Modern Astronomy, Vol.~3, Reviews in Modern
  Astronomy, ed. {G.~Klare}, 234--265

\bibitem[{{Campbell}(2009)}]{2009MNRAS.392..271C}
{Campbell}, C.~G. 2009, \mnras, 392, 271

\bibitem[{{Casse} \& {Ferreira}(2000)}]{2000A&A...353.1115C}
{Casse}, F. \& {Ferreira}, J. 2000, \aap, 353, 1115

\bibitem[{{Casse} \& {Keppens}(2002)}]{2002ApJ...581..988C}
{Casse}, F. \& {Keppens}, R. 2002, \apj, 581, 988

\bibitem[{{Casse} \& {Keppens}(2004)}]{2004ApJ...601...90C}
---. 2004, \apj, 601, 90

\bibitem[{{Colella} \& {Woodward}(1984)}]{1984JCoPh..54..174C}
{Colella}, P. \& {Woodward}, P.~R. 1984, Journal of Computational Physics, 54,
  174

\bibitem[{{Fendt} \& {Sheikhnezami}(2013)}]{2013ApJ...774...12F}
{Fendt}, C. \& {Sheikhnezami}, S. 2013, \apj, 774, 12

\bibitem[{{Ferreira}(1997)}]{1997A&A...319..340F}
{Ferreira}, J. 1997, \aap, 319, 340

\bibitem[{{Ferreira} \& {Casse}(2013)}]{2013MNRAS.428..307F}
{Ferreira}, J. \& {Casse}, F. 2013, \mnras, 428, 307

\bibitem[{{Ferreira} \& {Pelletier}(1995)}]{1995A&A...295..807F}
{Ferreira}, J. \& {Pelletier}, G. 1995, \aap, 295, 807

\bibitem[{{Fromang}(2013)}]{2013EAS....62...95F}
{Fromang}, S. 2013, in EAS Publications Series, Vol.~62, EAS Publications
  Series, 95--142

\bibitem[{{Gaibler} {et~al.}(2012){Gaibler}, {Khochfar}, {Krause}, \&
  {Silk}}]{2012MNRAS.425..438G}
{Gaibler}, V., {Khochfar}, S., {Krause}, M., \& {Silk}, J. 2012, \mnras, 425,
  438

\bibitem[{{Goedbloed} {et~al.}(2004){Goedbloed}, {Beli{\"e}n}, {Holst}, \&
  {Keppens}}]{2004PhPl...11.4332G}
{Goedbloed}, J.~P., {Beli{\"e}n}, A.~J.~C., {Holst}, B.~V.~D., \& {Keppens}, R.
  2004, Physics of Plasmas, 11, 4332

\bibitem[{{Gressel}(2010)}]{2010MNRAS.405...41G}
{Gressel}, O. 2010, \mnras, 405, 41

\bibitem[{{Hawley} {et~al.}(1995){Hawley}, {Gammie}, \&
  {Balbus}}]{1995ApJ...440..742H}
{Hawley}, J.~F., {Gammie}, C.~F., \& {Balbus}, S.~A. 1995, \apj, 440, 742

\bibitem[{{Johansen} \& {Levin}(2008)}]{2008A&A...490..501J}
{Johansen}, A. \& {Levin}, Y. 2008, \aap, 490, 501

\bibitem[{{Keppens} {et~al.}(2002){Keppens}, {Casse}, \&
  {Goedbloed}}]{2002ApJ...569L.121K}
{Keppens}, R., {Casse}, F., \& {Goedbloed}, J.~P. 2002, \apjl, 569, L121

\bibitem[{{King} {et~al.}(2007){King}, {Pringle}, \&
  {Livio}}]{2007MNRAS.376.1740K}
{King}, A.~R., {Pringle}, J.~E., \& {Livio}, M. 2007, \mnras, 376, 1740

\bibitem[{{K{\"o}nigl} \& {Salmeron}(2011)}]{2011ppcd.book..283K}
{K{\"o}nigl}, A. \& {Salmeron}, R. {The Effects of Large-Scale Magnetic Fields
  on Disk Formation and Evolution}, ed. {Garcia, P.~J.~V.}, 283--352

\bibitem[{{Lesur} \& {Longaretti}(2009)}]{2009A&A...504..309L}
{Lesur}, G. \& {Longaretti}, P.-Y. 2009, \aap, 504, 309

\bibitem[{{Li}(1995)}]{1995ApJ...444..848L}
{Li}, Z. 1995, \apj, 444, 848

\bibitem[{{Londrillo} \& {del Zanna}(2004)}]{2004JCoPh.195...17L}
{Londrillo}, P. \& {del Zanna}, L. 2004, Journal of Computational Physics, 195,
  17

\bibitem[{{Lubow} {et~al.}(1994){Lubow}, {Papaloizou}, \&
  {Pringle}}]{1994MNRAS.268.1010L}
{Lubow}, S.~H., {Papaloizou}, J.~C.~B., \& {Pringle}, J.~E. 1994, \mnras, 268,
  1010

\bibitem[{{Meliani} {et~al.}(2006){Meliani}, {Casse}, \&
  {Sauty}}]{2006A&A...460....1M}
{Meliani}, Z., {Casse}, F., \& {Sauty}, C. 2006, \aap, 460, 1

\bibitem[{{Mignone} {et~al.}(2007){Mignone}, {Bodo}, {Massaglia}, {Matsakos},
  {Tesileanu}, {Zanni}, \& {Ferrari}}]{2007ApJS..170..228M}
{Mignone}, A., {Bodo}, G., {Massaglia}, S., {Matsakos}, T., {Tesileanu}, O.,
  {Zanni}, C., \& {Ferrari}, A. 2007, \apjs, 170, 228

\bibitem[{{Murphy} {et~al.}(2010){Murphy}, {Ferreira}, \&
  {Zanni}}]{2010A&A...512A..82M}
{Murphy}, G.~C., {Ferreira}, J., \& {Zanni}, C. 2010, \aap, 512, A82+

\bibitem[{{Parkin} \& {Bicknell}(2013)}]{2013ApJ...763...99P}
{Parkin}, E.~R. \& {Bicknell}, G.~V. 2013, \apj, 763, 99

\bibitem[{{Pelletier} \& {Pudritz}(1992)}]{1992ApJ...394..117P}
{Pelletier}, G. \& {Pudritz}, R.~E. 1992, \apj, 394, 117

\bibitem[{{Pudritz} \& {Norman}(1983)}]{1983ApJ...274..677P}
{Pudritz}, R.~E. \& {Norman}, C.~A. 1983, \apj, 274, 677

\bibitem[{{Pudritz} {et~al.}(2007){Pudritz}, {Ouyed}, {Fendt}, \&
  {Brandenburg}}]{2007prpl.conf..277P}
{Pudritz}, R.~E., {Ouyed}, R., {Fendt}, C., \& {Brandenburg}, A. 2007,
  Protostars and Planets V, 277

\bibitem[{{Sauty} \& {Tsinganos}(1994)}]{1994A&A...287..893S}
{Sauty}, C. \& {Tsinganos}, K. 1994, \aap, 287, 893

\bibitem[{{Shakura} \& {Sunyaev}(1973)}]{1973shakuraetal}
{Shakura}, N.~I. \& {Sunyaev}, R.~A. 1973, \aap, 24, 337

\bibitem[{{Sheikhnezami} {et~al.}(2012){Sheikhnezami}, {Fendt}, {Porth},
  {Vaidya}, \& {Ghanbari}}]{2012ApJ...757...65S}
{Sheikhnezami}, S., {Fendt}, C., {Porth}, O., {Vaidya}, B., \& {Ghanbari}, J.
  2012, \apj, 757, 65

\bibitem[{{Simon} {et~al.}(2012){Simon}, {Beckwith}, \&
  {Armitage}}]{2012MNRAS.422.2685S}
{Simon}, J.~B., {Beckwith}, K., \& {Armitage}, P.~J. 2012, \mnras, 422, 2685

\bibitem[{{Stone} {et~al.}(1996){Stone}, {Hawley}, {Gammie}, \&
  {Balbus}}]{1996ApJ...463..656S}
{Stone}, J.~M., {Hawley}, J.~F., {Gammie}, C.~F., \& {Balbus}, S.~A. 1996,
  \apj, 463, 656

\bibitem[{{Tzeferacos} {et~al.}(2009){Tzeferacos}, {Ferrari}, {Mignone},
  {Zanni}, {Bodo}, \& {Massaglia}}]{2009MNRAS.400..820T}
{Tzeferacos}, P., {Ferrari}, A., {Mignone}, A., {Zanni}, C., {Bodo}, G., \&
  {Massaglia}, S. 2009, \mnras, 400, 820

\bibitem[{{Tzeferacos} {et~al.}(2013){Tzeferacos}, {Ferrari}, {Mignone},
  {Zanni}, {Bodo}, \& {Massaglia}}]{2013MNRAS.428.3151T}
---. 2013, \mnras, 428, 3151

\bibitem[{{Zanni} {et~al.}(2007){Zanni}, {Ferrari}, {Rosner}, {Bodo}, \&
  {Massaglia}}]{2007A&A...469..811Z}
{Zanni}, C., {Ferrari}, A., {Rosner}, R., {Bodo}, G., \& {Massaglia}, S. 2007,
  \aap, 469, 811

\end{thebibliography}
\end{document}